\documentclass[useAMS,usenatbib]{mn2e}
\usepackage{amsmath}
\usepackage{amssymb}
\usepackage{amsfonts}
\usepackage{xcolor}
\usepackage{graphicx}
\usepackage{enumerate}
\usepackage[hyperindex,breaklinks=true, colorlinks, citecolor=blue]{hyperref} 
\usepackage{float}

\newcommand{\wmap}{{\it WMAP\/}}  
\def\deg{\ifmmode^\circ\else$^\circ$\fi}
\def\ba{\begin{eqnarray}}
\def\ea{\end{eqnarray}}
\def\be{\begin{equation}}
\def\ee{\end{equation}}

\def\bdelta{{\delta }}

\makeatletter
\def\ref@jnl#1{{\rmfamily#1}}%
\newcommand\aj{\ref@jnl{AJ}}%
\newcommand\araa{\ref@jnl{ARA\&A}}%
\newcommand\apj{\ref@jnl{ApJ}}%
\newcommand\apjl{\ref@jnl{ApJ}}%
\newcommand\apjs{\ref@jnl{ApJS}}%
\newcommand\apss{\ref@jnl{Ap\&SS}}%
\newcommand\aap{\ref@jnl{A\&A}}%
\newcommand\aapr{\ref@jnl{A\&A~Rev.}}%
\newcommand\aaps{\ref@jnl{A\&AS}}%
\newcommand\baas{\ref@jnl{BAAS}}%
\newcommand\memras{\ref@jnl{MmRAS}}%
\newcommand\mnras{\ref@jnl{MNRAS}}%
\newcommand\pra{\ref@jnl{Phys.~Rev.~A}}%
\newcommand\prb{\ref@jnl{Phys.~Rev.~B}}%
\newcommand\prc{\ref@jnl{Phys.~Rev.~C}}%
\newcommand\prd{\ref@jnl{Phys.~Rev.~D}}%
\newcommand\pre{\ref@jnl{Phys.~Rev.~E}}%
\newcommand\prl{\ref@jnl{Phys.~Rev.~Lett.}}%
\newcommand\pasp{\ref@jnl{PASP}}%
\newcommand\pasj{\ref@jnl{PASJ}}%
\newcommand\ssr{\ref@jnl{Space~Sci.~Rev.}}%
\newcommand\nat{\ref@jnl{Nature}}%
\newcommand\iaucirc{\ref@jnl{IAU~Circ.}}%
\newcommand\aplett{\ref@jnl{Astrophys.~Lett.}}%
\newcommand\apspr{\ref@jnl{Astrophys.~Space~Phys.~Res.}}%
\newcommand\nphysa{\ref@jnl{Nucl.~Phys.~A}}%
\newcommand\physrep{\ref@jnl{Phys.~Rep.}}%
\newcommand\planss{\ref@jnl{Planet.~Space~Sci.}}%
\newcommand\procspie{\ref@jnl{Proc.~SPIE}}%

\makeatletter
\newcommand\footnoteref[1]{\protected@xdef\@thefnmark{\ref{#1}}\@footnotemark}
\makeatother

\title[An improved source-subtracted and destriped 408\,MHz all-sky map]{An improved source-subtracted and destriped 408\,MHz all-sky map}
\author[M. Remazeilles et al.]{M. Remazeilles,$\!$\thanks{\url{E-mail: mathieu.remazeilles@manchester.ac.uk}}$^1$ C. Dickinson,$\!$\thanks{\url{E-mail: clive.dickinson@manchester.ac.uk}}$^1$ A. J. Banday,$\!^{2,3}$ M.-A. Bigot-Sazy,$\!^{1}$ T. Ghosh$^{4}$ \\
$^1$Jodrell Bank Centre for Astrophysics, Alan Turing Building, School of Physics \& Astronomy, The University of Manchester, \\
Oxford Road, Manchester, M13 9PL, U.K. \\
$^2$Universit\'{e} de Toulouse, UPS-OMP, IRAP, F-31028 Toulouse cedex 4, France \\
$^3$CNRS, IRAP, 9 Av. colonel Roche, BP 44346, F-31028 Toulouse cedex 4, France \\
$^4$Institut d'Astrophysique Spatiale, CNRS (UMR8617) Universit\'{e} Paris-Sud 11, B\^{a}timent 121, Orsay, France
}
\begin{document}


\setlength{\topmargin}{-15mm}

\pagerange{\pageref{firstpage}--\pageref{lastpage}} \pubyear{2002}

\maketitle

\label{firstpage}

\begin{abstract}
The all-sky 408\,MHz map of Haslam et al. is one the most important
total-power radio surveys. It has been widely used to study diffuse
synchrotron radiation from our Galaxy and as a template to remove
foregrounds in cosmic microwave background data. However, there are a
number of issues associated with it that must be dealt with, including
large-scale striations and contamination from extragalactic radio
sources. 

We have re-evaluated and re-processed the rawest data available to
produce a new and improved 408\,MHz all-sky map. We first quantify the
positional accuracy ($\approx 7$\,arcmin) and effective beam
($56.0\pm1.0$\,arcmin) of the four individual surveys from which it
was assembled. Large-scale striations associated with $1/f$ noise in
the scan direction are reduced to a level $\ll 1$\,K using a
Fourier-based filtering technique. The most important improvement
results from the   removal of extragalactic sources. We have used an
iterative combination of two techniques -- two-dimensional Gaussian
fitting and minimum curvature spline surface inpainting -- to remove the
brightest sources ($\gtrsim 2$\,Jy), which provides a significant
improvement over previous versions of the map. We quantify the impact
with power spectra and a template fitting analysis of foregrounds to the {\it WMAP} data.

The new map is publicly available and is recommended as the template
of choice for large-scale diffuse Galactic synchrotron emission. We
also provide a higher resolution
map with small-scale fluctuations added, assuming a power-law angular
power spectrum down to the pixel scale (1.7\,arcmin). This should prove useful in simulations used for studying the feasibility of detecting HI
fluctuations from the Epoch of Reionization.    
\end{abstract}


\section{Introduction}\label{sec:intro}

The total-power radio sky has been surveyed at a number of frequencies
using single-dish radio telescopes.  A compilation of large-area radio
maps from 10\,MHz to 94\,GHz was produced  by
\cite{deOliveiraCosta2008}, who used them to produce a Global Sky
Model (GSM) of diffuse Galactic radio emission. These maps can be used
to study the diffuse Galactic synchrotron radiation from the
interstellar medium
\citep{Beuermann1985,Reich1988,Koo1992,Davies1996,Platania1998,Platania2003,Bennett2003a,Finkbeiner2004,Jaffe2011,Strong2011,Orlando2013,Mertsch2013},
as well as Galactic sources such as supernova remnants
\citep{Sushch2014,Xiao2014} and HII regions
\citep{Wendker1991,Foster2001,Xu2013}. Perhaps more importantly, radio
maps have been used extensively as a synchrotron template for removing
foreground emission in cosmic microwave background (CMB) data 
\citep{Netterfield1997,Bennett1992,Bennett2003b,Kogut1996,Banday2003,Bennett2003a,Bennett2013}. More
recently, they have been used in simulations for HI intensity mapping
experiments, both at moderate redshifts
\citep{Ansari2012,Battye2013,Wolz2014} and at high redshift from the
Epoch of Reionization
\citep{Jelic2008,Parsons2012,Moore2013,Shaw2014}.

\begin{table*}
\caption{Summary of the different versions of the $408$\,MHz Haslam map.}
\label{tab:haslam_versions}
\begin{tabular}{|l|l|l|l|l|}
\hline
Map version map & Location  & Characteristics & Coordinates & Post-processing \\
\hline
Bonn ECP & MPIfR Survey Sampler website\footnoteref{url:bonn}  & $1080\times 540$ pixel grid & Celestial B1950 & None\\
LAMBDA ECP map & NCSA Digital Image Library  & $1024\times 512$ pixel grid &  Galactic & None\\
Unfiltered LAMBDA map  (HAS82) & LAMBDA website\footnoteref{url:lambda} & {\tt HEALPix} $N_{\rm side}=512$ & Galactic & None\\
LAMBDA dsds map (HAS03) & LAMBDA website\footnoteref{url:lambda} & {\tt HEALPix} $N_{\rm side}=512$ & Galactic & Destriping/desourcing\\
\protect\cite{Davies1996} map &R.\,A.~Watson (priv. comm.)  &$1080\times540$ pixel grid &Celestial B1950 & Destriping \\
\protect\cite{Platania2003} map &VizieR hosted by CDS\footnote{\label{footnote:cds}\url{http://cdsarc.u-strasbg.fr}} &{\tt HEALPix} $N_{\rm side}=512$ &Galactic & Destriping/desourcing\\
Finkbeiner map &D. Finkbeiner (priv. comm.)  &{\tt HEALPix} $N_{\rm side}=256$ &Galactic & Destriping/desourcing\\
This work (HAS14) & LAMBDA website\footnoteref{url:lambda} & {\tt HEALPix} $N_{\rm side}=512$ & Galactic & Destriping/desourcing\\  
\hline
\end{tabular}
 \end{table*}

The most famous and widely used radio map, is the 408\,MHz all-sky map
of \cite{Haslam1981,Haslam1982}, hereafter the Haslam map, which has
accrued over 1000 citations to-date.  The data were taken in the 1960s
and 1970s as part of four separate surveys using three of the worlds
largest single-dish telescopes. However, only a processed combined map
produced by \cite{Haslam1982} is readily available. Remarkably, it is
still widely used
\citep{Macellari2011,Planck2011_ame,Guzman2011,Ghosh2012,Peel2012,Lu2012,
  Iacobelli2013, Planck2013_haze} over 30 years since the full-sky
version was first released. Although total-power radio surveys have
been made more recently at 1.4\,GHz \citep{Reich1986,Reich2001} and
2.3\,GHz \citep{Jonas1998}, and new data at higher frequencies are
forthcoming \citep{King2014,Hoyland2012,Tello2013}, the Haslam map is
likely to remain a widely used template of synchrotron emission.

However, the sky at frequencies of $\sim 100$-1400\,MHz is still
poorly understood, and the fidelity of the synchrotron template is
rather poor compared to modern CMB data. Despite the application of
various post-processing techniques, the current Haslam map still
suffers from strong source residuals and other artefacts including
significant striations caused by $1/f$ noise in the scan direction. 
Several attempts have been made to correct for these effects, as described in \cite{Davies1996,Platania2003,Bennett2003a}. 

The most widely used version of the Haslam map is the destriped and source-subtracted version produced by
the {\it WMAP} team \citep{Bennett2003a} using data from the NCSA ADIL
and made available through NASA's Legacy Archive for Microwave
Background Analysis (LAMBDA)
website\footnote{\url{http://lambda.gsfc.nasa.gov}\label{url:lambda}.}; specifically, the map contained in the file {\tt lambda\_haslam408\_dsds.fits}, available from the LAMBDA website{\footnote{\url{http://lambda.gsfc.nasa.gov/product/foreground/fg_haslam_get.cfm}.}, which hereafter we refer to as the HAS03 map. However, closer inspection of this map
reveals significant residuals from the source removal process, which
add significant small-scale power at high Galactic
latitudes. Furthermore, additional artefacts have been introduced due
to multiple regridding steps during post-processing (see
Section~\ref{subsec:lambda}). 

In this paper, we produce a new clean 408\,MHz radio map with fewer
artefacts and reduced systematics. In particular, the strong source
residuals have been minimised and additional pixelization artefacts
have been negated by reprocessing the Equidistant Cylindrical
Projection (ECP) raw map downloaded from the Max Planck Institute for
Radioastronomy Survey Sampler
website\footnote{\label{url:bonn}\url{http://www3.mpifr-bonn.mpg.de/survey.html}.}. We
use a multi-stage source processing based on both two-dimensional
Gaussian fitting and minimum curvature spline surface inpainting. We
also determine the effective beam of the 408\,MHz sky map and estimate
the positional accuracy of the data. The goal is that the new 408\,MHz
template will be used both for accurate synchrotron modelling and as a
robust template for component separation. 

The paper is organised as follows. In Section~\ref{sec:data}, we
describe the current version of the Haslam map and its associated
problems. In Section~\ref{sec:proc} we perform a new characterisation
and processing of the raw map, including an estimation of the beam in
Section~\ref{subsec:beam} and positional offsets in
Section~\ref{subsec:offsets}. We then discuss the destriping of the
map in Section~\ref{sec:destriping}, and in
Section~\ref{sec:desourcing} we present our two desourcing approaches
and the corresponding results of source removal on both simulations
and data. 
We discuss the scientific exploitation of our product in
Section~\ref{sec:discuss}, and conclude in Section~\ref{sec:concl},
where we describe a website dedicated to the release of the new Haslam
map.

\section{Data description}\label{sec:data}

\subsection{The raw data}\label{subsec:raw}

\begin{figure*}
  \begin{center}
    \includegraphics[width=\textwidth]{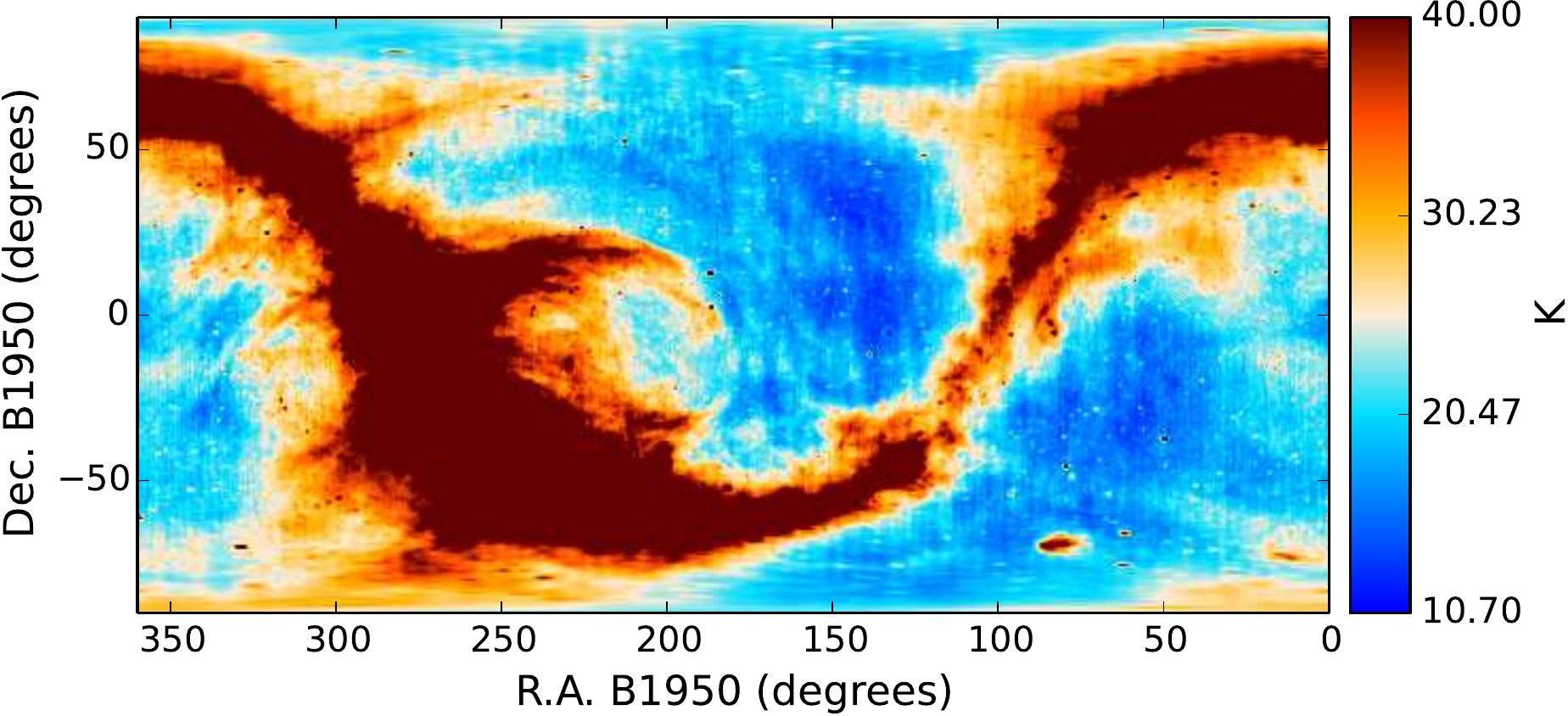}\\
 \end{center}
\caption{
  Mapping of the raw Haslam map at
  408\,MHz from the Bonn Survey Sampler website ($1080\times 540$
  pixel ECP map in B1950 Celestial coordinates). The map covers the
  entire sky with RA$=0^{\circ}$ on the far right-hand side, increasing
  to the left, and the celestial poles at the top
  (Dec.$=+90^{\circ}$) and bottom (Dec.$=-90^{\circ}$). Vertical
  striations as well as numerous extragalactic radio sources can be
  seen to contaminate the diffuse Galactic synchrotron emission.}
\label{Fig:ecp}
\end{figure*}

The 408\,MHz all-sky map of \citet{Haslam1982} is an atlas combining
data from four different partial surveys of the sky, using the Jodrell
Bank Mk-I 76-m, the Effelsberg 100-m, the Parkes 64-m, and the Jodrell
Bank Mk-IA 76-m telescopes at four different epochs between 1965 and
1978 \citep{Haslam1970,Haslam1974,Haslam1981}. The surveys were made
by scanning the telescopes in elevation, which, through the rotation
of the earth, allowed the zero levels to be set in a consistent
manner. The other advantage is that contamination from the sky and
ground will be similar for a given declination, allowing their
removal. The angular resolution of each survey ranges from 37\,arcmin
to 51\,arcmin, with the full-sky map smoothed to a common resolution,
stated to be 51\,arcmin ($0.\!^{\circ}85$). Positional accuracies are
claimed to be accurate to $\approx 1$\,arcmin. We believe this to be
an underestimate given the discussions in the aforementioned papers
and the likely pointing accuracy of telescope control systems in those
years. The overall calibration, including zero levels, was achieved by
comparison with a previous 404\,MHz survey
\citep{Pauliny-Toth1962}. The absolute calibration is thought to be
accurate to better than 10\,\% and the overall zero level to $\pm
3$\,K.

Several electronic versions of the Haslam map are in existence, as
listed in Table~\ref{tab:haslam_versions}. As far as we are
aware\footnote{The original raw data are thought to be available on
  tape and attempts are being made to retrieve these data via Glyn
  Haslam's family members.}, the rawest available data are provided
in the form of an Equidistant Cylindrical Projection (ECP) map, which
can be downloaded from the Bonn Survey Sampler
website\footnoteref{url:bonn} of the Max Planck Institute for
Radioastronomy. The ECP map is a $1080 \times 540$ pixel
grid\footnote{When downloading this map, it is possible to create a
  $1081\times541$ pixel map due to the way the script creates the
  pixel grid. In this case, the additional pixels are simply a
  repetition of some pixels and should be removed.} in Celestial
coordinates at Epoch B1950, with a pixel size of $0.\!^{\circ}33$. We
believe this is the original map created by \cite{Haslam1982} because
the minimum/maximum intensity scale is the same, and the pixel size is
identical. No additional interpolation has been applied, as verified
by the fact that the rotation matrix used in the Bonn survey sampler
script is diagonal. However, we note that this version has been corrected for the zero level mismatch between the Jodrell Bank telescope survey and the Effelsberg telescope survey, as described in \cite{Reich1988}. Other versions include destriped and desourced versions by \cite{Davies1996} and \cite{Platania2003}, as well as an unpublished version from D.~Finkbeiner (priv. comm.).
A {\tt HEALPix} interpolation, at $N_{\rm side} = 512$, of the ECP
grid is available on the  LAMBDA website under the name of ``Haslam
408 MHz map with no filtering'', which hereafter we refer to as the HAS82 map. According to the documentation
header, the beam full width at half maximum (FWHM) of the map is $51$
arcmin. The scanning technique used to make the original observations is well suited to mapping extended emission, but dos not provide a precise representation of point sources, as expained succinctly in \citet{Reich1988}. Because of additional smoothing from scanning and pointing errors, and the repixelization of the ECP grid to a {\tt HEALPix} map, we expect that the effective beam FWHM of the {\tt HEALPix} projection map will be slightly  larger than the original 51\,arcmin resolution.

However, an additional version of the raw data is available from the
NCSA (National Center for Supercomputing Applications) Image
Library. This is a $1024 \times 512$ pixel grid in Galactic
coordinates that assumes that the data corresponds to the Celestial
Epoch B1975. This is the version that has been post-processed and
interpolated by the {\it WMAP} team \citep{Bennett2003a} and provided
as the destriped and desourced Haslam map released on the LAMBDA
website (HAS03). This version of the 408\,MHz sky map has been widely used in
recent years, particularly as a synchrotron foreground template, and
will be discussed further in Section~\ref{subsec:lambda}.

In this paper, we start from the $1080 \times 540$ ECP map of the Bonn
survey sampler (see  Fig.~\ref{Fig:ecp}), hereafter referred to as the
"raw map", which we believe this to be the least manipulated data
available. It is mainly dominated by diffuse Galactic synchrotron
emission on large angular scales coming from the Galactic plane and
large radio loops and spurs at high Galactic latitudes. However, it is
clearly contaminated by strong extragalactic radio sources on the beam
scale, and exhibits strong baseline striping, due to correlated
low-frequency $1/f$ instrumental noise, with a typical amplitude of
$\pm 1$\,K \citep{Davies1996}.

\begin{figure}
  \begin{center}
    \includegraphics[width=0.5\columnwidth]{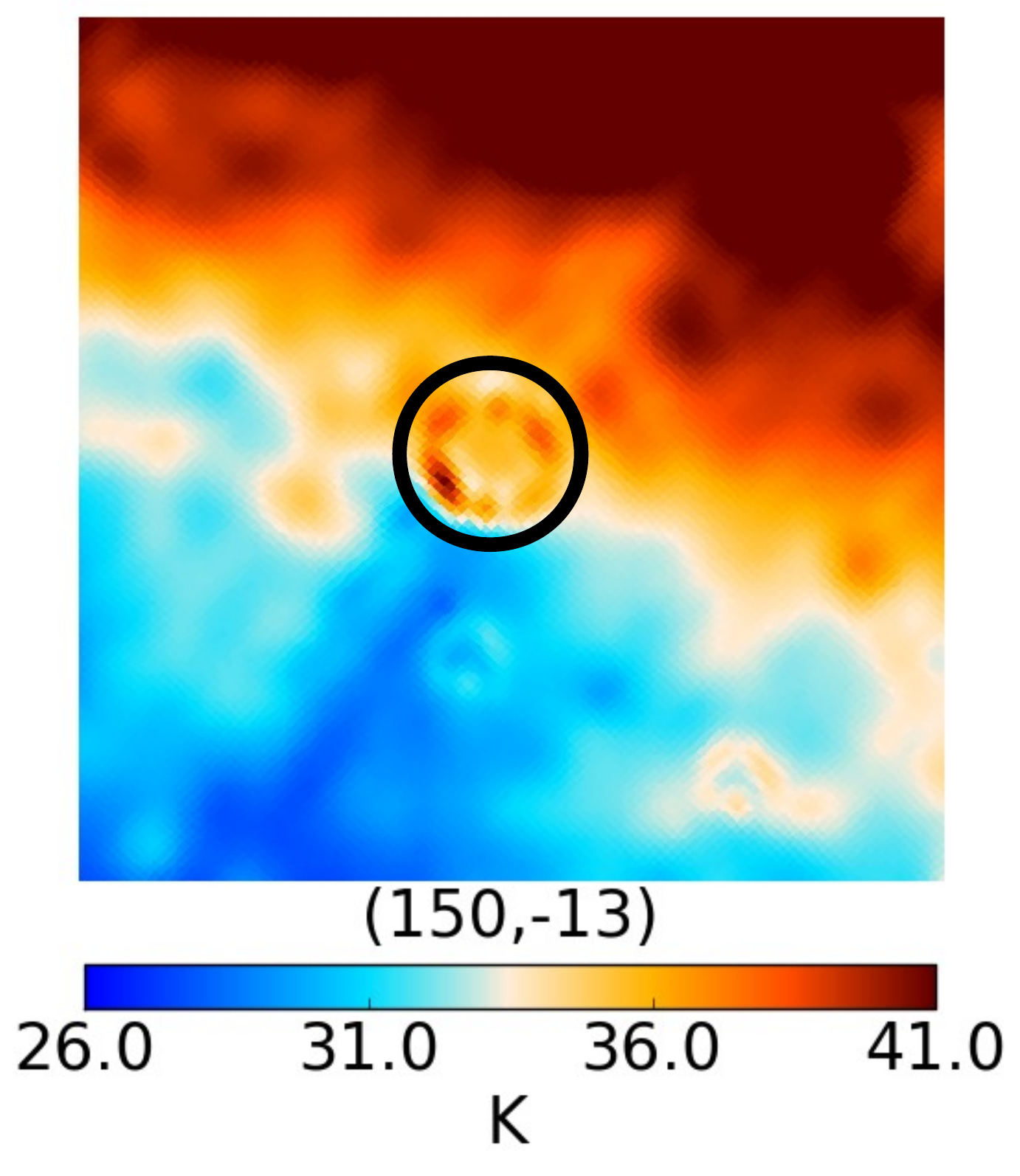}~
    \includegraphics[width=0.5\columnwidth]{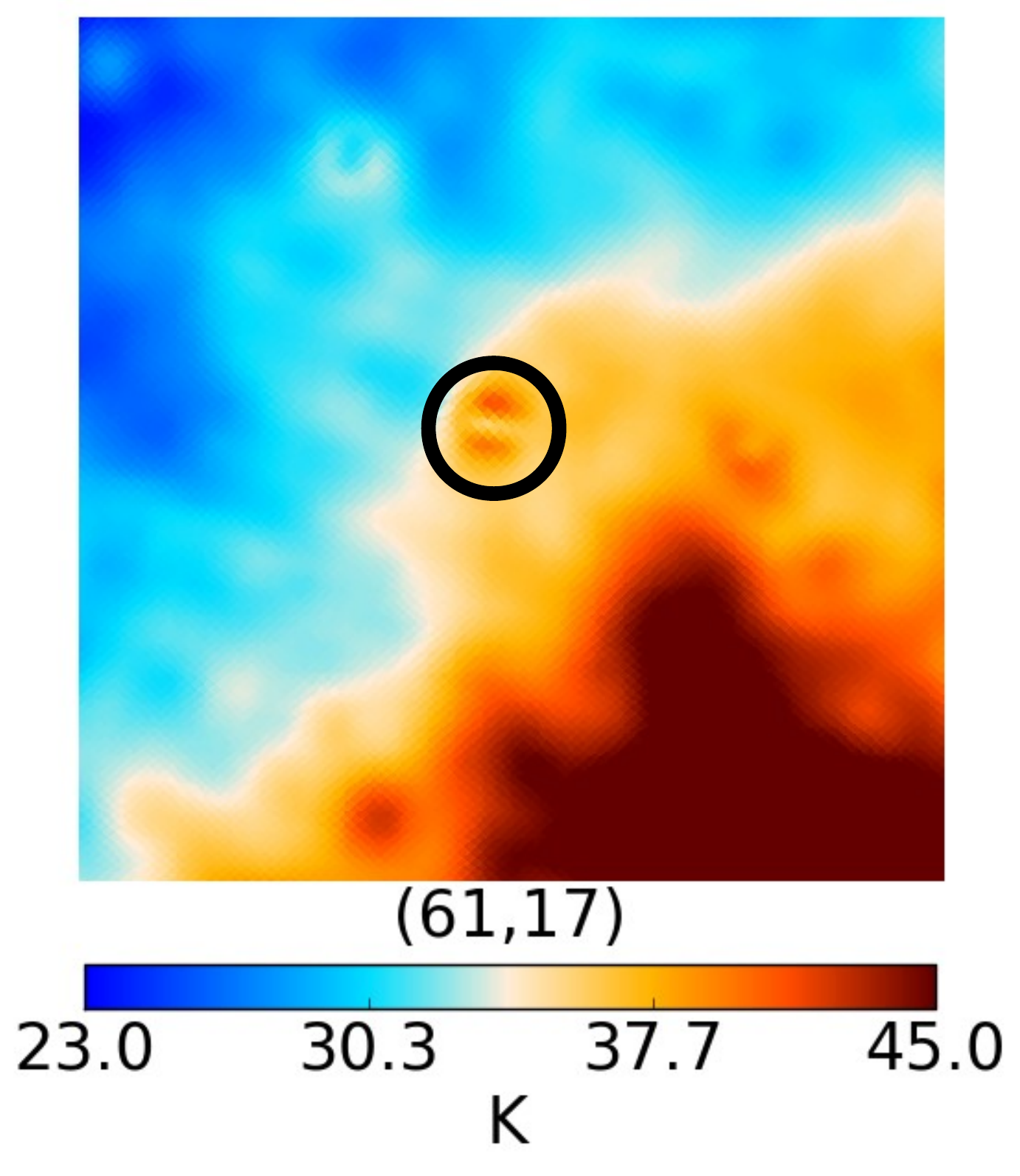}\\
    \includegraphics[width=0.5\columnwidth]{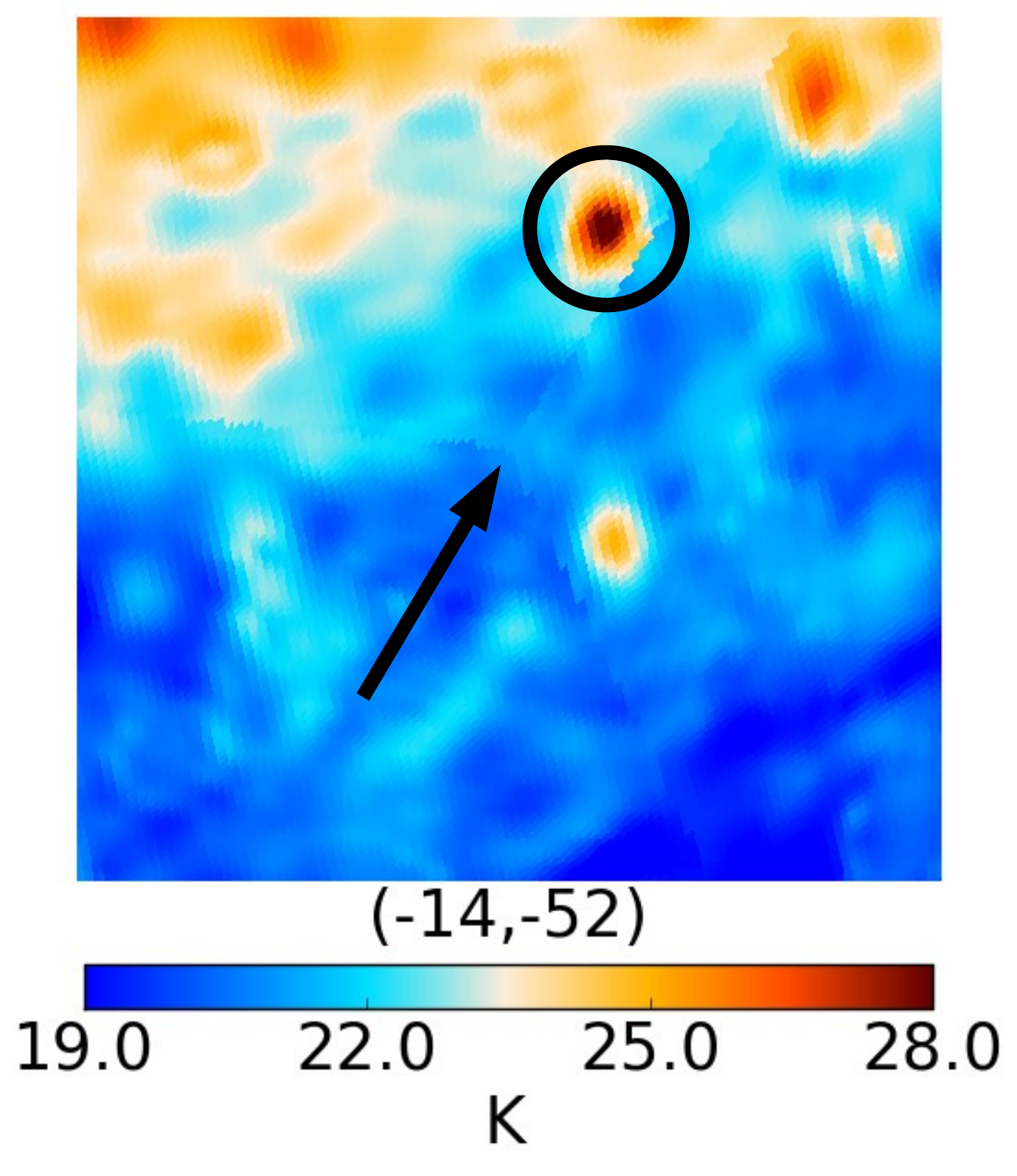}~
    \includegraphics[width=0.5\columnwidth]{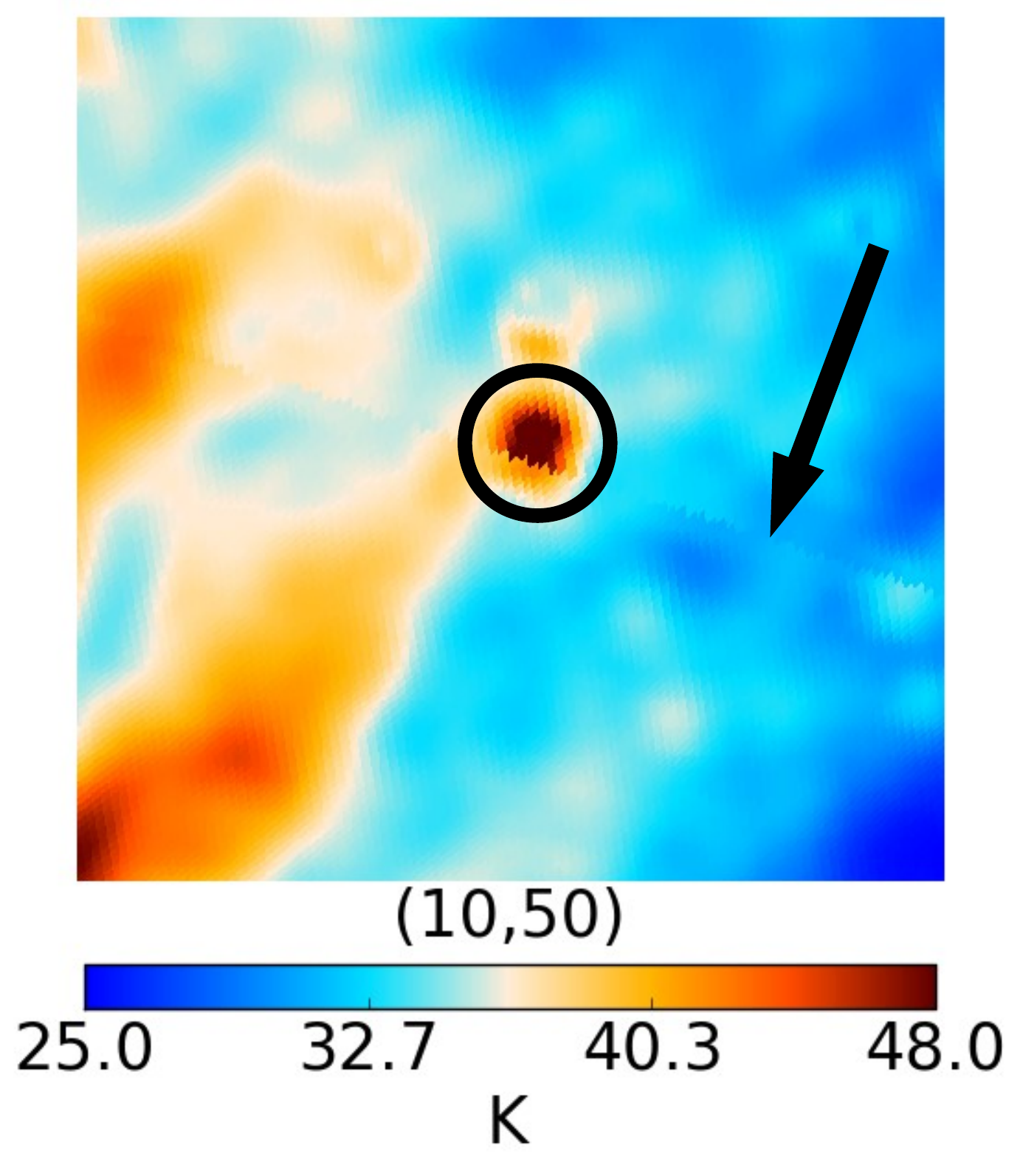}~
\end{center}
\caption{Four $12.5^\circ\times 12.5^\circ$ gnomonic projections of
  the LAMBDA post-processed (HAS03) version of the 408\,MHz Haslam
  map. All the gnomonic projections shown in the present work have Galactic latitude increasing upward and longitude to the left. Artefacts are evident, including source
  residuals (indicated by circles in the \emph{top left} and \emph{top
    right} panels), unsubtracted sources (indicated by circles in the
  \emph{bottom left} and \emph{bottom right}), and line features due to
  remapping to a {\tt quadcube} pixelization (indicated by arrows in
  the \emph{bottom left} and \emph{bottom right}).
}
\label{Fig:lambda-artefacts}
\end{figure}


\subsection{The LAMBDA post-processed Haslam map (HAS03)}
\label{subsec:lambda}

The WMAP post-processing consists of three steps: cubic interpolation
of the $1024 \times 512$ NCSA map from Galactic coordinates to
Celestial B1975 coordinates, destriping of the map in the Fourier
domain, and interpolation over the location of the strong sources in
the spatial domain (``inpainting'') for source removal. The destriped
and desourced map was then converted to a {\tt HEALPix} grid in
Galactic coordinates with $N_{\rm side} = 512$. Since it was assumed
for the first step that the mean epoch of observations was B1975, it
implies that the original $408$\,MHz data must have been precessed
from the B1950 epoch to B1975 in the image extracted from the NCSA
Digital Image library.

In the second step, the two-dimensional Fourier transform of the $1024
\times 512$ map was computed in order to remove in the Fourier domain
the low-frequency pattern that is responsible for the vertical stripes
in the spatial domain in Celestial coordinates. In order to avoid that the sources affect the Fourier transform, a basic source subtraction was applied on the data before destriping by median filtering the source pixels. In the third step, the
destriped map was reprojected onto a Quadrilateralized Sky Cube ({\tt
  quadcube}), i.e. the edges of a cube are projected onto a sphere so
that the sky is divided into six equal area faces
\citep{1992ASPC...25..379W}. The strong radio sources were then
isolated using iterative median filtering: pixels exceeding the local
background by more than $2\sigma$ were replaced by the local median,
and the process repeated iteratively. An inpainting technique was then
applied, using a locally weighted polynomial regression (LOESS) method
to interpolate through the locations of the sources.   

However, Fig.~\ref{Fig:lambda-artefacts} indicates that the HAS03
version still shows significant artefacts. These include source
residuals from imperfect desourcing, baseline offsets, and artefacts
due to a remapping of the raw data to the {\tt quadcube}
projection. In the top panels of the figure, we have drawn circles to
indicate spurious small-scale residual patterns arising from the
imperfect source processing. Moreover, some of the strongest radio
sources have not been filtered at all in the HAS03 map, as
shown in the bottom panels of the figure. In the bottom left panel, an
arrow has been drawn to highlight a cube vertex due to the {\tt
  quadcube} projection of the raw data utilised in the post-processing.

These numerous deficiencies in the HAS03 version of the Haslam map have motivated the present work: a new reprocessed version of the Haslam map is needed to better characterize the synchrotron foreground in the CMB/radio observations.

\begin{figure*}
  \begin{center}
    \includegraphics[width=\columnwidth]{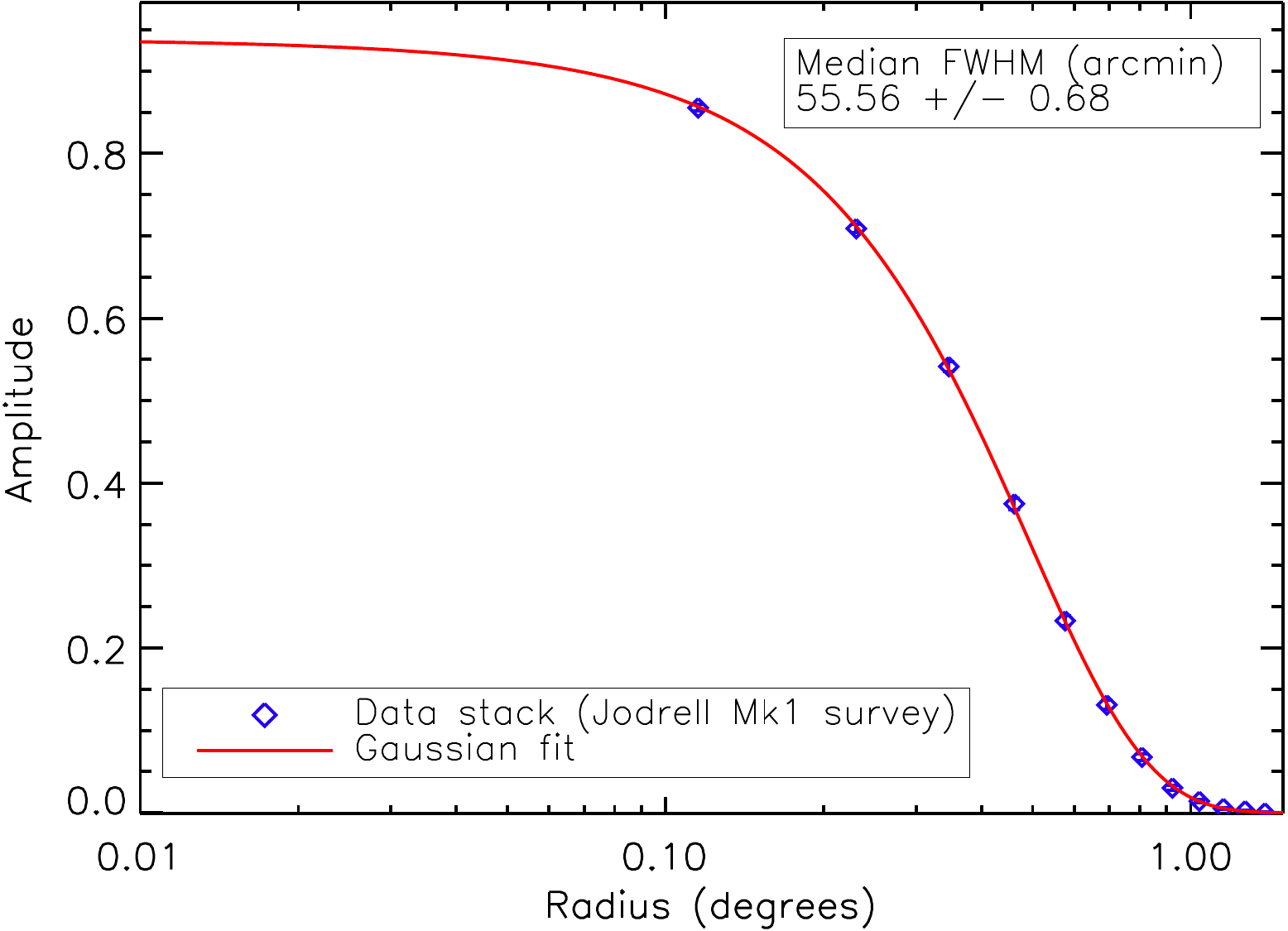}~
    \includegraphics[width=\columnwidth]{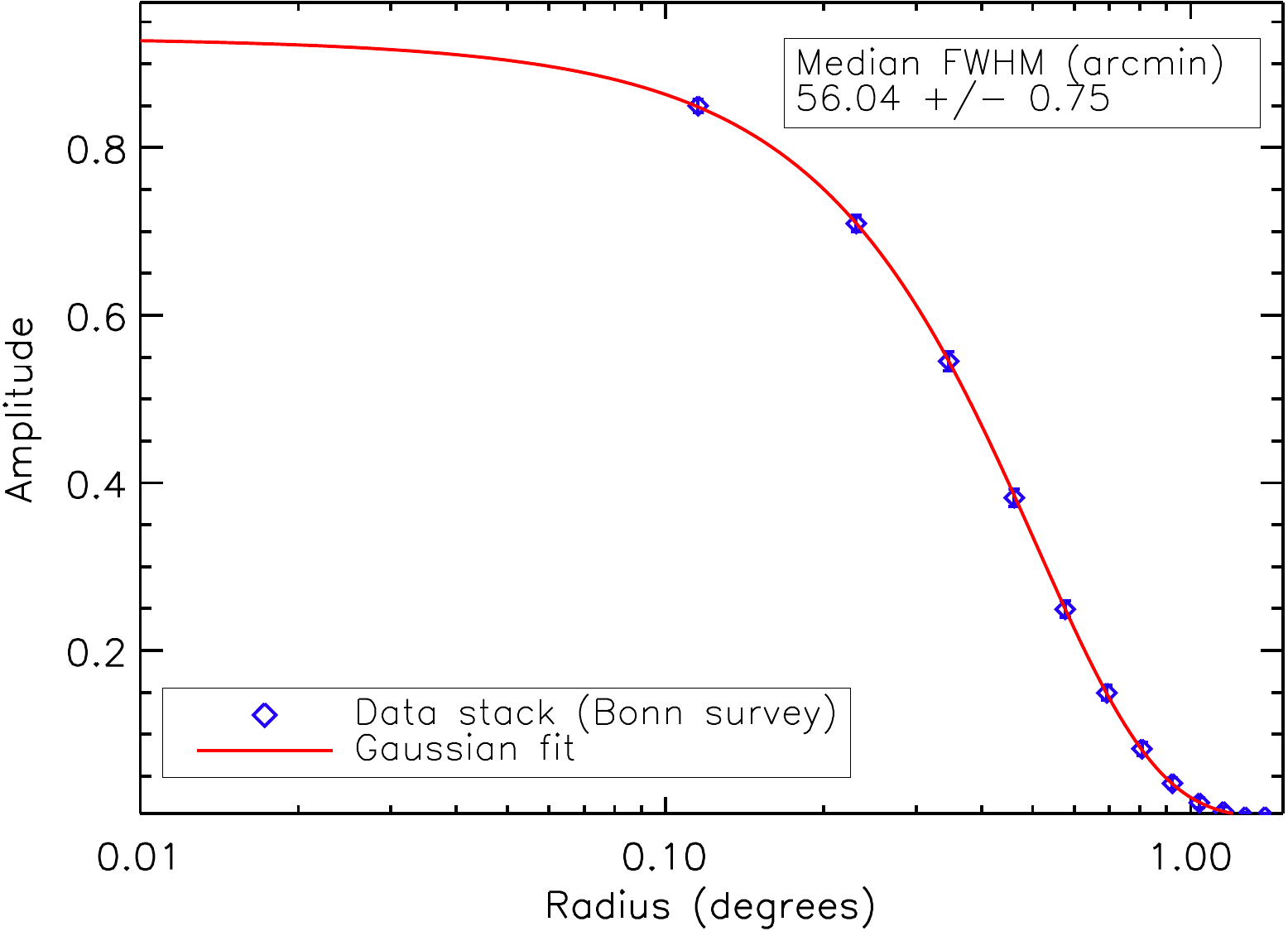}~\\
    \includegraphics[width=\columnwidth]{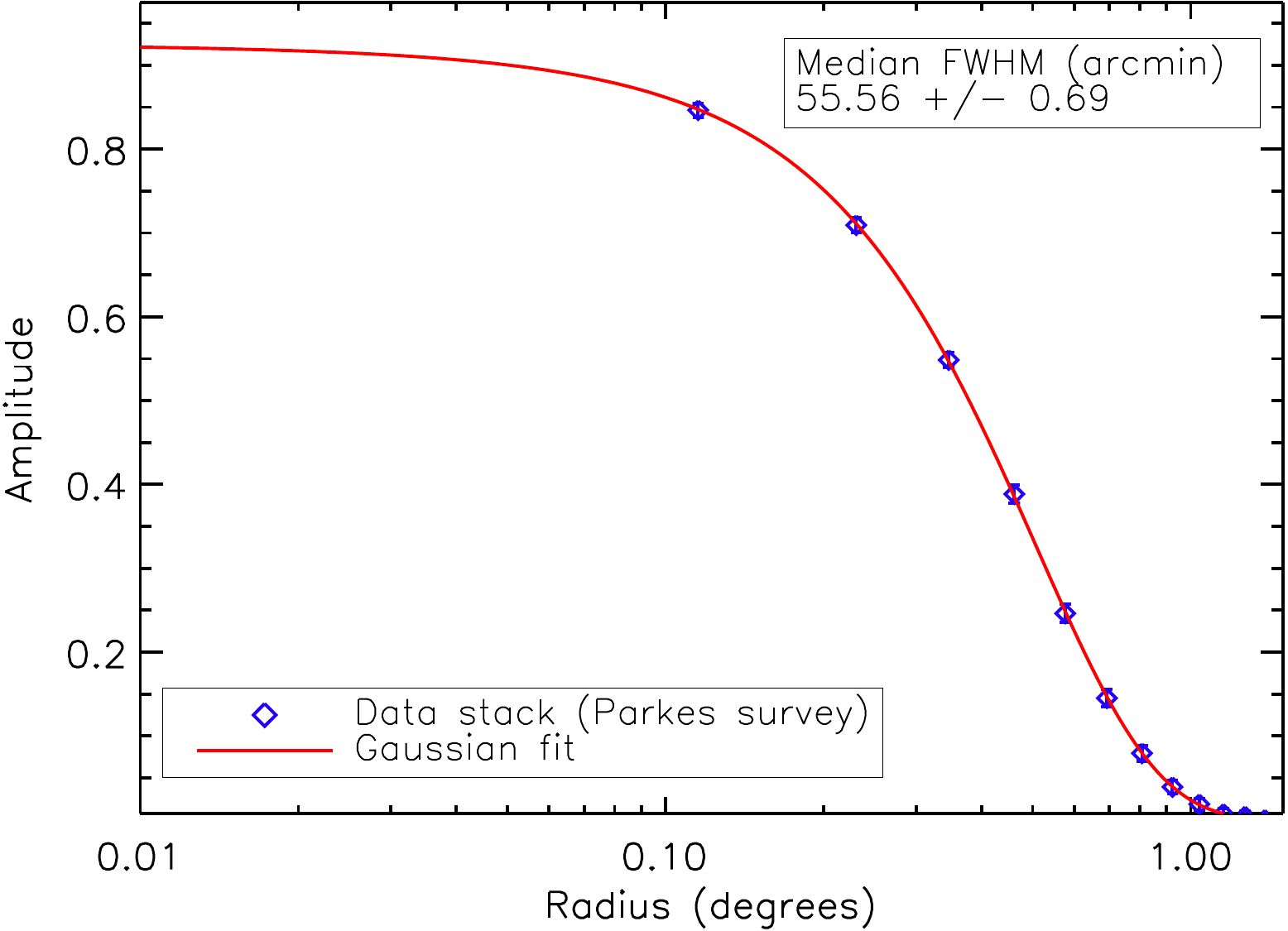}~
    \includegraphics[width=\columnwidth]{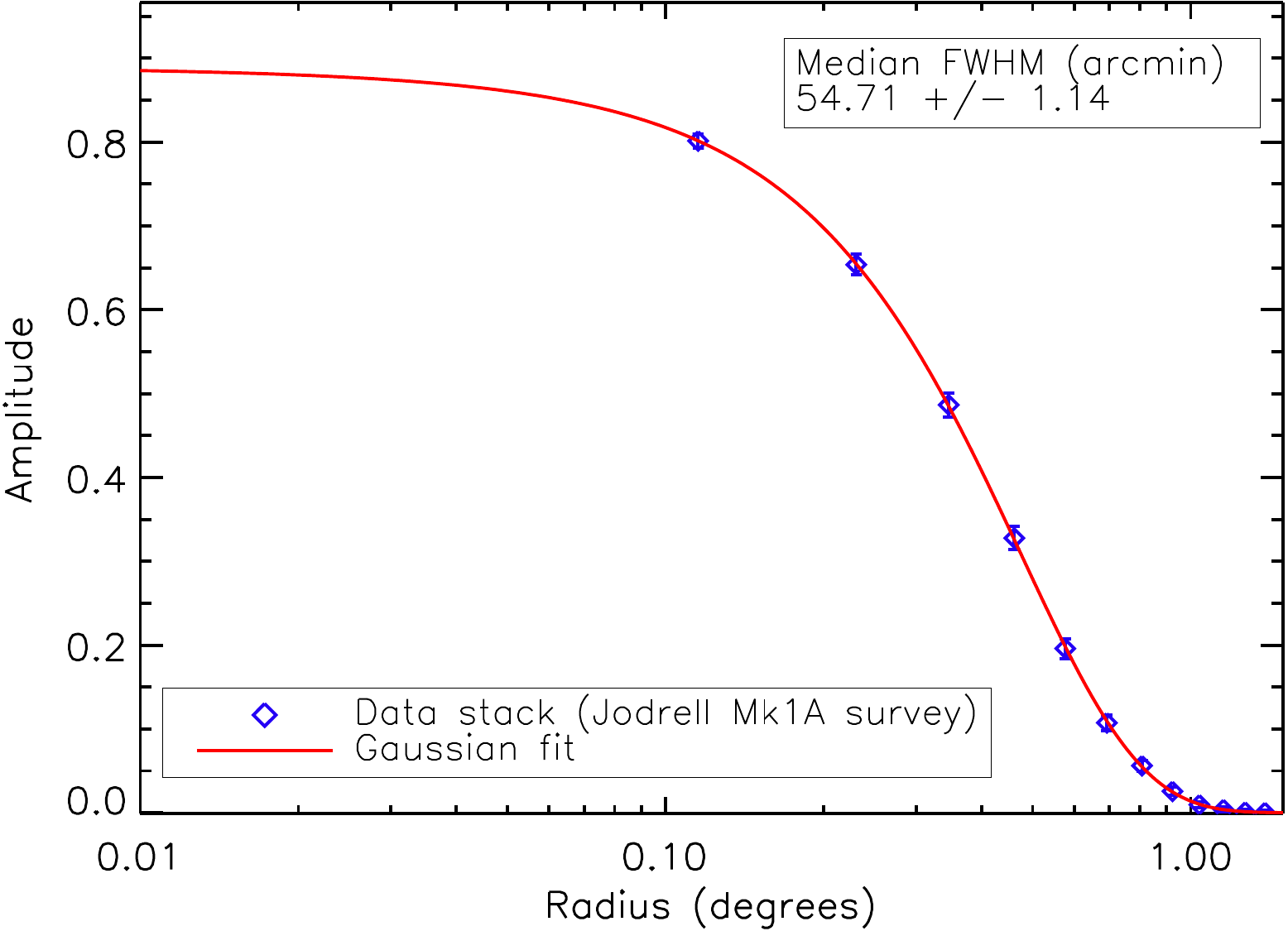}~
 \end{center}
\caption{Estimation of the effective beam for the four independent surveys --
  Jodrell Mk-I, Bonn, Parkes south, and Jodrell Mk-IA -- that comprise
  the Haslam map.}
\label{Fig:surveybeam}
\end{figure*}

\section{Reprocessing the Haslam map}\label{sec:proc}

The residual artefacts in the HAS03 version of the Haslam
map may introduce an excess of power at small angular scales in the
computation of the angular power spectrum of the Galactic synchrotron
at $408$\,MHz. Moreover, these artefacts may be wrongly interpreted as
a synchrotron signal by component separation algorithms, such as
template fitting methods \citep{Davies2006} and Bayesian
parametric fitting methods \citep{2008ApJ...672L..87E}. In the context
of multi-frequency CMB data analysis, such residuals could influence
the final estimation of some cosmological parameters. In this section,
we perform a reprocessing of the Haslam map to reduce, as far as
possible, those artefacts.

We begin by quantifying the beam of the 408\,MHz Haslam map. The beam
FWHM of the Haslam map is claimed to be 51\,arcmin in the original
survey, and according to the documentation header of the unfiltered
HAS03 map. However, the transformations/interpolations between
different pixelisation schemes and the applied filtering processes
could modify the effective beam. In fact, the HAS03 map specifies a resolution of $\sim 1^{\circ}$. 

In addition, the exact epoch of the data survey is not well defined
since the overall Haslam map is an atlas obtained from four different
partial sky surveys with data taken at different epochs. The imperfect
knowledge of the epoch of the survey may generate positional
(pointing) offsets in the Haslam map that we quantify by examining the
position on the map of known point-like radio sources.

\subsection{Estimation of the effective beam of the {\tt HEALPix} interpolation of the Haslam map}\label{subsec:beam}

After interpolating the ECP grid to a {\tt HEALPix} map in Celestial
B1950 coordinates, we first perform independent estimations of the
effective local beam in the sky regions corresponding to each of the four original surveys.  Specifically, we consider the $75$ strongest radio
sources in each of the Jodrell Mk-I Galactic Anticentre survey, the
Bonn survey, the Parkes South survey, and the Jodrell Mk-IA North Polar
survey. These point-like sources are isolated using a median filtering
with a  $1^{\circ}$ radius on the {\tt HEALPix} sphere. We compute the
profiles of the $75$ sources of the survey, in a $1.\!^{\circ}5$
aperture by averaging the amplitudes of the pixels in successive rings
of one pixel width, and then perform a Gaussian fit for each source
profile (with four parameters including the maximum value of the
Gaussian function, the central value, the sigma value, and the
background offset). We renormalise the $75$ source profiles and stack
them by using both the median average and the median absolute
deviation to reject outliers in the determination of the mean profile:
those deviating at more than $3\sigma$ are rejected. We then perform a
Gaussian fit of the stack profile, which provides the FWHM of the
median stack. For completeness, we also provide the FWHM computed from
the median average of the FWHM of each independent profile. We have
tested the robustness of our beam estimation on a Planck Sky
Model\footnote{\label{notepsm}\url{http://www.apc.univ-paris7.fr/\~delabrou/PSM/psm.html}}
({\tt PSM}, \citet{2013A&A...553A..96D}) simulation of a radio source
map with an input Gaussian beam of $51$ arcmin, for which we recovered
FWHM=$51.09\pm0.25$ arcmin.

Fig.~\ref{Fig:surveybeam} shows the effective beam profile (blue dots) and
Gaussian fit (red line) for each of the four surveys in separate
panels. In each case, the beam is well approximated by a Gaussian
profile up to $1^{\circ}$ radius. Moreover, the beam FWHM is stable
across the four surveys, with the inverse variance weighted average of the four beams,
$\sum_{i=1}^4 \left(b_i / \sigma_i^2\right) / \sum_{i=1}^4 \left(1/
  \sigma_i^2\right)$, yielding \mbox{${\rm FWHM} = 55.59$ arcmin}
while the mean value of the four beams, $(1/4) \sum_{i=1}^4 b_i$, and
the standard error on the mean, $(1/4) \sqrt{\sum_{i=1}^4
  \sigma_i^2}$, give \mbox{${\rm FWHM} = 55.47\pm 0.42$ arcmin}.

We then repeat the analysis on the full sky $408$\,MHz map, and
determine \mbox{${\rm FWHM} = 56.02\pm 0.56$ arcmin}, which is consistent with
the average value obtained from the four partial surveys. We conclude
that the effective beam FWHM of the $408$\,MHz Haslam map is $56.0\pm
0.6$ arcmin, which is larger than the typically quoted value.

\subsection{Pointing offsets}\label{subsec:offsets}

We now investigate the positional accuracy of the Haslam map. Pointing
errors are expected at a level of a few arcmin, due to positional
inaccuracies of the telescopes. Furthermore, the reprojection and
assumed epoch of the individual surveys can generate additional
positional errors. We quantify the positional accuracy of the data by
examining the locations of the strongest radio sources in the
$408$\,MHz raw map.

For this purpose we isolate the $22$ strongest sources in the Haslam
map (i.e., a subset of those used in Section~\ref{subsec:beam}), at
latitude $\vert b \vert > 20^{\circ}$ outside the Galactic plane, by
implementing a median filter on the sphere. Here, the medians are
calculated over a $1.\!^{\circ}5$ radius disc. We identify these
sources in the NASA Extragalactic Database
(NED)\footnote{\url{http://ned.ipac.caltech.edu/}\label{url:ned}} to
obtain their true R.A./Dec. positions in the B1950 Celestial
coordinate system, which are known to $\ll 1$ arcmin accuracy. The
name of each radio source and its location on the sky are listed in
Table~\ref{tab:raw-offset2}. We then perform a two-dimensional (2D)
Gaussian fitting of each source profile in the $408$\,MHz raw map to
determine their positions ($\widehat{\mbox{R.A.}}$,
$\widehat{\mbox{Dec.}}$) in B1950 coordinates. The 2D Gaussian fitting
of the data is performed over $3^{\circ}$ radius discs centred on the
NED locations of the sources on the sky. The pointing offsets of the
sources are computed by measuring the angular distance on the sphere
between their measured and expected coordinates.

The pointing offsets measured are listed in the fourth column of
Table~\ref{tab:raw-offset2}. The average value of the offsets is
$7.4\pm 3.2$ arcmin, which is comparable to the pixelization size, but
is not a significant fraction of the beamwidth. Nevertheless, it is
considerably larger than the quoted positional error ($\approx
1$\,arcmin) of the original data. Furthermore, any desourcing of known
sources based on their true coordinates could be affected by such
offsets (Section~\ref{sec:desourcing}).  

To investigate these offsets further, we plot them on the sky in
Fig.~\ref{Fig:offsets}, multiplied by a factor of 100 to make them
visible. They appear to be mostly random, yet there are specific
directions which are peculiar to the area of the sky considered. It is
interesting to note that the different areas correspond to the four
distinct surveys at $408$\,MHz: for example, in the area of the sky
corresponding to the Parkes south survey (${\rm R.A.} \in
[00^h,24^h]$, ${\rm Dec.} \in [-90^\circ,-5^\circ]$), all the offsets
are pointing toward the south pole direction.  With access to the data
from the individual surveys, this could be investigated further and
possibly corrected for. However,  for now we consider that they are
small enough to be acceptable given the beam width. We have verified
that a simple error in precessing the maps from their original epochs
to B1950 can not be responsible for the large-scale patterns observed.

\begin{table}
\caption{Source name (first column), Right Ascension in degrees
  (second column), Declination in degrees (third column), and pointing
  offsets in arcmin (fourth column) of the 22 strongest radio sources
  of the raw 408\,MHz map.}
\label{tab:raw-offset2}
\begin{tabular}{|lrr|r|}
\hline
Source & B1950 R.A.   & B1950 Dec. & $\Delta\theta$ \\
                  & (deg.) & (deg.) & (arcmin)\\
\hline
M87 & $187.1$ & $12.7$ & $3.6$ \\
NGC1316 & $50.2$ & $-37.4$ & $8.8$ \\
ESO 252-G A018 & $79.6$ & $-45.8$ & $7.7$ \\
PKS 0540-693 & $85.1$ & $-69.3$ & $8.0$ \\
PKS 2356-61 & $359.1$ & $-61.2$ & $9.4$ \\
ESO 075-G 041 & $328.2$ & $-69.9$ & $7.7$ \\
Hydra A & $138.9$ & $-11.9$ & $10.8$ \\
3C273 & $186.6$ & $2.3$ & $2.3$ \\
3C295 & $212.4$ & $52.4$ & $4.2$ \\
PKS 1932-46 & $293.1$ & $-46.5$ & $12.5$ \\
PKS 1814-63 & $273.7$ & $-63.8$ & $5.3$ \\
NGC4261 & $184.2$ & $6.1$ & $2.8$ \\
3C033 & $16.6$ & $13.1$ & $8.5$ \\
NGC7018 & $316.1$ & $-25.6$ & $12.6$ \\
ESO 362- G 021 & $80.3$ & $-36.5$ & $7.7$ \\
IC 4296 & $203.4$ & $-33.7$ & $1.3$ \\
3C433 & $320.4$ & $24.8$ & $9.9$ \\
3C048 & $23.7$ & $32.9$ & $9.8$ \\
PKS 0410-75 & $62.5$ & $-75.2$ & $7.6$ \\
3C196 & $122.5$ & $48.4$ & $7.4$ \\
3C444 & $332.9$ & $-17.3$ & $11.0$ \\
3C098 & $59.0$ & $10.3$ & $4.7$ \\
\hline
\end{tabular}
 \end{table}

\begin{figure}
  \begin{center}
    \includegraphics[width=\columnwidth]{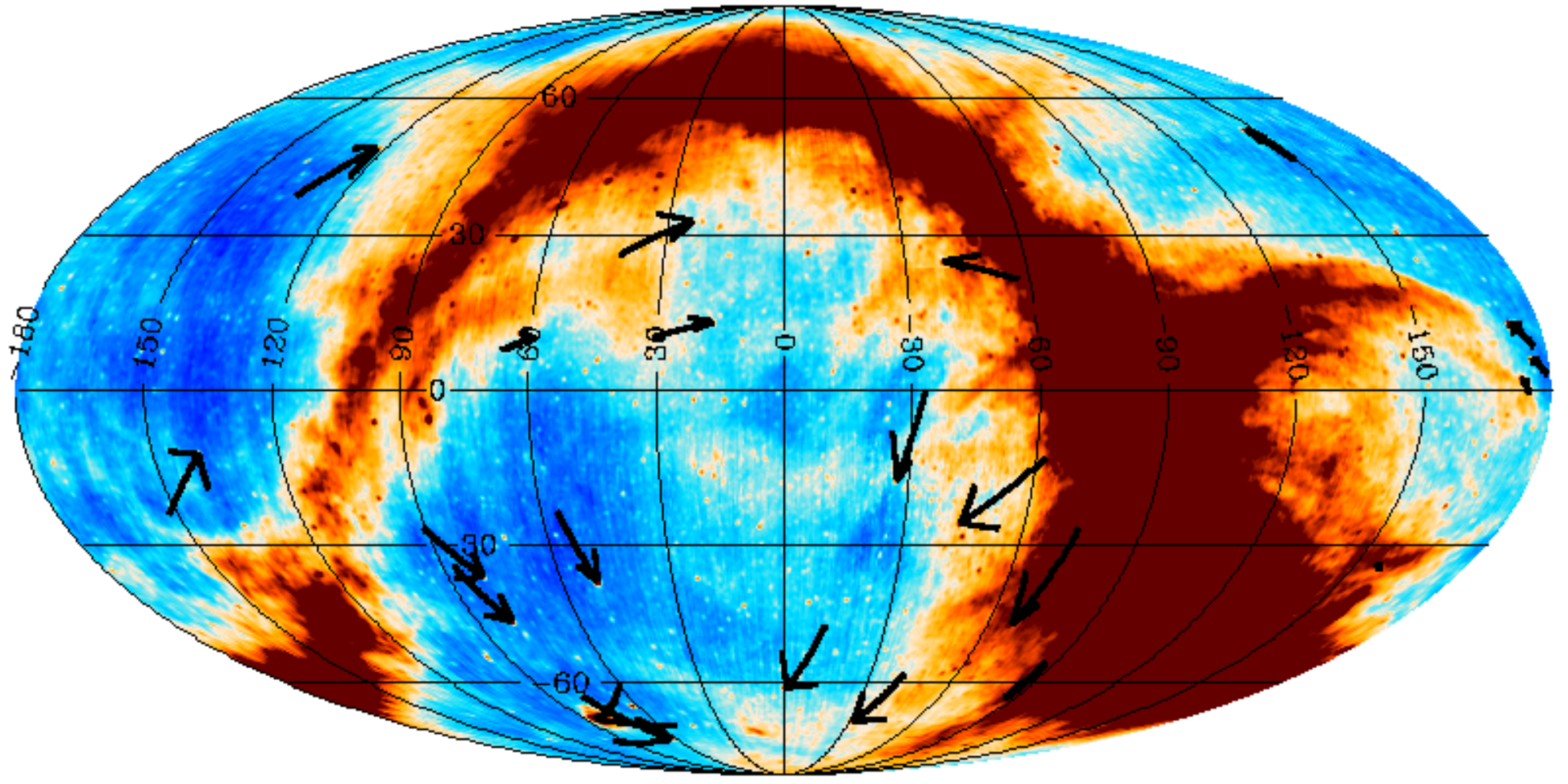}~
  \end{center}
\caption{Pointing offsets, multiplied by a factor of $100$, of the 22 strongest radio sources from Table~\ref{tab:raw-offset2} on the raw map. The head of the arrows corresponds to the effective position of the source on the map.
R.A.$=0^{\circ}$ is in the centre of the map and increases to the left.}
\label{Fig:offsets}
\end{figure}

\subsection{Destriping}
\label{sec:destriping}

The reduction of the striping in the Haslam map can be performed by
implementing a destriping algorithm. We apply the WMAP approach (code
provided by Janet Weiland, priv. comm.) that we adapted
  to the size of the ECP grid ($1080\times 540$), which is different
  from the size of NCSA grid ($1024\times 512$) processed by the {\it WMAP} team. We
  also improved the destriping by removing a slightly wider range of
  wavelengths in Fourier space\footnote{Note that a simple source subtraction by median filtering is performed prior to destriping to avoid strong source pixels affect the Fourier transform.}. The improvement is shown in
  Fig.~\ref{Fig:destr_comp}: the HAS03 map (left panel)
  clearly shows residual stripes near the north celestial pole where
  our newly processed map (right panel) does not. The Galactic plane
is masked such that the destriping is applied to $98$\% of the sky
(bottom panel of Fig.~\ref{Fig:ncsa-destr}). In the Galactic plane,
the diffuse background level is so large compared to the level of
striations that destriping is not required. In the top panel of
Fig.~\ref{Fig:ncsa-destr}, the raw ECP map is represented 
 before destriping: the
striations are clearly visible near the south pole. The middle panel
presents the destriped ECP map,  in which we can see, particularly
near the south pole, that the striations have been reduced. The bottom
panel of the figure shows the difference between the raw and destriped
ECP maps, exhibiting the level of striations that have been removed
from the raw Haslam map. 

In order to quantify the amount of residual striations in the
destriped map, we inspect a low-background $2^{\circ} \times
2^{\circ}$ region of the sky, centred at $(\mbox{R.A.}, \mbox{Dec.}) =
(0^\circ,-40^\circ)$ in Celestial coordinates. We make a horizontal
cut of the region, i.e. a band
\mbox{$\{\mbox{R.A.}\in[-0.5^\circ,0.5^\circ],
  \mbox{Dec.}=-40^\circ\}$}, that is orthogonal to the striations. In
Fig.~\ref{Fig:striations}, we plot the resulting profile of the pixel
amplitude in that band for both the raw map (solid black) and the
destriped map (solid red), from which we have removed the background
zero level to highlight the fluctuations due to the striations. The
r.m.s. of the striping fluctuations in this region of the sky is found
to be $\sigma_{raw} = 0.35$\,K for the raw ECP map and $\sigma_{destr}
= 0.10$\,K for the destriped ECP map. The destriping algorithm reduces
the amplitude of the striations in the Haslam map by more than
$70$\,\%. Nevertheless, residual striations can still be seen at a low-level, in regions devoid of strong Galactic emission.

\begin{figure}
  \begin{center}
    \includegraphics[width=0.5\columnwidth]{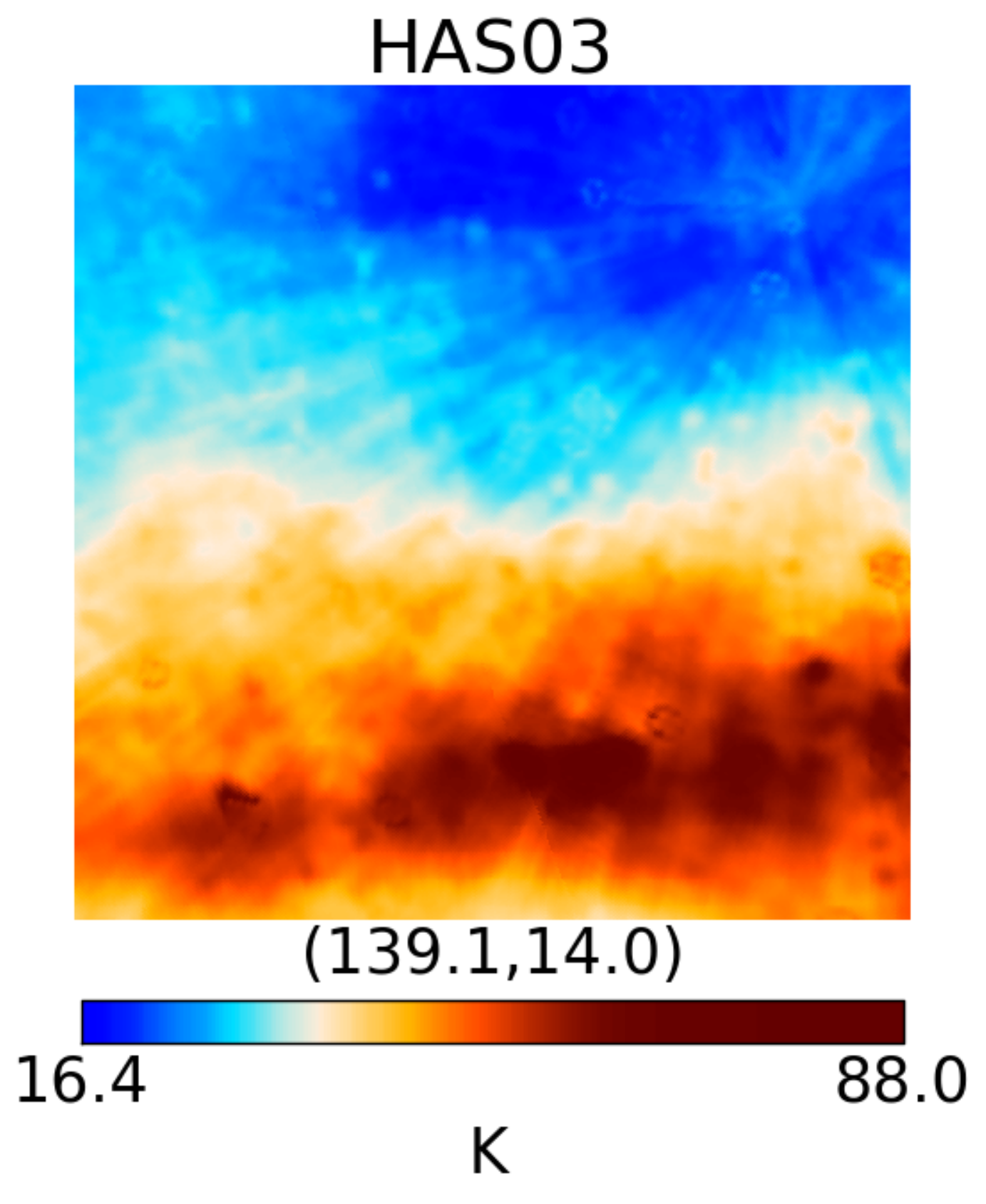}~
    \includegraphics[width=0.5\columnwidth]{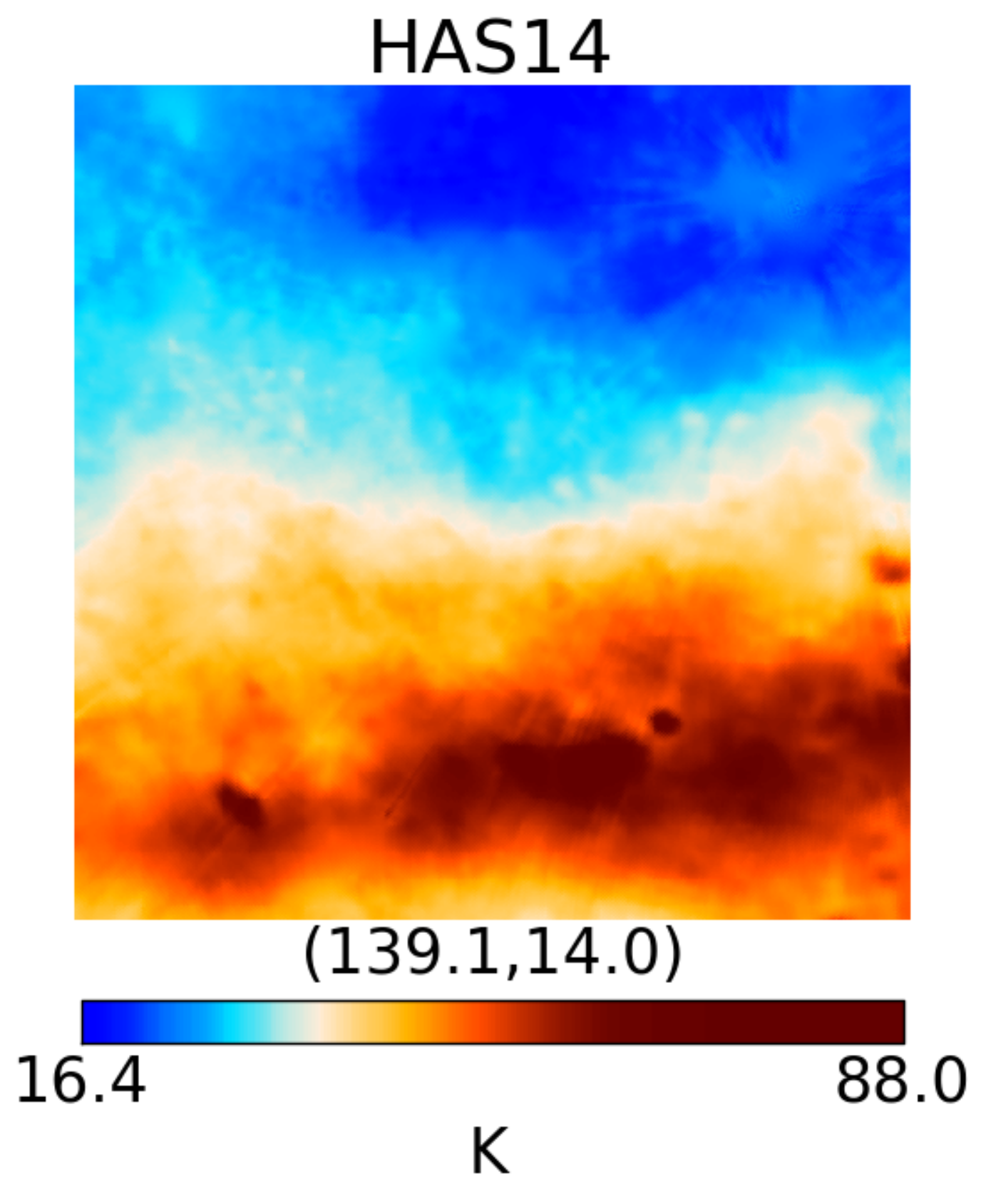}
 \end{center}
\caption{
  Improvement in destriping shown in a ${33^\circ\times 33^\circ}$
  gnomonic projection of the 408\,MHz map centred at
  ${(l,b)=(139^{\circ}, 14^{\circ})}$, 
near the north celestial pole. {\it Left} panel: the HAS03 destriped map. {\it Right} panel: our newly destriped map (HAS14).  
}
\label{Fig:destr_comp}
\end{figure}

\begin{figure}
  \begin{center}
    \includegraphics[width=\columnwidth]{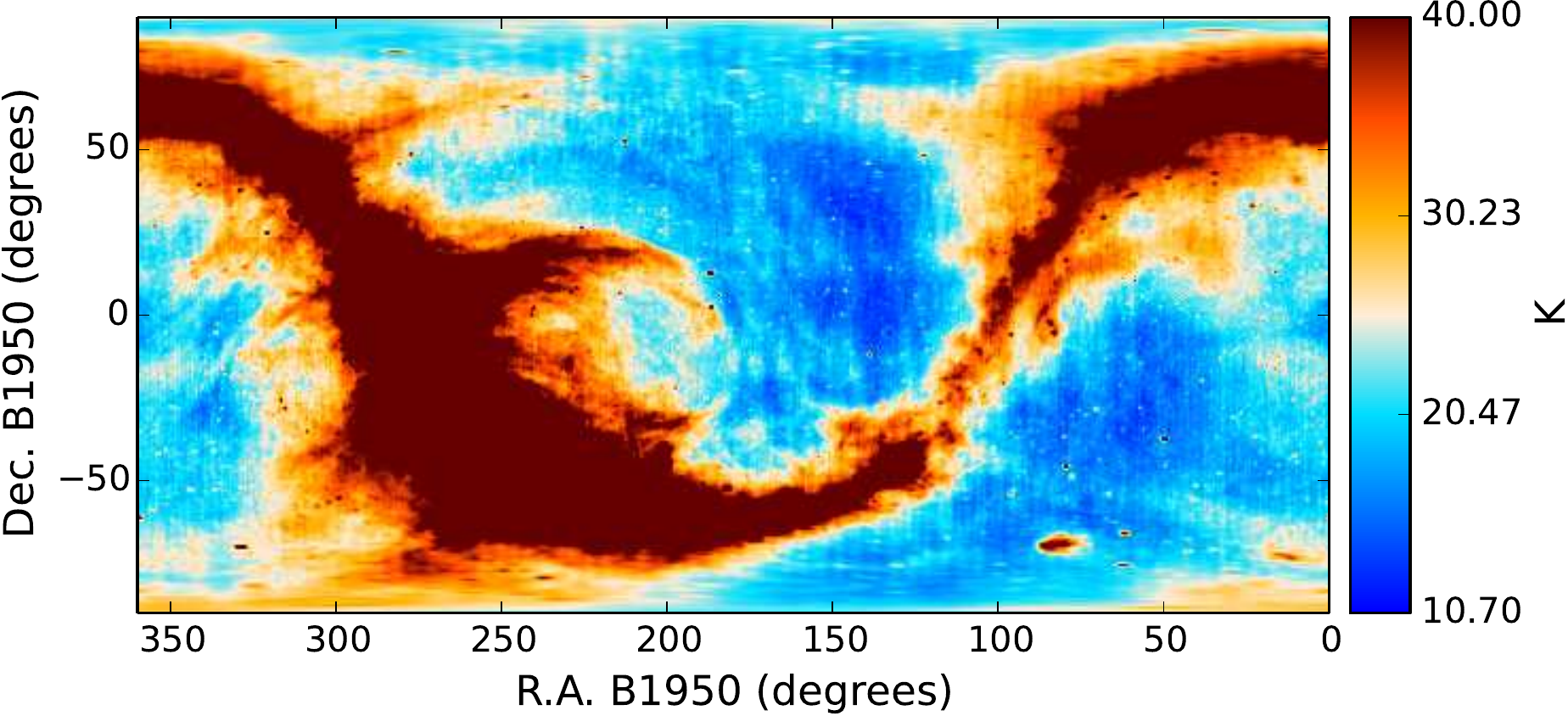}\\
    \includegraphics[width=\columnwidth]{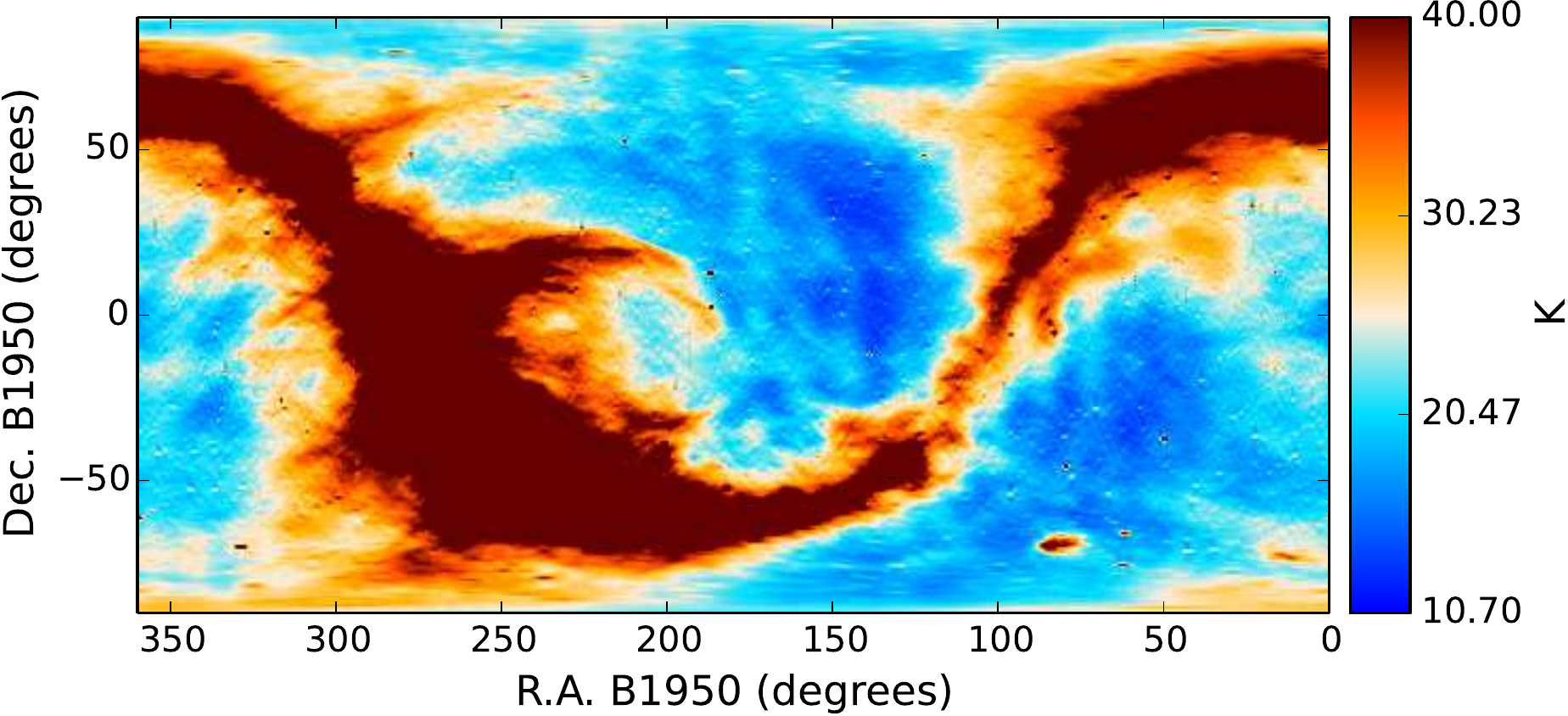}\\
    \includegraphics[width=\columnwidth]{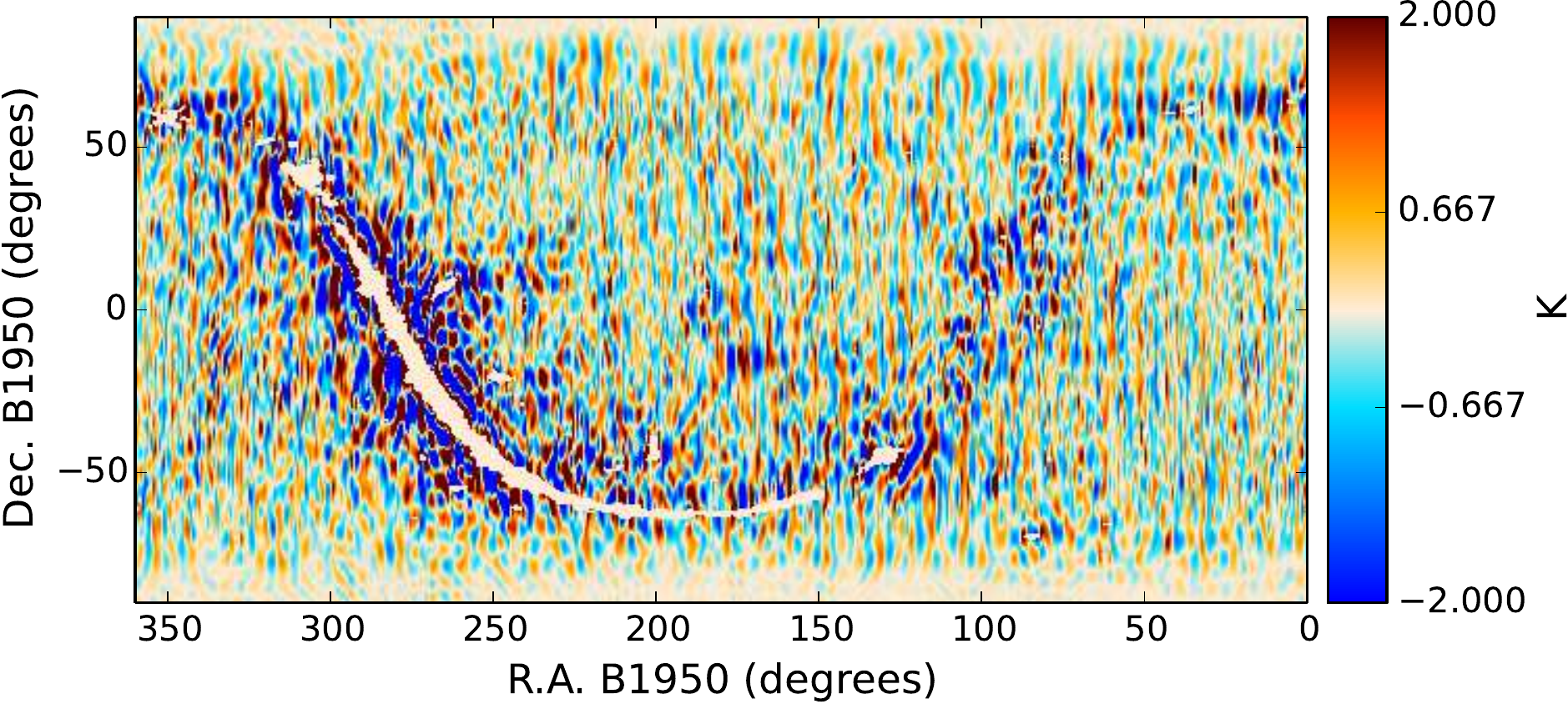}\\
 \end{center}
\caption{
Destriping of the 408\,MHz sky map: raw Haslam map (\emph{top panel}),
destriped Haslam map (\emph{middle panel}), and the difference map
(\emph{bottom panel}).  
}
\label{Fig:ncsa-destr}
\end{figure}

\begin{figure}
  \begin{center}
    \includegraphics[width=\columnwidth]{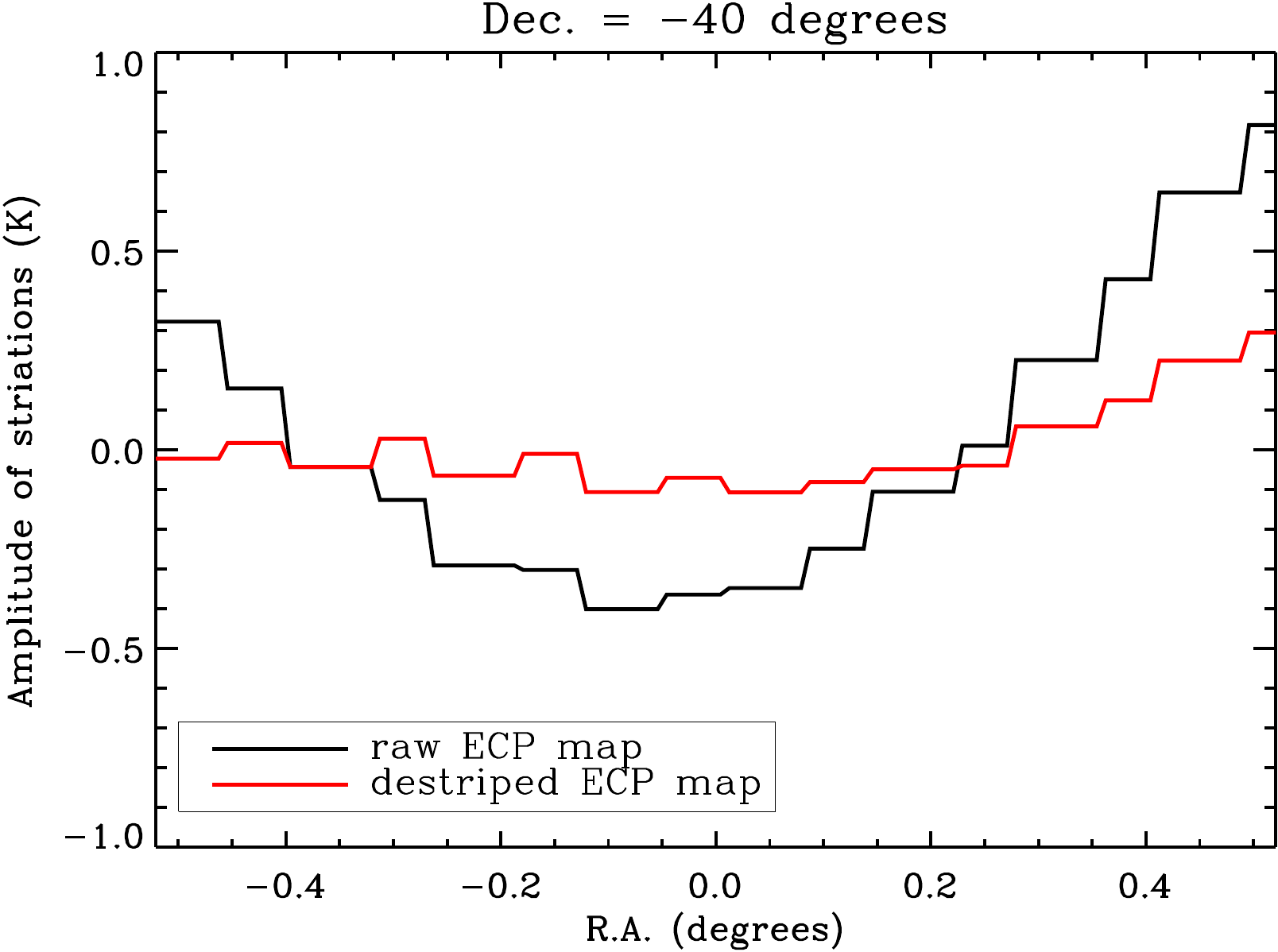} 
 \end{center}
\caption{
Amplitude of the striations in the band \mbox{$\{{\rm
    R.A.}\in[-0.5^\circ,0.5^\circ], {\rm Dec.}=-40^\circ\}$} of both
the raw map (solid black) and the destriped map (solid red).
}
\label{Fig:striations}
\end{figure}
                     
\subsection{Desourcing}\label{sec:desourcing}

The destriped ECP map (middle panel of Fig.~\ref{Fig:ncsa-destr}) is
clearly contaminated by strong extragalactic radio sources. In this
section we remove these from the destriped $408$\,MHz map by
implementing two different desourcing algorithms. 

We detect the sources by implementing a median filter on the map. The
median average value at each pixel is determined from those pixels
within a $1.\!^{\circ}5$ radius disc. An alternative would be to use
the coordinate information from a source catalogue. However, we have
seen in Section~\ref{subsec:offsets} that the ECP Haslam map suffers
from positional offsets, typically $\sim 7$\,arcmin, hence it is
better to process sources at their effective location rather than at
their true location on the sky. The radio sources are subsequently identified simply by subtracting
the median filtered map from the input Haslam map.

Once the sources are detected, we process only those with a magnitude
larger than a certain threshold with respect to the local
background. In the following, we consider three different thresholds:
$6$\,K, $3$\,K, and $1.5$\,K over the background. In terms of flux
density, these correspond to $9$\,Jy, $5$\,Jy, and $2$\,Jy at
$408$\,MHz and $56$ arcmin resolution. Our approach for desourcing the
Haslam map then combines two methods, either Gaussian fitting (Section~\ref{subsec:gaussfit}) or minimum curvature inpainting (Section~\ref{subsec:inpainting}), depending both on the source considered and
on the local diffuse background surrounding the source.

\subsubsection{Two-dimensional Gaussian fitting}\label{subsec:gaussfit}

In order to remove a given source from the diffuse 408\,MHz map, we
fit a 2D Gaussian profile at the location of the source, and then
subtract the resulting fit from the data. There are then nine
parameters to fit, corresponding to the background plane geometry
($A_0$,$A_1$,$A_2$), the amplitude of the Gaussian profile ($A_3$),
the centre coordinates ($A_4$, $A_5$), the orientation angle ($A_6$),
and the Gaussian FWHM in both dimensions ($A_7$,$A_8$). It is
necessary to fit for the background profile surrounding the source but
only the Gaussian part of the fit is then subtracted from the data. 

The expression for the nine-parameter fitting function is given by the
equations \ref{eq:gaussian} and \ref{eq:gaussian2}:
\begin{align}\label{eq:gaussian}
F(X,Y;\left\{A_i\right\}) &= A_0 + A_1X + A_2Y\cr
& +   A_3\,\exp\left[-{1\over 2}\left({\widetilde{X}\over A_7}\right)^2 -{1\over 2}\left({\widetilde{Y}\over A_8}\right)^2\right],
\end{align} 
where
\begin{align}\label{eq:gaussian2}
\widetilde{X} & =  \cos\left(A_6{\pi\over 180}\right)\left(X-A_4\right) - \sin\left(A_6{\pi\over 180}\right)\left(Y-A_5\right)\cr
\widetilde{Y} & =  \sin\left(A_6{\pi\over 180}\right)\left(X-A_4\right) + \cos\left(A_6{\pi\over 180}\right)\left(Y-A_5\right).
\end{align}

The data are fitted within a disc of $1.\!^{\circ}5$ radius centred on
the location of a given detected source, and the Gaussian part of the
fit is then subtracted from the data within a disc of $5^{\circ}$
radius. The goodness-of-fit is measured as:
\begin{align}\label{eq:chi2}
\chi^2 = \sum_{k=1}^N {\left(Z_k - F(X_k,Y_k;\left\{A_i\right\})\right)^2 \over \sigma^2},
\end{align}
where $Z_k$ is the data point at the location $(X_k,Y_k)$,
$F(X_k,Y_k;\left\{A_i\right\})$ is the fit at the same location,
$\sigma$ is the noise r.m.s. value (here we set $\sigma=800$\,mK), $N$
is the number of data points that are fitted. The number of degrees of
freedom is $N_{\rm dof} = N - 9$, although the data points are not
completely independent (the absolute value of $\chi^2$ is unimportant
for this application).

\subsubsection{Minimum curvature spline surface inpainting}\label{subsec:inpainting}

As an alternative to Gaussian fitting, we also implement an inpainting
technique, based on the interpolation of a set of points with a
minimum curvature spline surface.

The interpolating function $F(X,Y)$ minimises the Lagrangian
\begin{align}
&\mathcal{L}\left(F,\lambda\right)\cr
&=\sum_{i=1}^N \vert\vert Z_i - F(X_i,Y_i)\vert\vert^2 \cr
&+\! \lambda\!\iint\!\left[\left({\partial^2 F\over \partial X^2}\right)^2\!+\! \left({\partial^2 F\over \partial Y^2}\right)^2 \!+\!2\left({\partial^2 F\over \partial X\partial Y}\right)^2\right]\!dXdY,
\end{align}
where $Z_i$ is the data value at the interpolation point pixel
$(X_i,Y_i)$. The integral expression under the Lagrange multiplier
$\lambda$ is the ``bending energy'' that constrains the interpolated
surface to be as smooth as possible. In the case of three
interpolation points, the interpolating function is a plane
surface. In the case of more than three interpolation points, the
solution for the interpolating function is given by the Thin Plate
Spline \citep{1982Franke}:
\begin{align}
F(X,Y) &=  \sum A_i d_i^2\log d_i + a + bX + cY,
\end{align}
where
\begin{align}
d_i^2 &= (X-X_i)^2+(Y-Y_i)^2
\end{align}
is the distance to the interpolation point.

A mask with $1.\!^{\circ}5$ radius discs centred on the locations of
the sources is constructed, then inpainting is performed on the masked
pixels.

\begin{figure}
  \begin{center}
    \includegraphics[width=0.5\columnwidth]{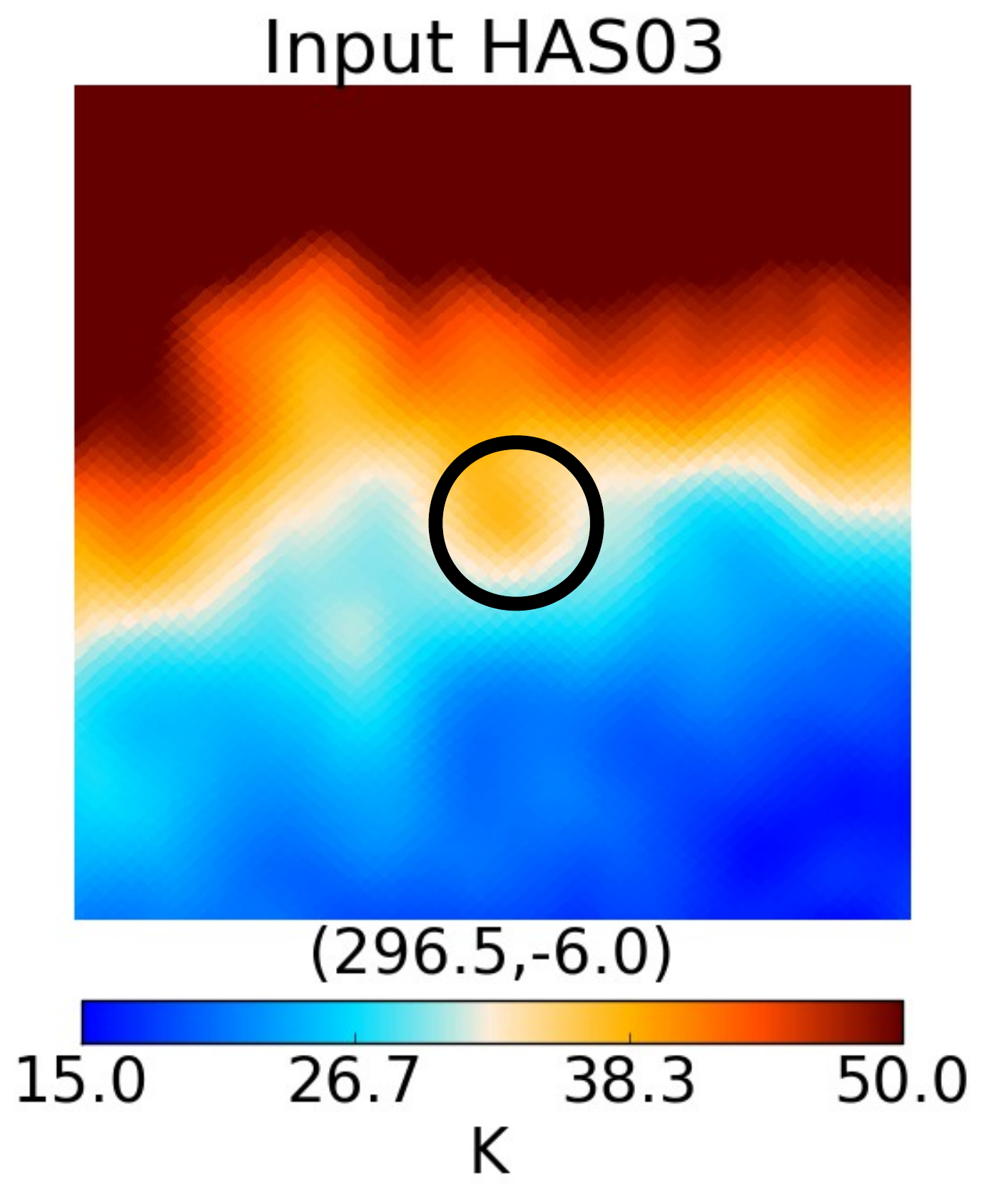}~
    \includegraphics[width=0.5\columnwidth]{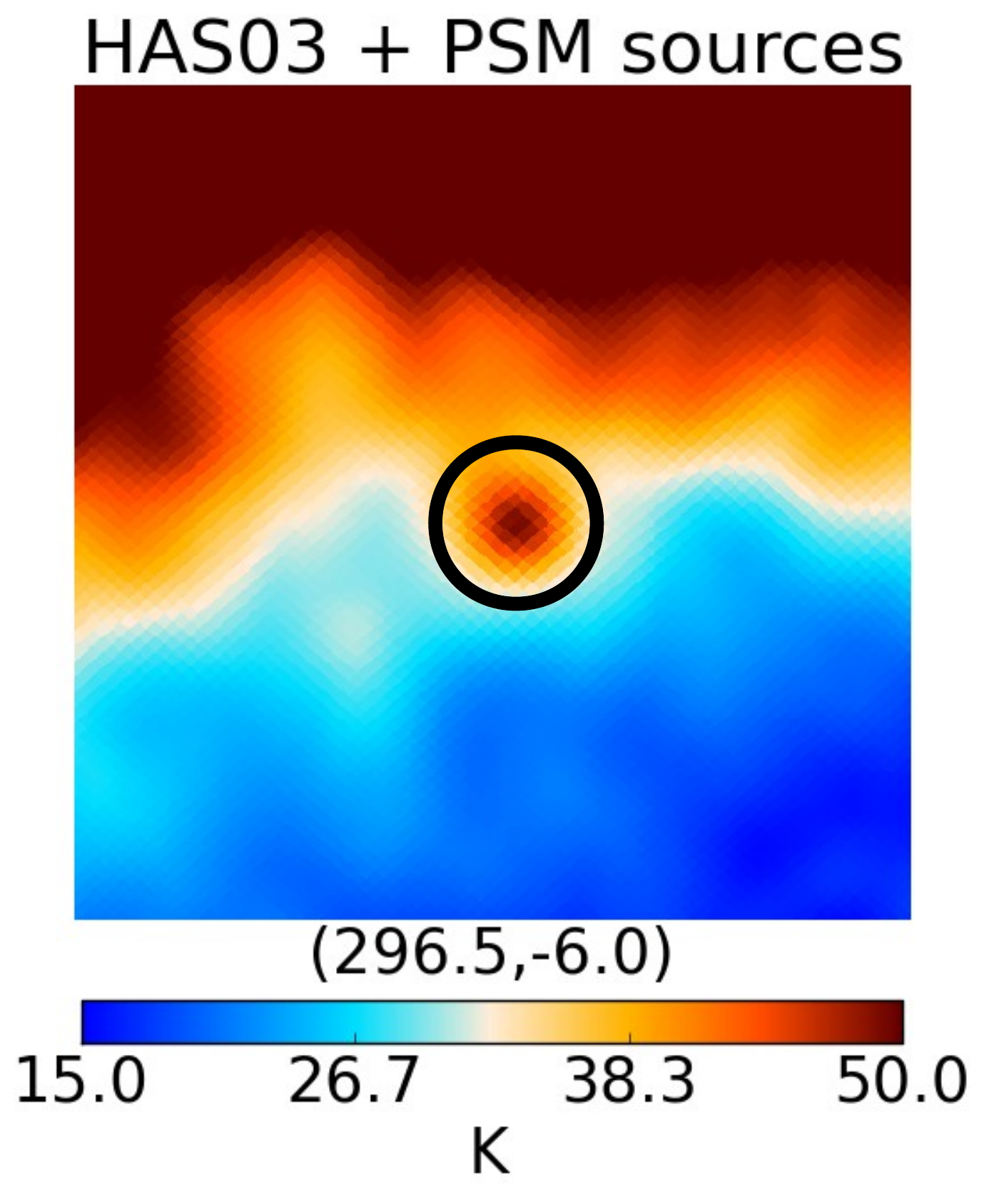}~\\
    \includegraphics[width=0.5\columnwidth]{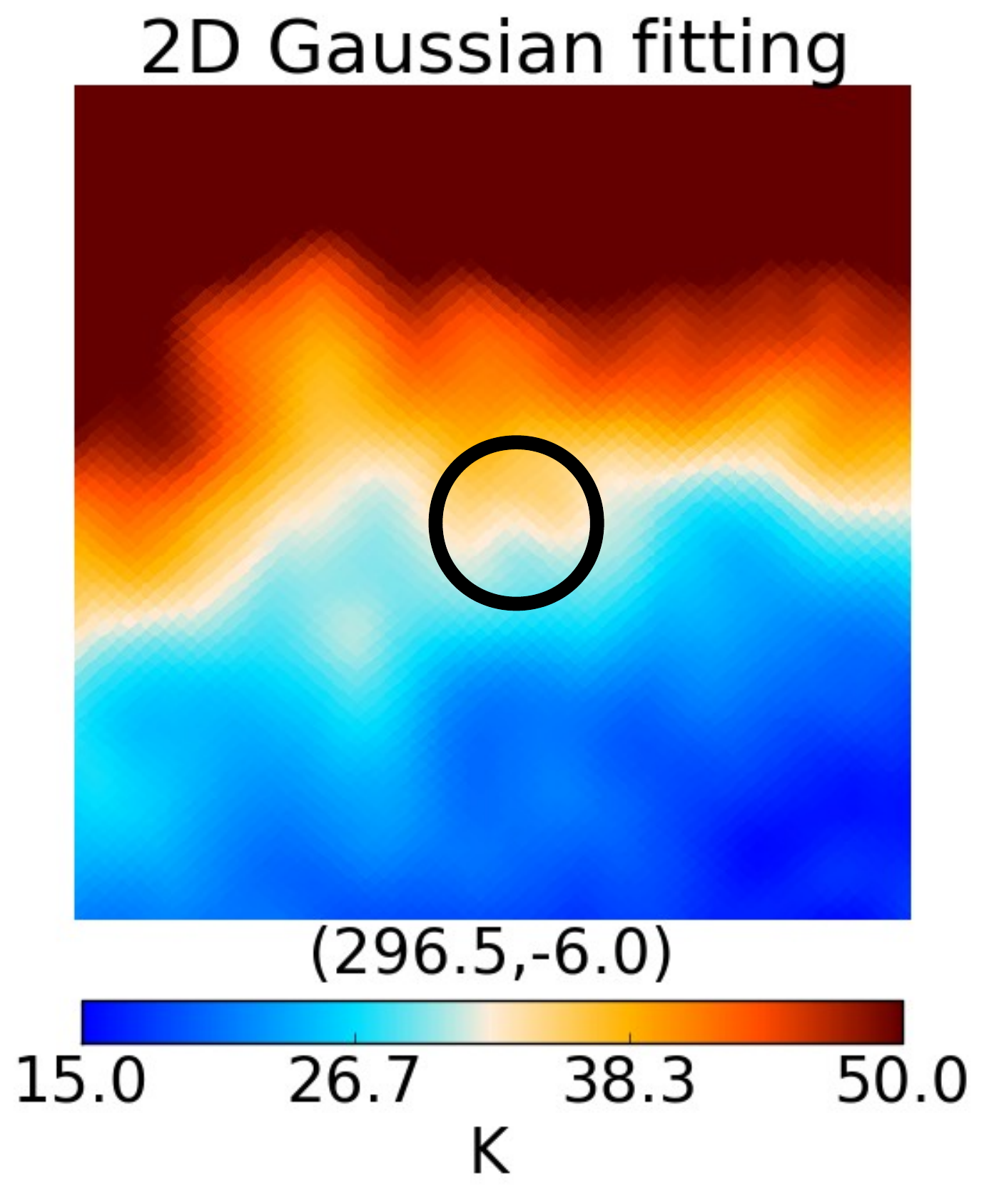}~
    \includegraphics[width=0.5\columnwidth]{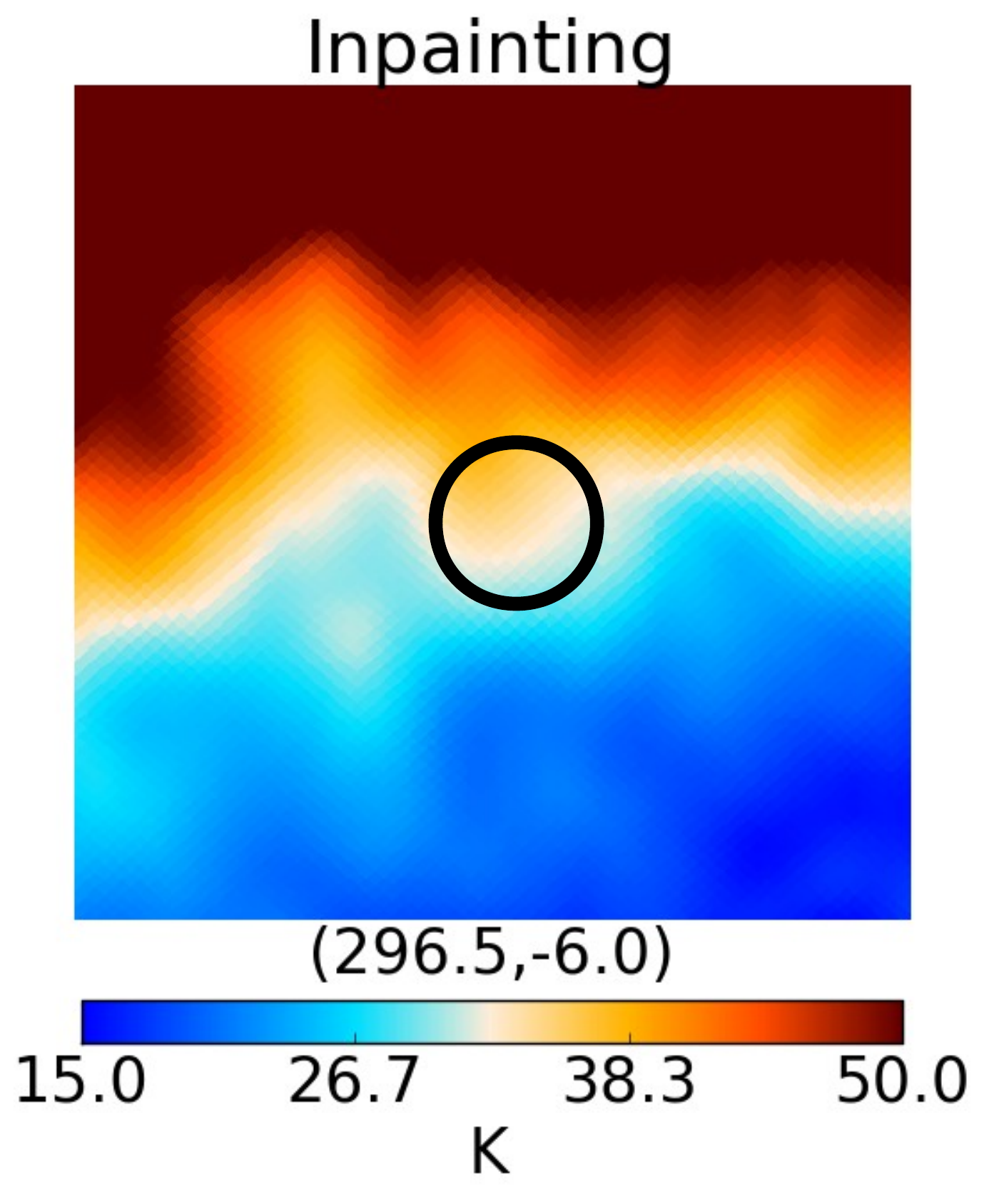}~
 \end{center}
\caption{Desourcing of {\tt PSM} simulations: an $8^\circ\times
  8^\circ$ gnomonic projection centred on $(l,b) =
  (296.5^\circ,-6.0^\circ)$.  \emph{Top left}: input map
  (HAS03). \emph{Top right}: total map (input HAS03 map + {\tt PSM}
  source map). \emph{Bottom left}: 2D Gaussian fitting. \emph{Bottom
    right}: inpainting.}
\label{Fig:simppps1}
\end{figure}

\subsubsection{Results on {\tt PSM} simulations}
\label{subsec:simu}

We first validate both methods, inpainting and 2D Gaussian fitting, on
simulations. The input diffuse $408$\,MHz map is the HAS03 map. The radio sources are simulated using the {\tt PSM}
software to generate a map containing strong sources with a flux
larger than $10$\,Jy. Both methods are then tested on the coaddition
of the HAS03 and the {\tt PSM} source map. For the
inpainting, the source mask is constructed by blanking the pixels of
the total map where the amplitude of the {\tt PSM} source map is
larger than $0.8$\,K. For the Gaussian fitting, the fit is performed
on the total map within $1.\!^{\circ}5$ radius discs and the
subtraction applied within $5^{\circ}$ radius discs.  The results are
shown in Fig.~\ref{Fig:simppps1} and Fig.~\ref{Fig:simps1}.

The top left panel of Fig.~\ref{Fig:simppps1} shows the input diffuse
map, i.e. the HAS03 map. This input map exhibits spurious small scale
patterns, as indicated by the circle, because of the imperfect
desourcing achieved in the HAS03 version of the 408\,MHz data. The
top right panel presents the total map, where we have added the
simulated {\tt PSM} sources to the input map. In the bottom right
panel we then show the result of inpainting on this map, while in the
bottom left panel the result of the Gaussian fit and subtraction
process can be seen. In both cases, the recovered map is similar to
the input diffuse map: the spurious small scale structures within the
drawn circle have been restored whereas the {\tt PSM} source has been
removed. 

\begin{figure}
  \begin{center}
    \includegraphics[width=\columnwidth]{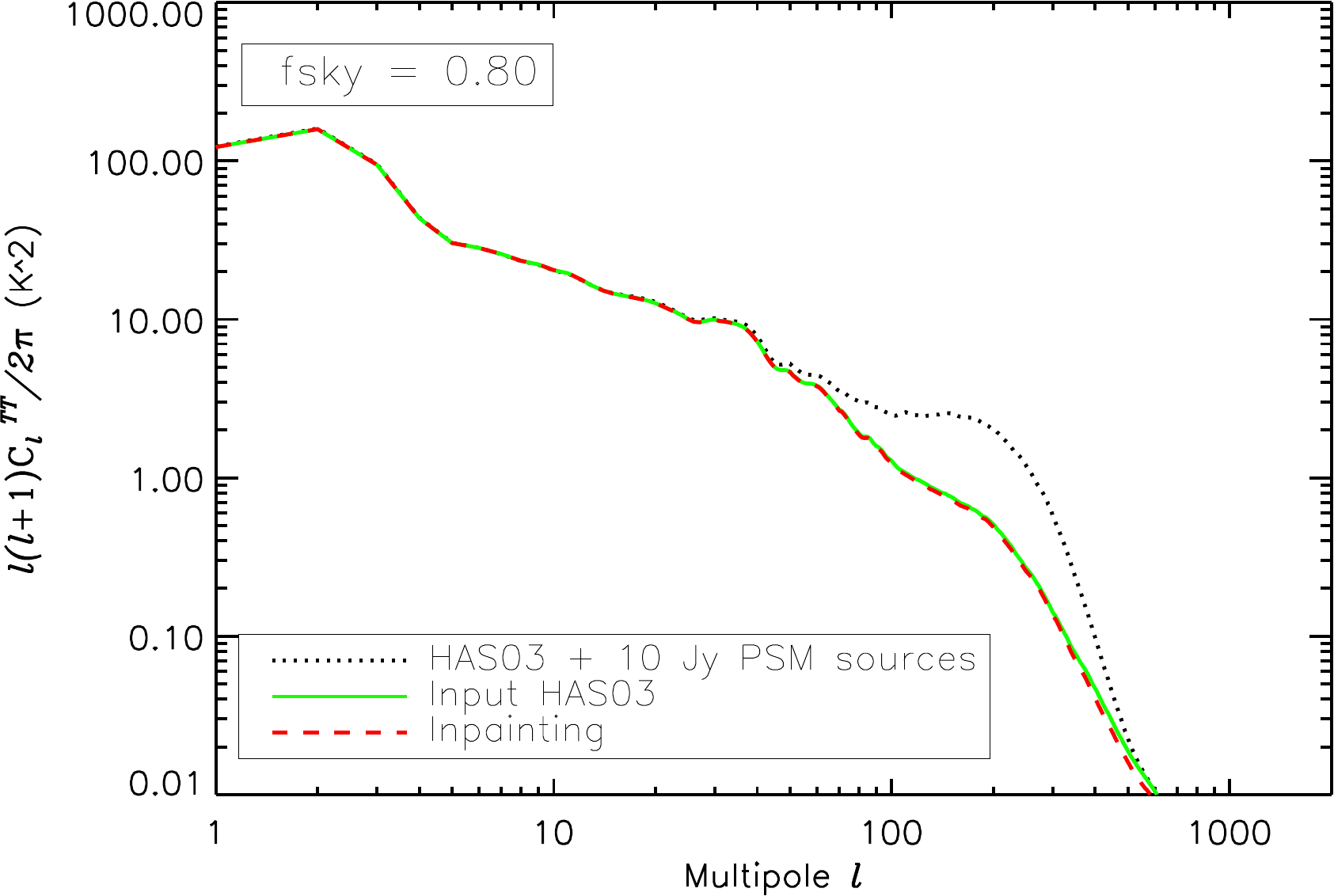}~\\
    \includegraphics[width=\columnwidth]{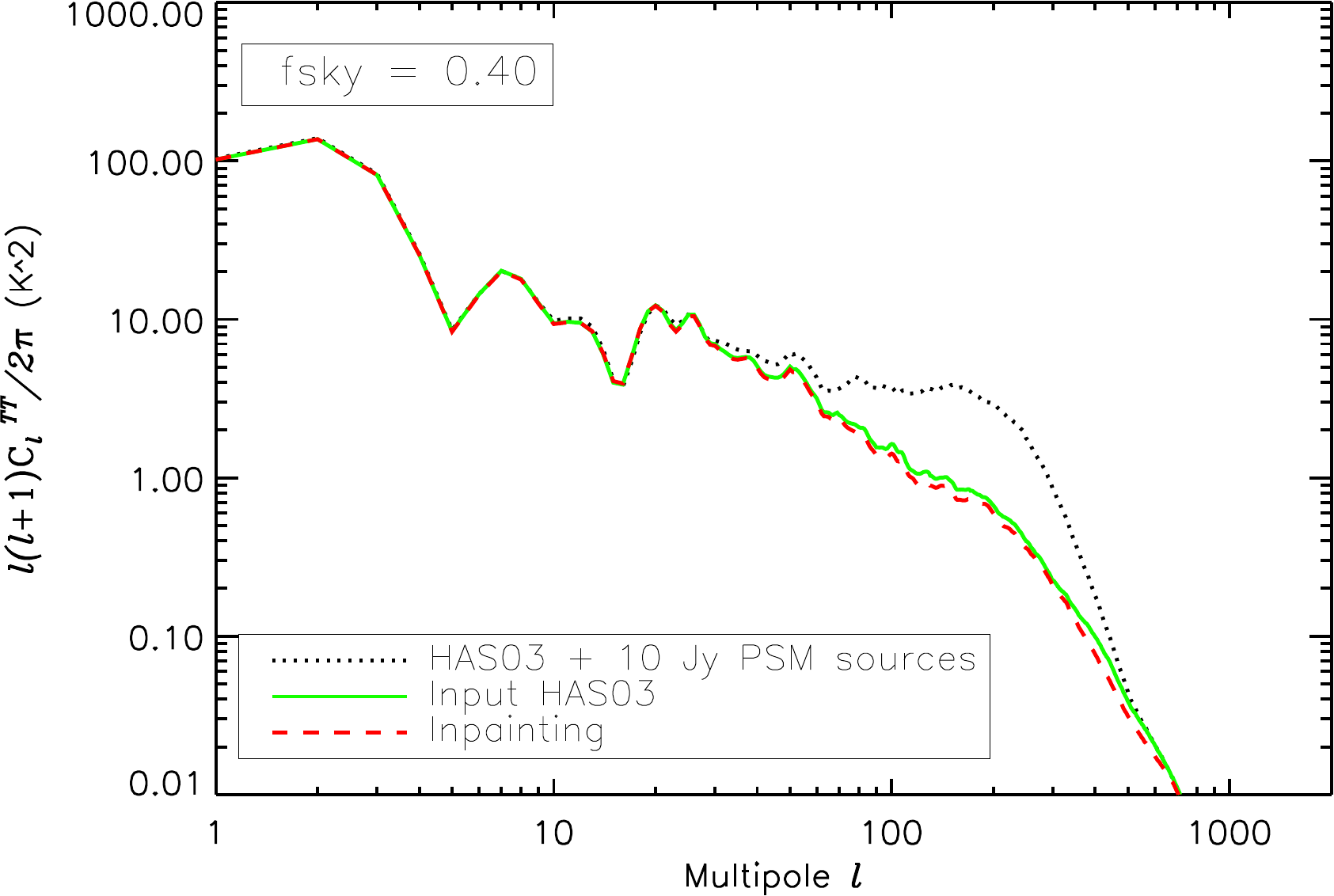}
 \end{center}
\caption{Angular power spectrum of the {\tt PSM} simulation
  (\emph{top}: $f_{sky}=0.80$, \emph{bottom}: $f_{sky}=0.40$): input
  map (solid green), total map (dotted black), inpainted map (dashed
  red).}
\label{Fig:simps1}
\end{figure}

We also compute the angular power spectrum of the different maps using
the {\tt PolSpice} pseudo-$C_{\ell}$ code \citep{2004MNRAS.350..914C}. We produce
two different Galactic masks ($f_{sky}=0.80$ and $f_{sky}=0.40$) based
on two successive intensity thresholds of the input map. In the top
panel of Fig.~\ref{Fig:simps1} we plot the angular power spectra for
$80$\% sky coverage of the input map (solid green), the total map with
the {\tt PSM} sources added (dotted black), and the inpainted map
(dashed red). The power spectrum of the total map shows an excess of
power on small scales due to the presence of the {\tt PSM} sources
whereas the power spectrum of the inpainted map successfully matches
the input power spectrum by removing the source contamination at small
scales. In the bottom panel of the figure, we also validate the
inpainting result at high Galactic latitude by computing the power
spectrum on 40\% of the sky, after masking the Galactic plane, so that
the Galactic signal is not dominating the power spectrum on scales of
$\sim 1^{\circ}$.

\subsubsection{Results on data: multi-stage approach}\label{subsec:multi}

Some of the radio sources at $408$\,MHz are faint or extended. In this
case the Gaussian fitting may fail because the number of fitted data
points becomes insufficient to compensate for the low signal-to-noise
ratio. Some other sources, especially at low Galactic latitudes, are
surrounded by a strong background signal, with a varying and complex
geometry. In this case also, the 2D Gaussian fitting approach may fail
because the removal of the source would require a much larger number
of parameters to be fitted. We therefore need to perform a multi-stage
approach on the data, where some of the sources are fitted with a 2D
Gaussian profile and then subtracted from the Haslam map whereas other
sources are inpainted. 

In the first stage, we perform a two-dimensional Gaussian fitting of
all the sources detected above $6$\,K over the background. This
threshold typically corresponds to sources with a flux density larger
than $9$\,Jy at $408$\,MHz for a $56$ arcmin beam.  We quantify the
goodness-of-fit by using the $\chi^2$ (Eq. \ref{eq:chi2}) measuring
the deviation of the Gaussian fit from the source profile. The median
value of the goodness-of-fit is $\chi^2_m= 6775.6$
for the $6$\,K-threshold. We visually validated on the map the
performance of the source removal for that particular case where the
goodness-of-fit reaches the limit $\chi^2 = \chi^2_m$. We therefore
choose as a conservative upper limit criterion of the goodness-of-fit
the median value $\chi^2_m$ in order to guarantee the exclusion of the
imperfect Gaussian fits: if the $\chi^2$ of the fit is smaller than
the median value $\chi^2_m$ of all the fits then we subtract the fit
from the data.  Otherwise, if $\chi^2 > \chi^2_m$ then we process the
source by inpainting instead of fitting as a second stage.

\begin{figure}
  \begin{center}
    \includegraphics[width=\columnwidth]{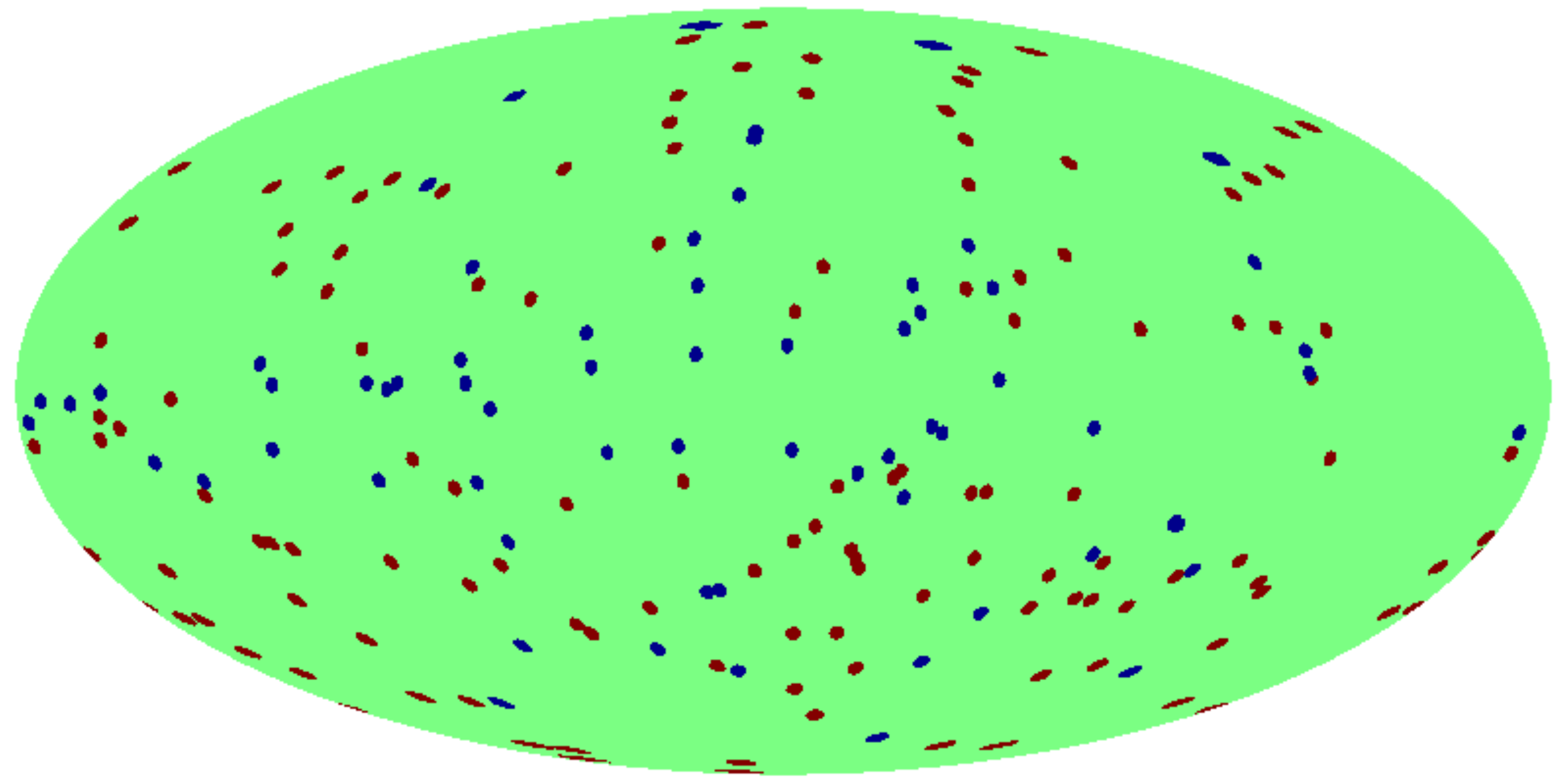}~\\
    \includegraphics[width=\columnwidth]{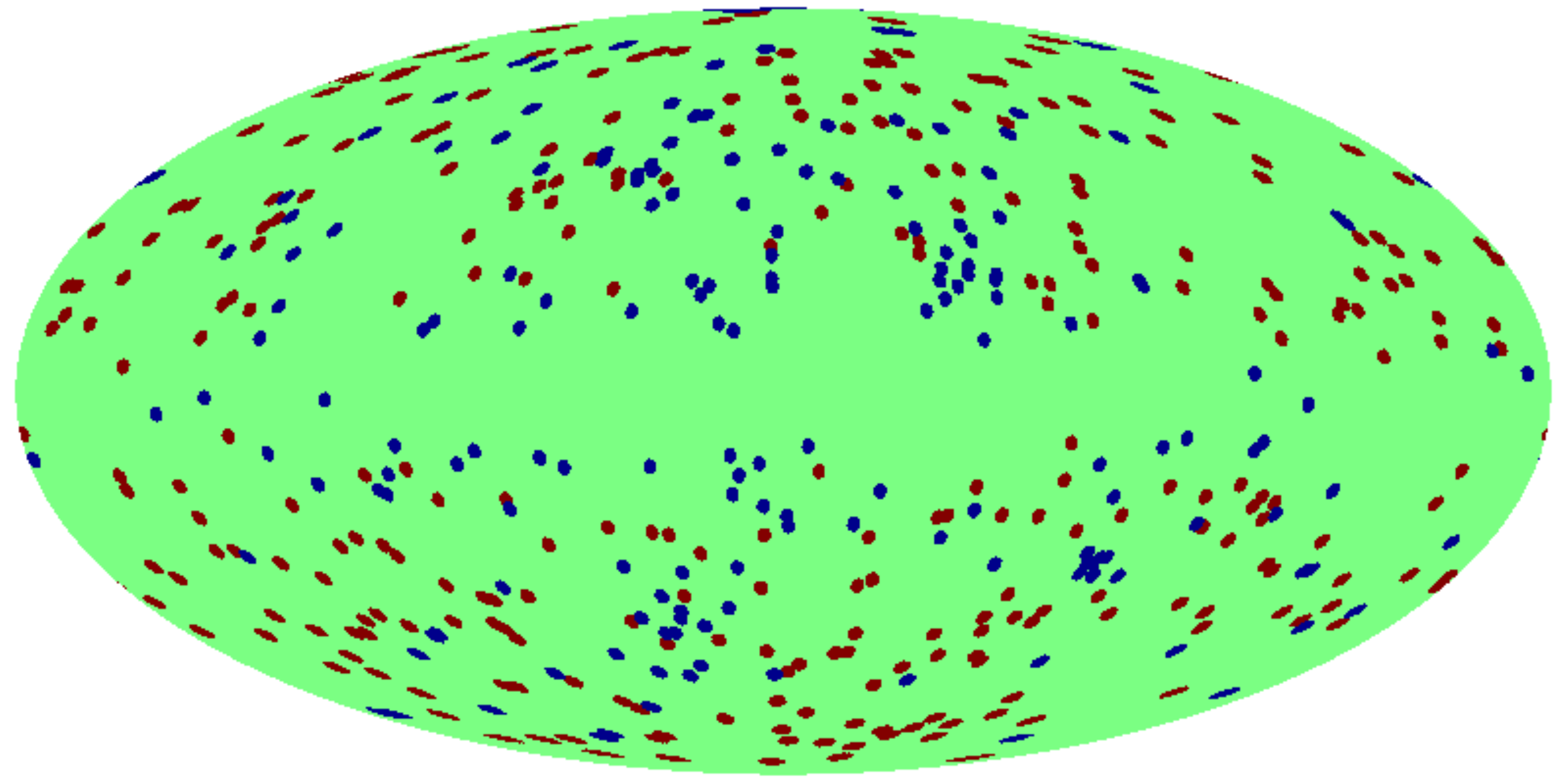}~\\
    \includegraphics[width=\columnwidth]{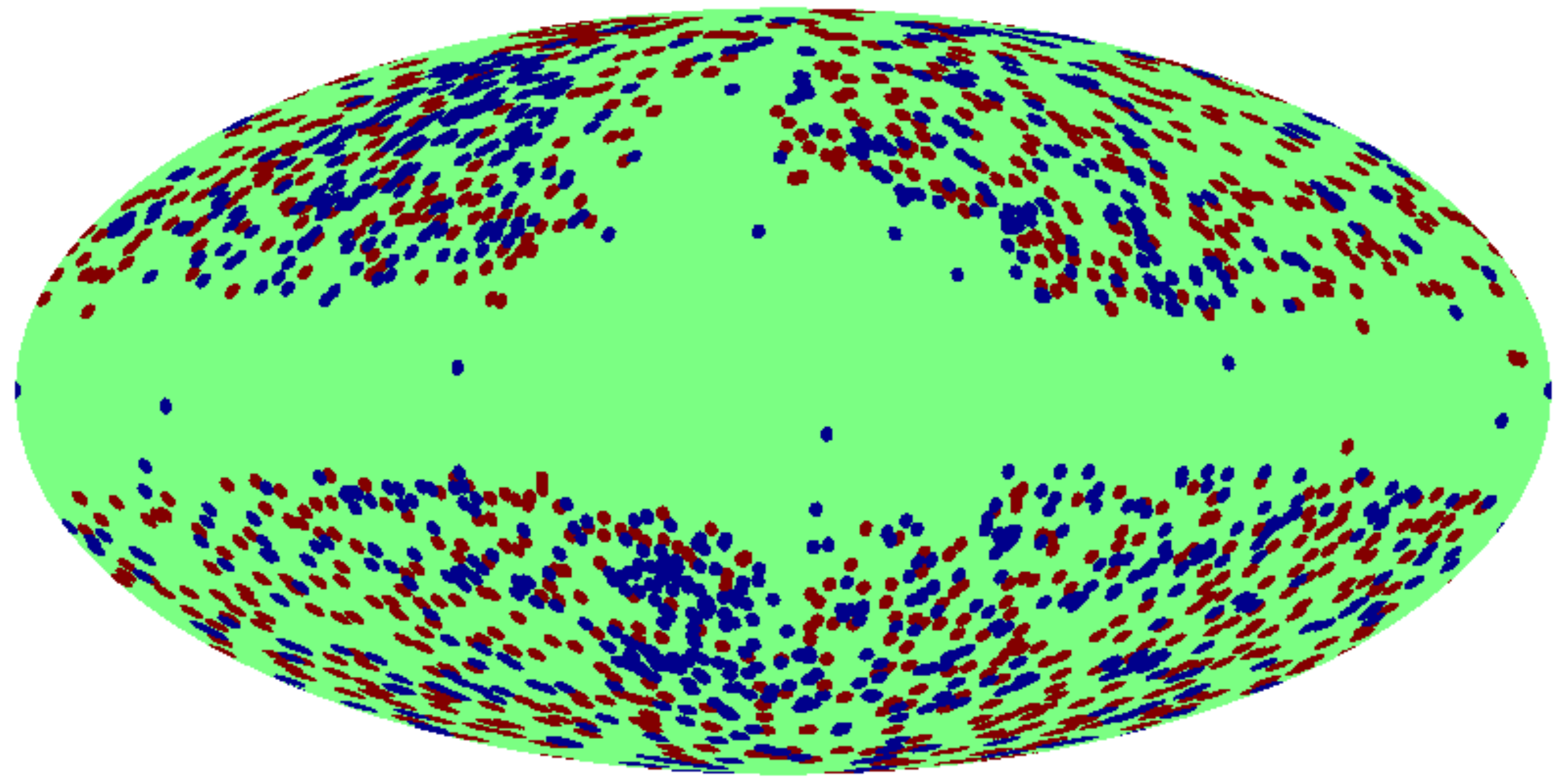}~\\
 \end{center}
\caption{Location of the extragalactic sources that have been fitted
  (red) or inpainted (blue).  \emph{Top}: set of sources processed at
  the first threshold ($>6$\,K over the background). \emph{Middle}:
  further set of sources subsequently processed at the second threshold
  ($>3$\,K over the background). \emph{Bottom}: additional set of sources finally
  processed at the third threshold ($1.5$\,K over the
  background). Note that we have attempted to retain the brightest
  Galactic sources in the map, hence their clear absence from the figure. 
}
\label{Fig:whichone}
\end{figure}

Figure~\ref{Fig:whichone} shows the location of the sources detected
in the Haslam map when processed using three different detection
thresholds, applied successively. We can see that there are more inpainted sources near the Galactic
plane (or regions of high background), whereas the Gaussian fitting
works better at higher Galactic latitude, as expected. 

A total number of 369 sources are  detected at the $6$\,K ($\sim
9$\,Jy) detection threshold with 202 extragalactic in origin. Since
most of the well-known Galactic sources are very bright, they should
be present in the list. We inspected each of the detected sources by
hand and compared it to sources in the NED\footnoteref{url:ned} and
SIMBAD\footnote{\url{http://simbad.u-strasbg.fr/simbad/}} online
databases. If a bright Galactic source was matched, we did not
subtract if from the map. At this threshold, 167 were found to be
Galactic.

After the first step at the threshold of 6\,K, fainter radio sources
are still found to contaminate the diffuse $408$\,MHz map. We
therefore need to use a deeper detection threshold for the
desourcing. However, instead of desourcing the raw Haslam map at a
deeper threshold, we repeat the processing directly on the
$6$\,K-processed map. Specifically, we search for the sources in the
$6$\,K-processed map that are $3$\,K ($\sim 5$\,Jy) over the
background, and we perform either Gaussian fitting or inpainting on
the $6$\,K-processed map. For the $3$\,K threshold,
$\chi^2_m = 2773.7$, and a total of 924 sources are
detected, with 462 corresponding to Galactic sources, so that only the
remaining 462 are then subtracted (middle panel of
Fig. \ref{Fig:whichone}).  We then repeat the multi-stage processing
again on the $3$\,K-processed map and locate 1613 more sources that
are $1.5$\,K ($\sim 2$\,Jy) over the background (2870 point-like
sources are detected at this threshold but 1257 were found to be
Galactic in origin). The use of successive detection thresholds for
desourcing is motivated by the necessity to avoid overlapping sources
in the processing. We stop the desourcing at the threshold of $1.5$\,K
since this is approaching the level of residual artefacts (striations)
and noise in the map. This is then the final processed map.

\begin{figure*}
  \begin{center}
    \includegraphics[width=0.33\textwidth]{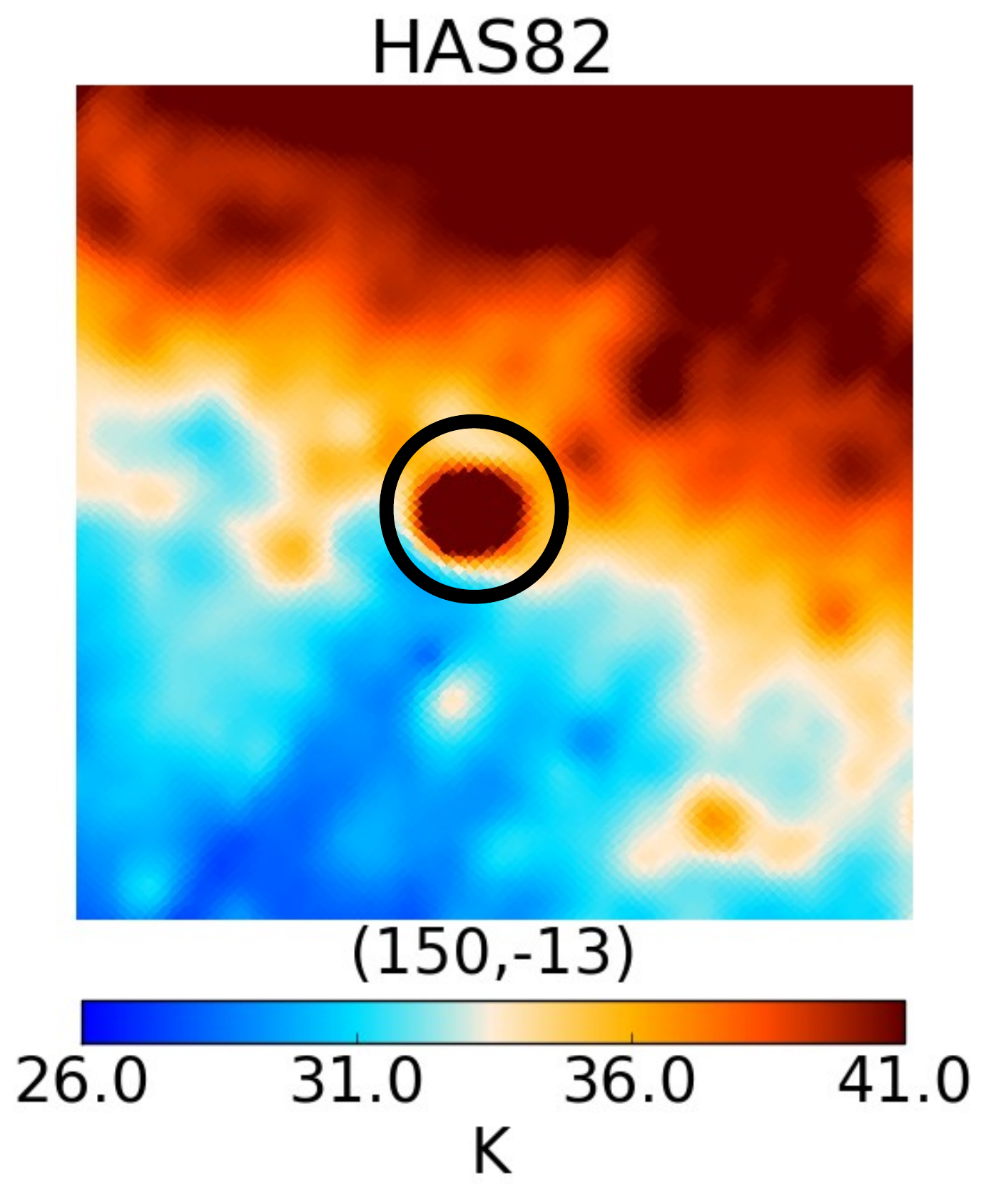}~
    \includegraphics[width=0.33\textwidth]{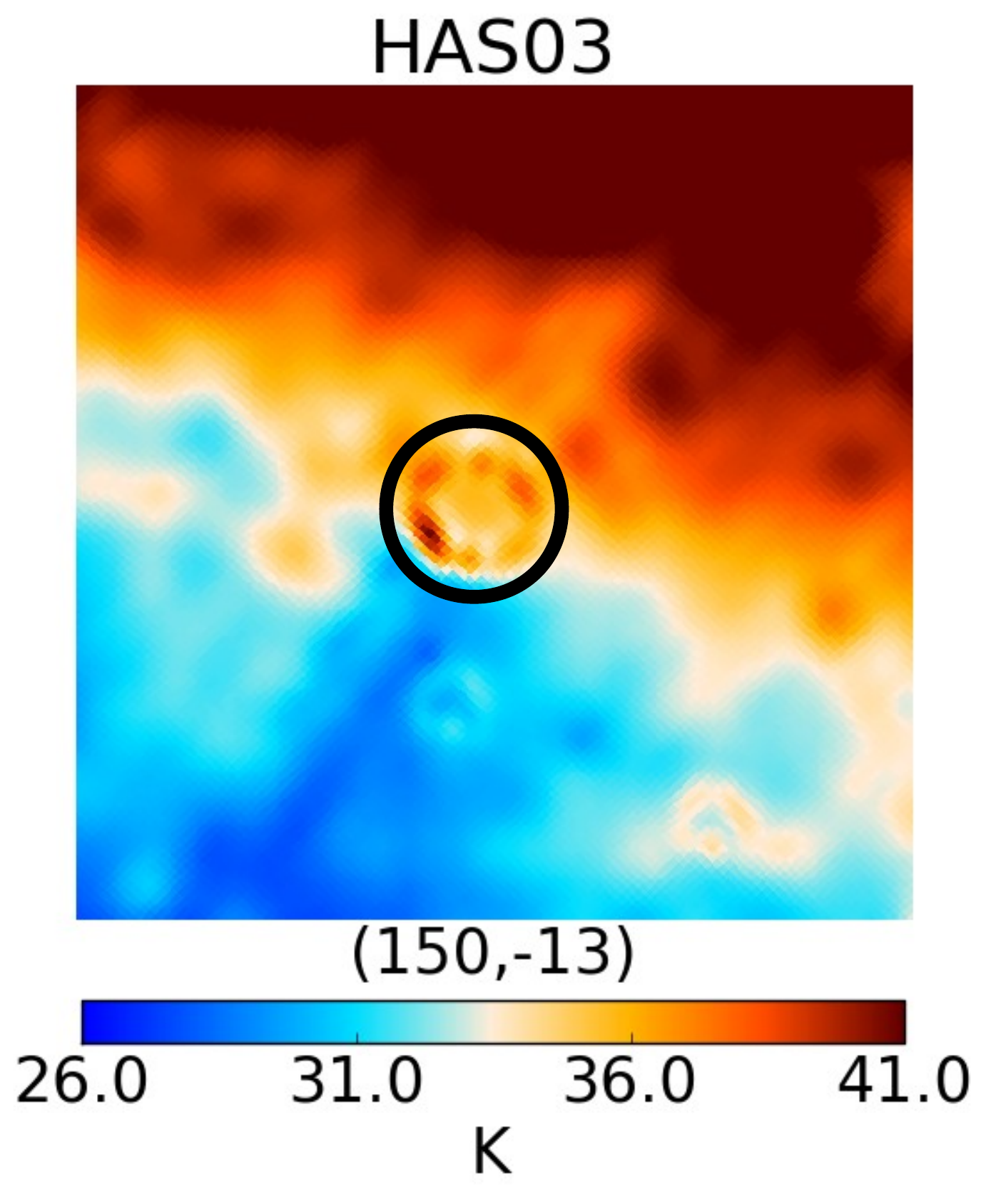}~
    \includegraphics[width=0.33\textwidth]{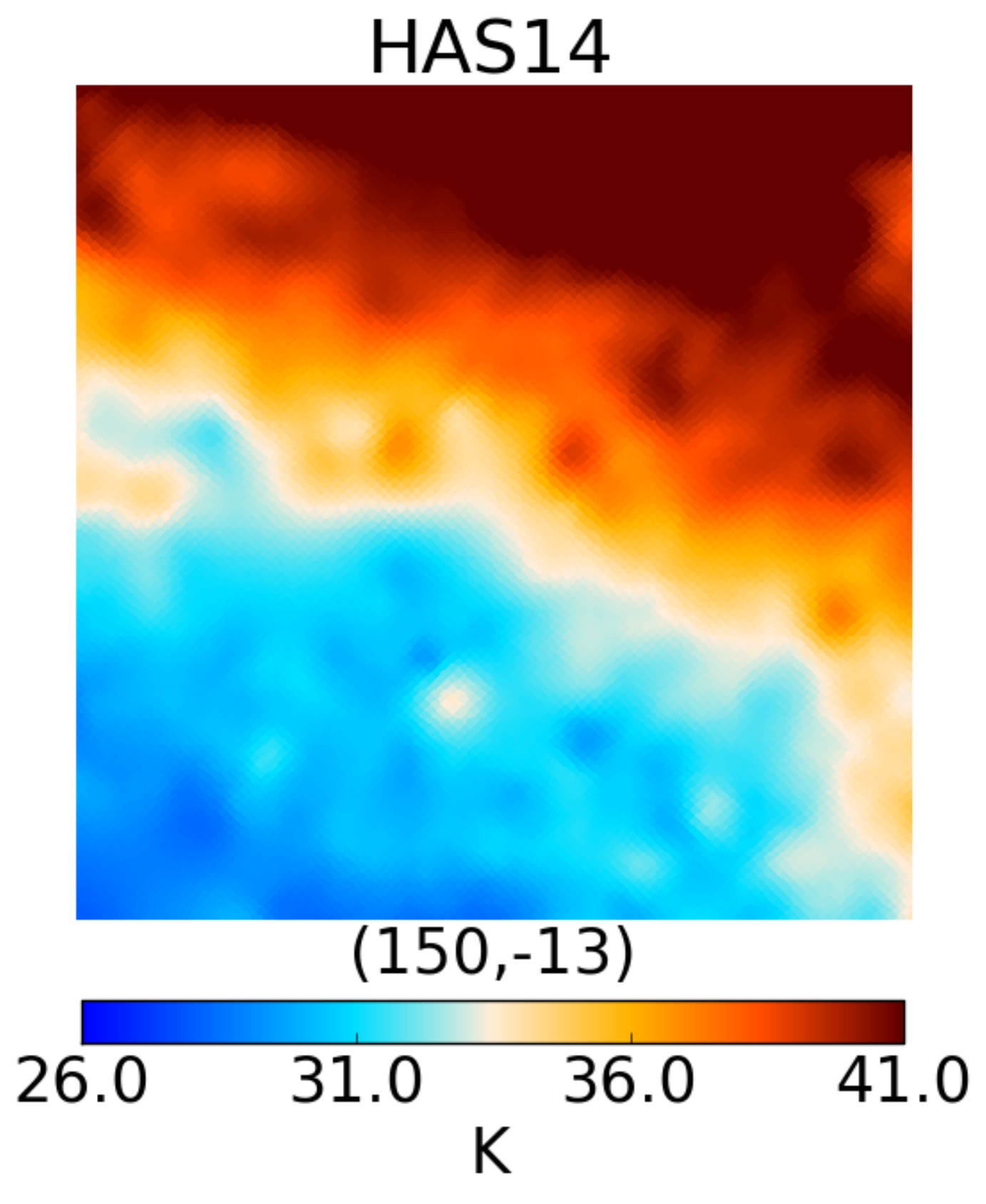}~   
    \\
    \includegraphics[width=0.33\textwidth]{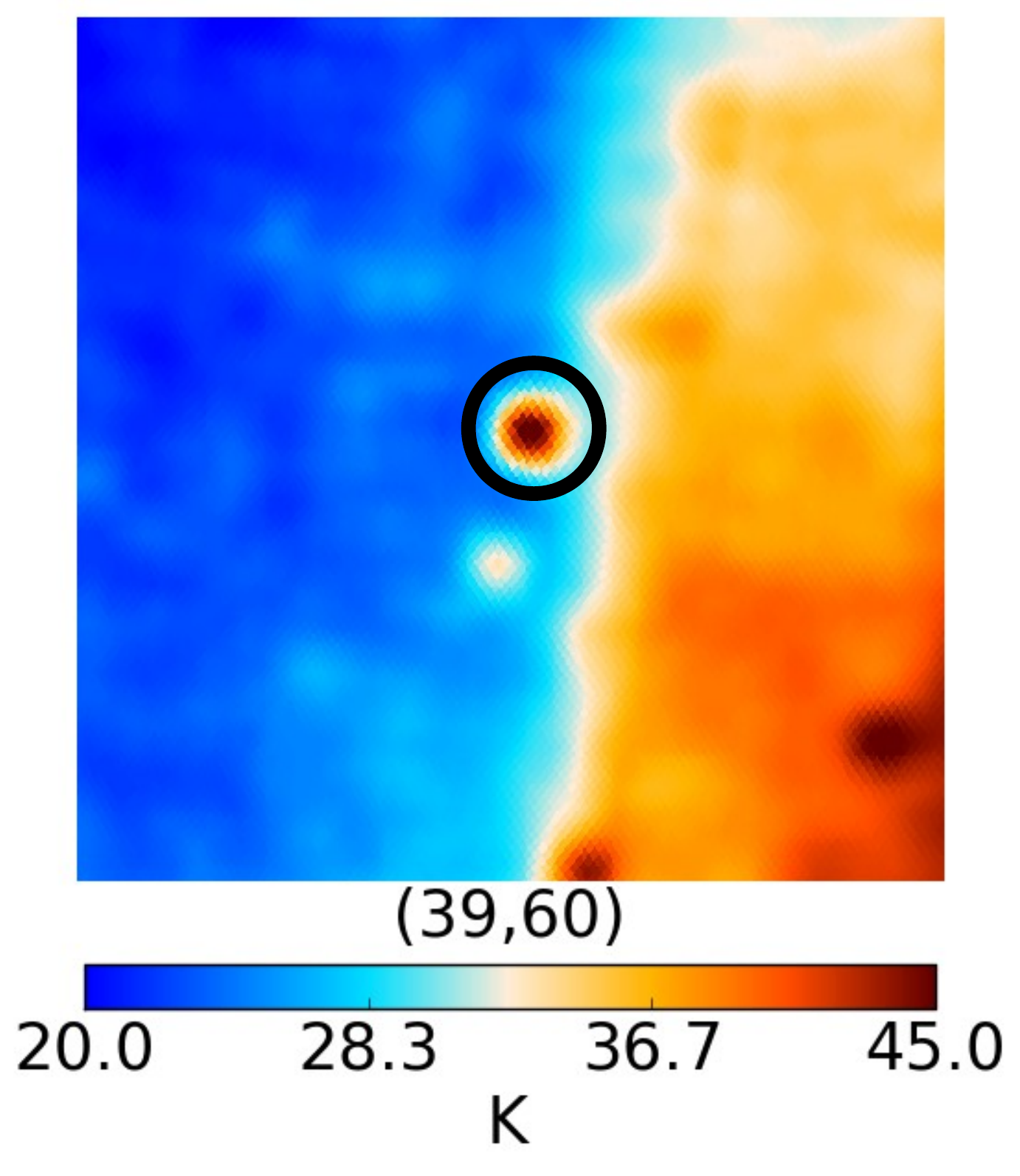}~
    \includegraphics[width=0.33\textwidth]{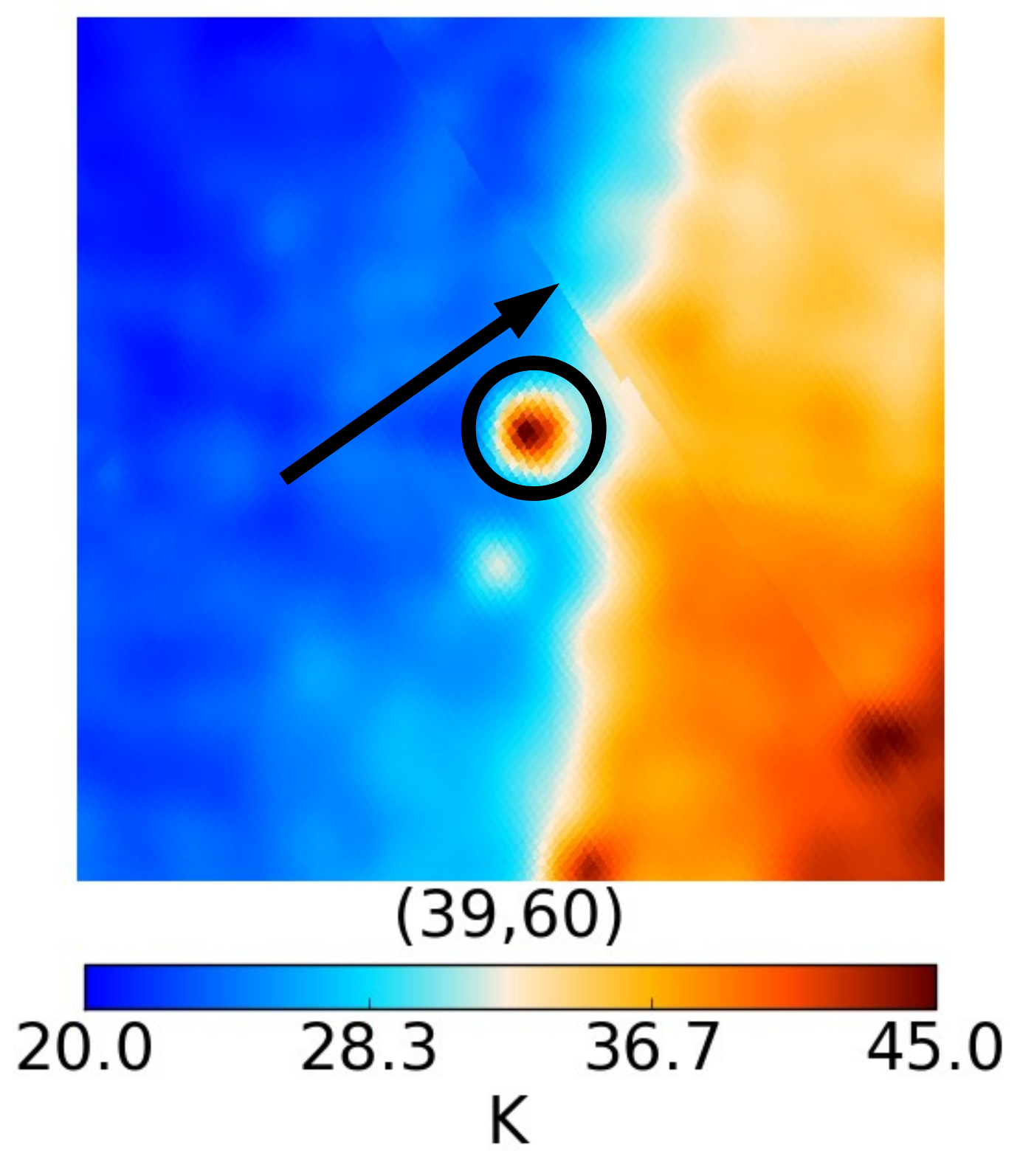}~
    \includegraphics[width=0.33\textwidth]{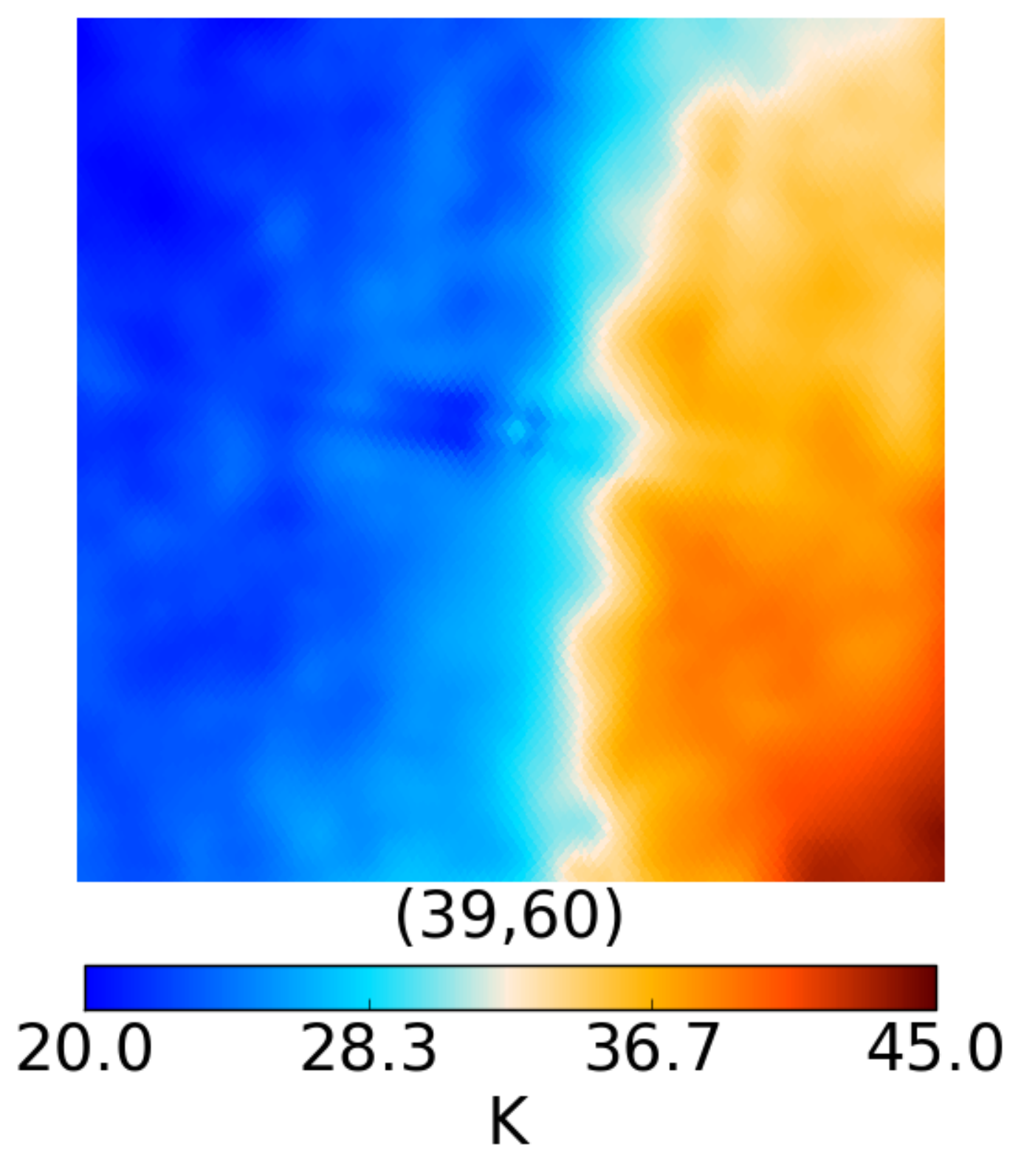}~   
    \\
    \includegraphics[width=0.33\textwidth]{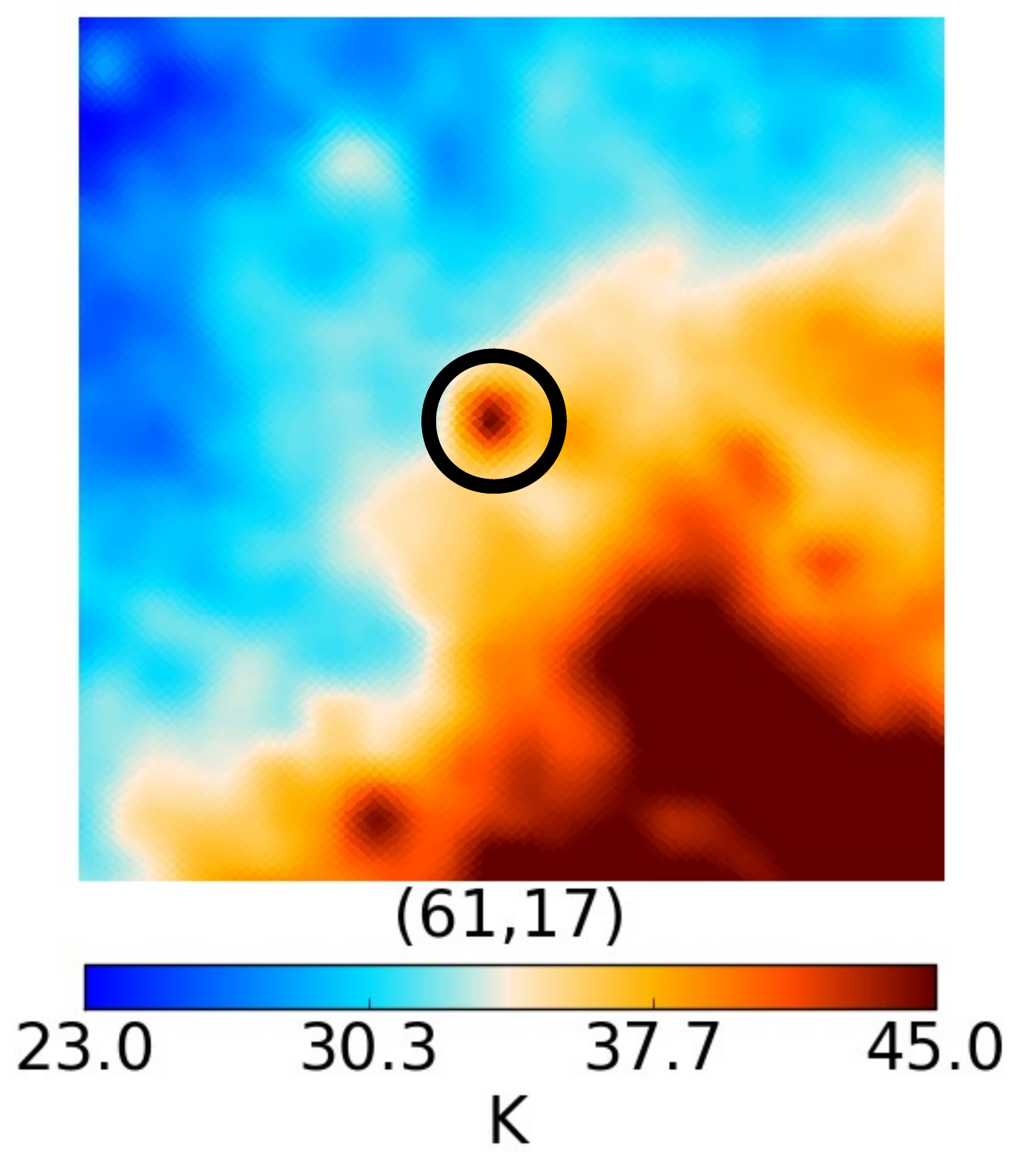}~    
    \includegraphics[width=0.33\textwidth]{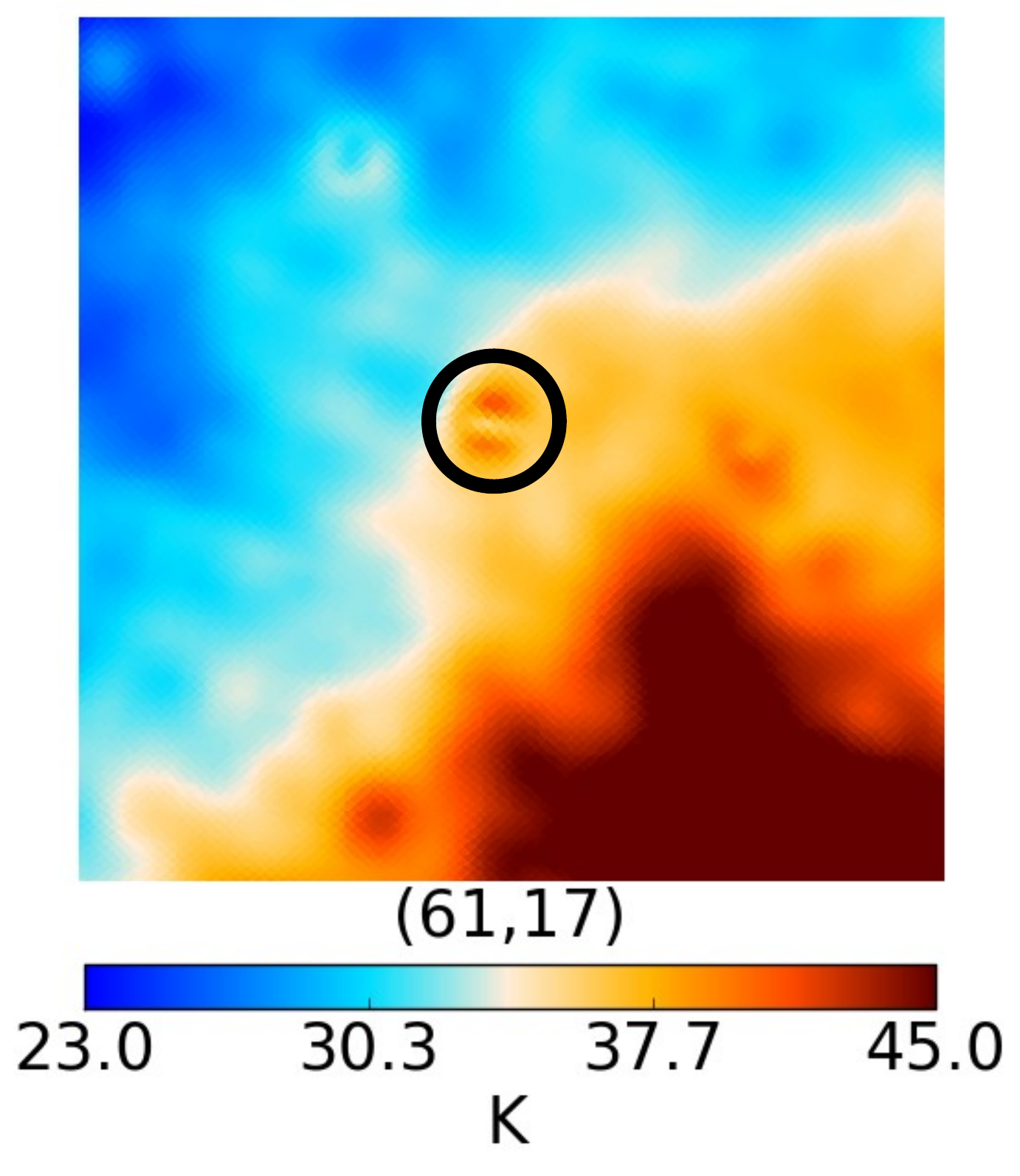}~
    \includegraphics[width=0.33\textwidth]{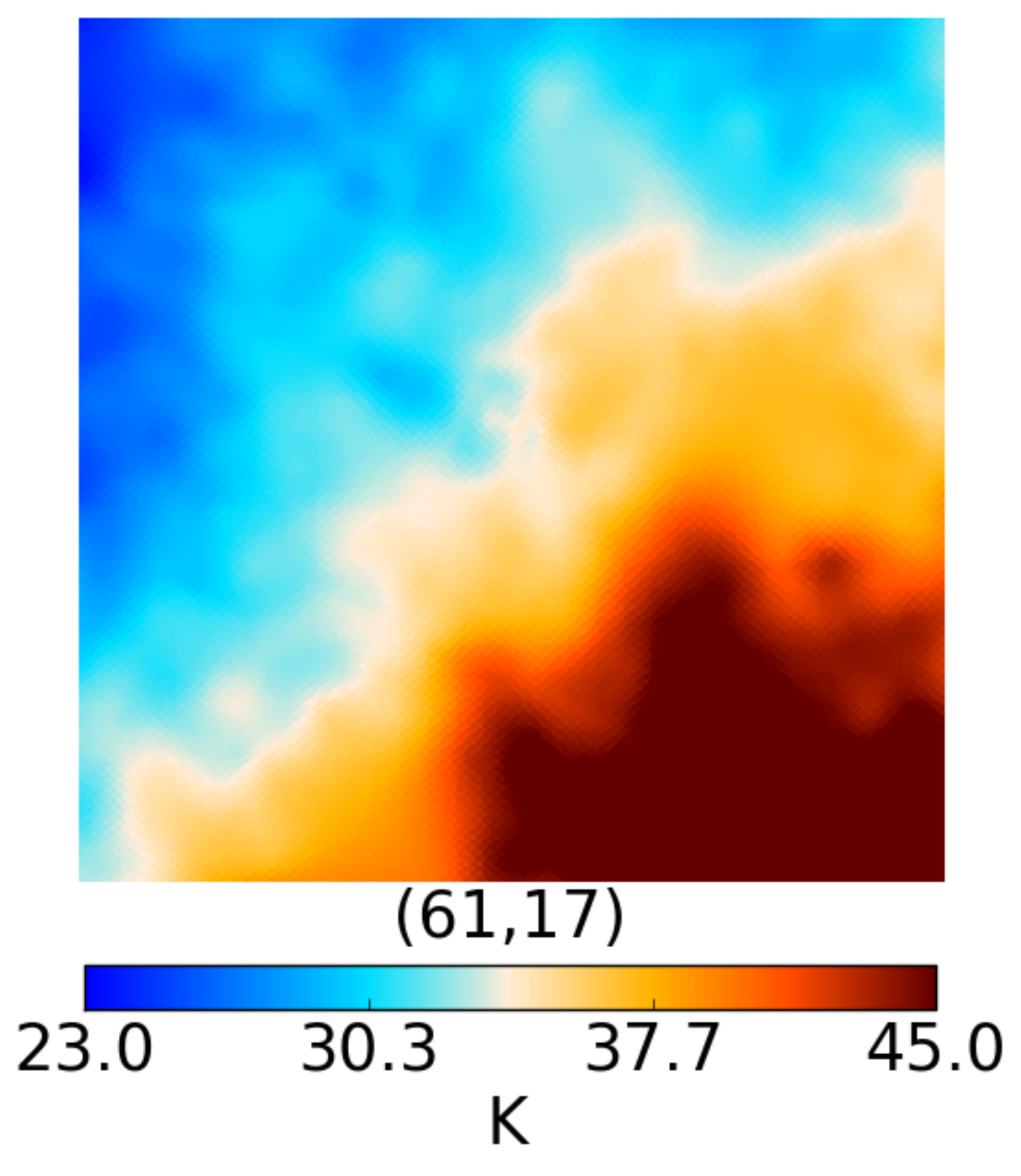}~  
\end{center}
\caption{Three $12.5^\circ\times 12.5^\circ$ gnomonic projections
  (\emph{first row} to \emph{third row}) of the raw Haslam map
  (\emph{left column}), of the HAS03 post-processed version
  (\emph{middle column}), and of our reprocessed 408\,MHz sky map (HAS14) 
  (\emph{right column}). Note that the HAS03 post-processing is based
  on the NCSA map whereas that described here utilises the ECP map.}
\label{Fig:final}
\end{figure*}

\begin{figure*}
  \begin{center}
    \includegraphics[width=0.33\textwidth]{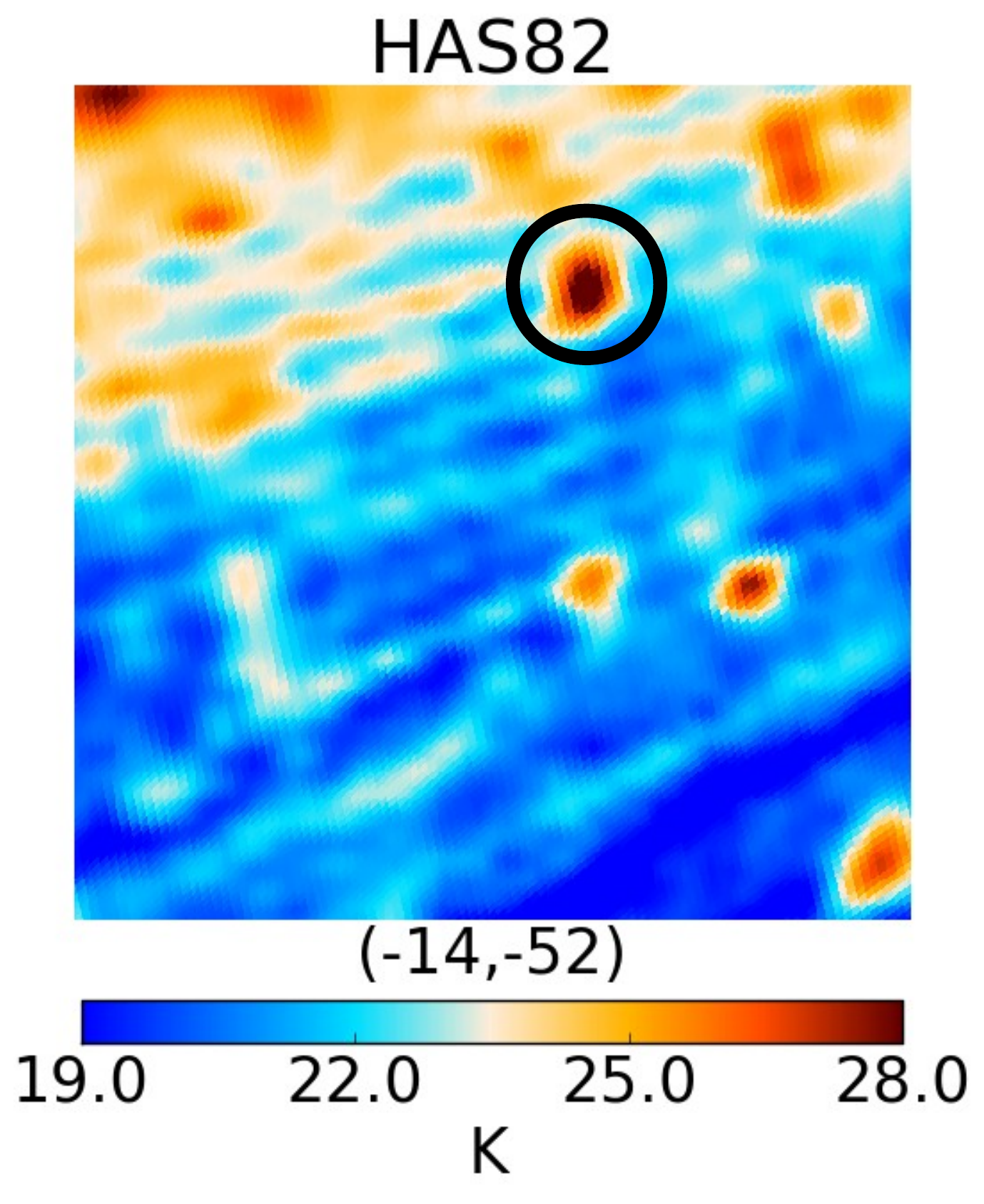}~    
    \includegraphics[width=0.33\textwidth]{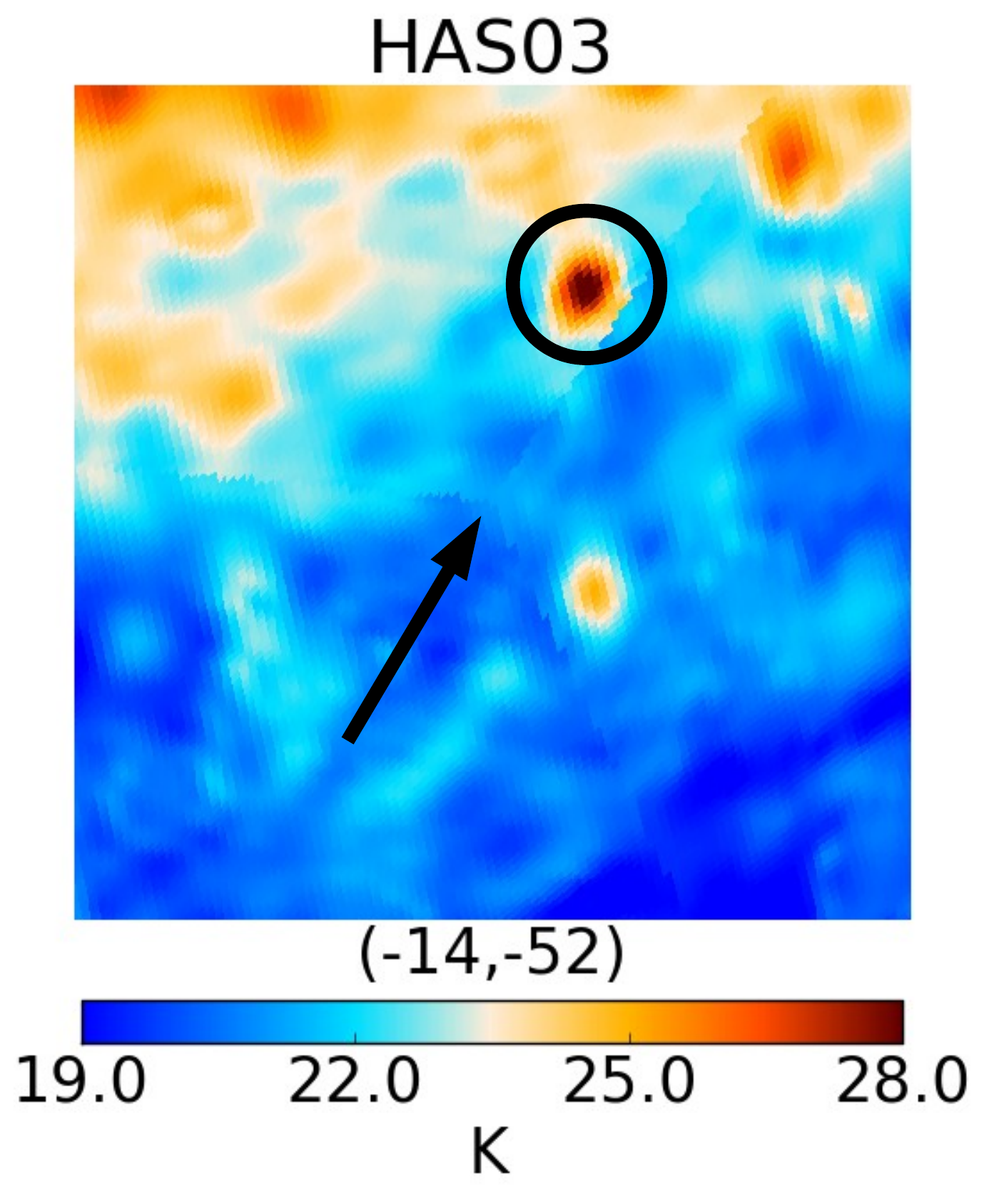}~
    \includegraphics[width=0.33\textwidth]{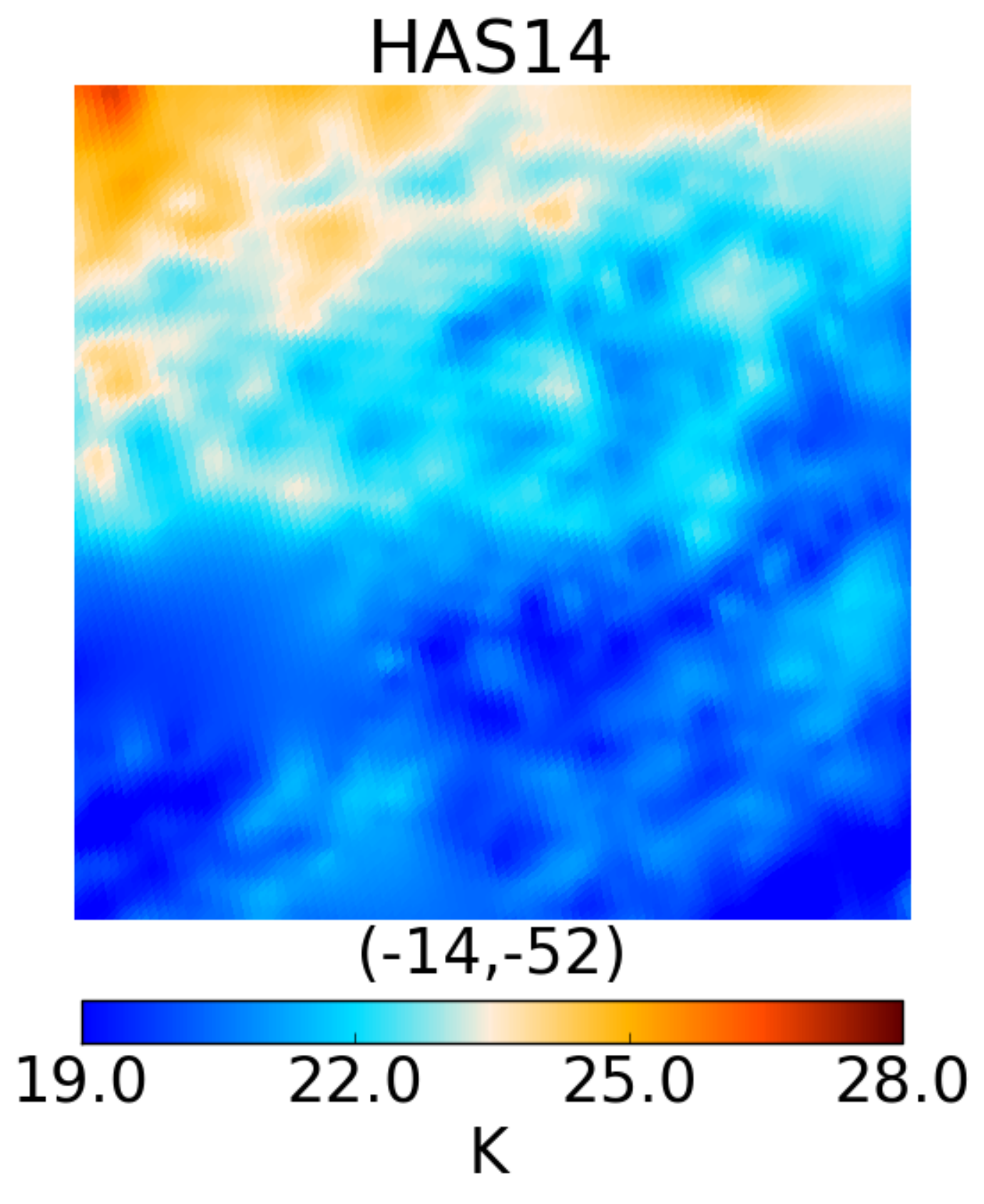}~    
    \\
    \includegraphics[width=0.33\textwidth]{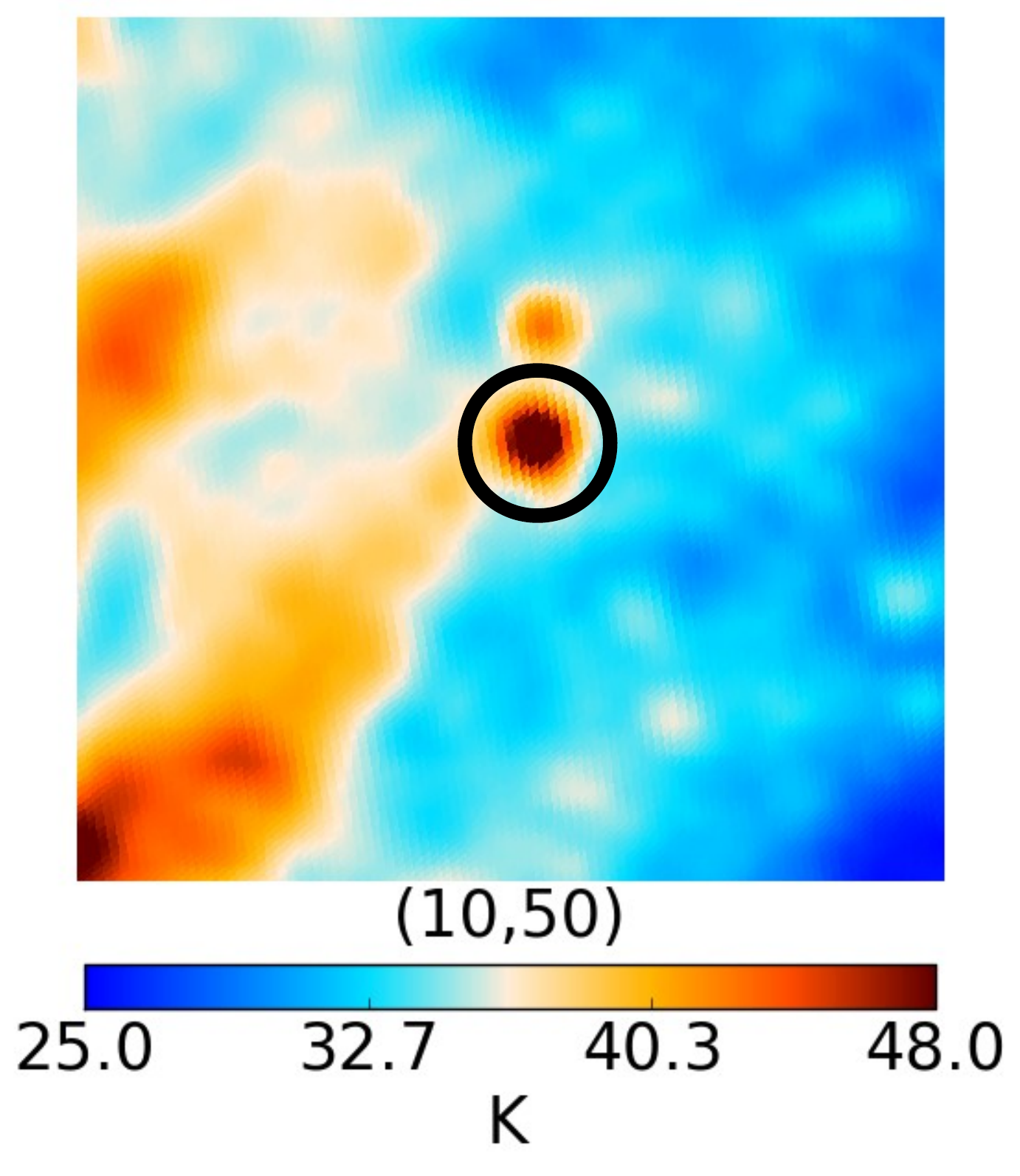}~ 
    \includegraphics[width=0.33\textwidth]{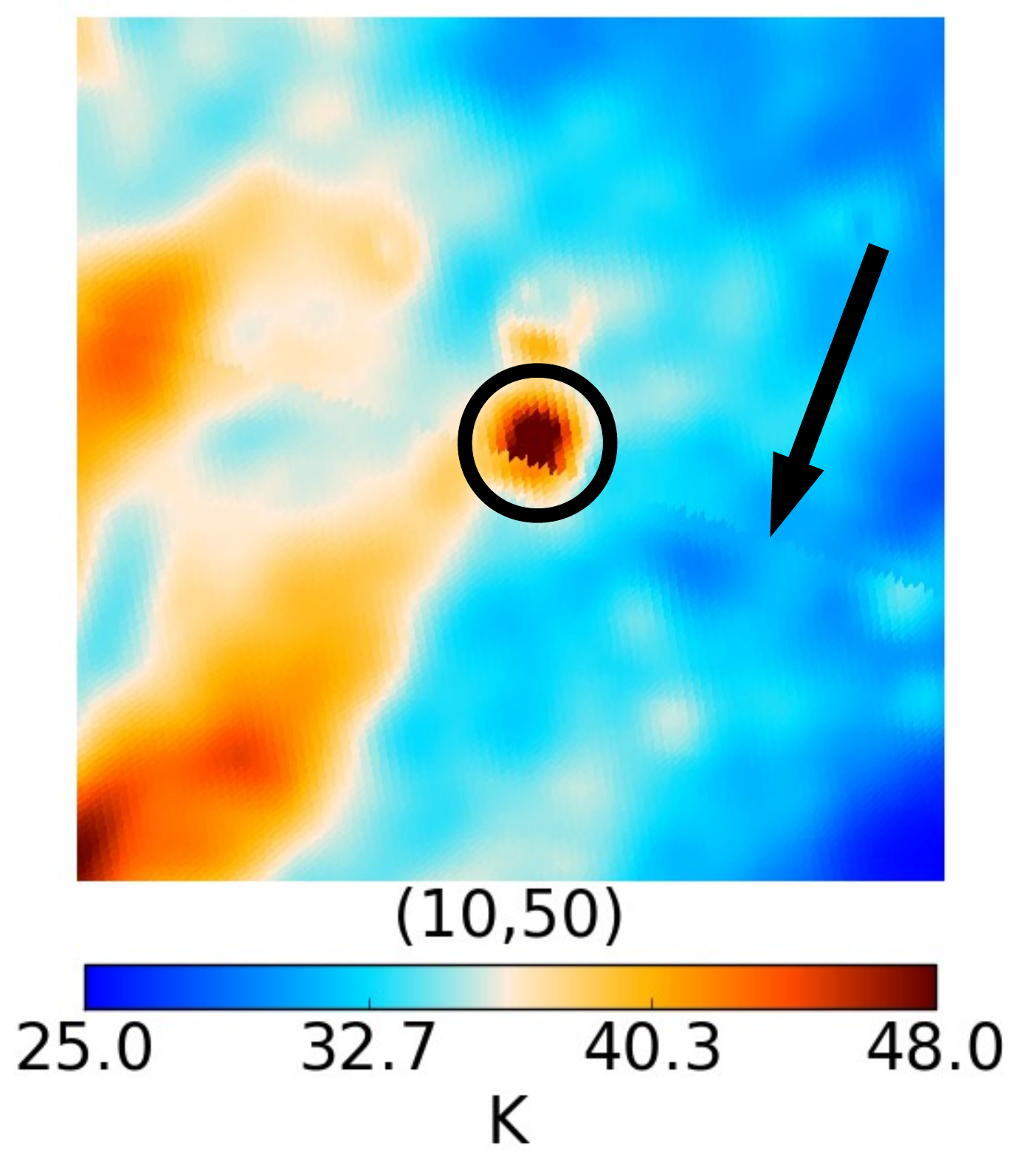}~
    \includegraphics[width=0.33\textwidth]{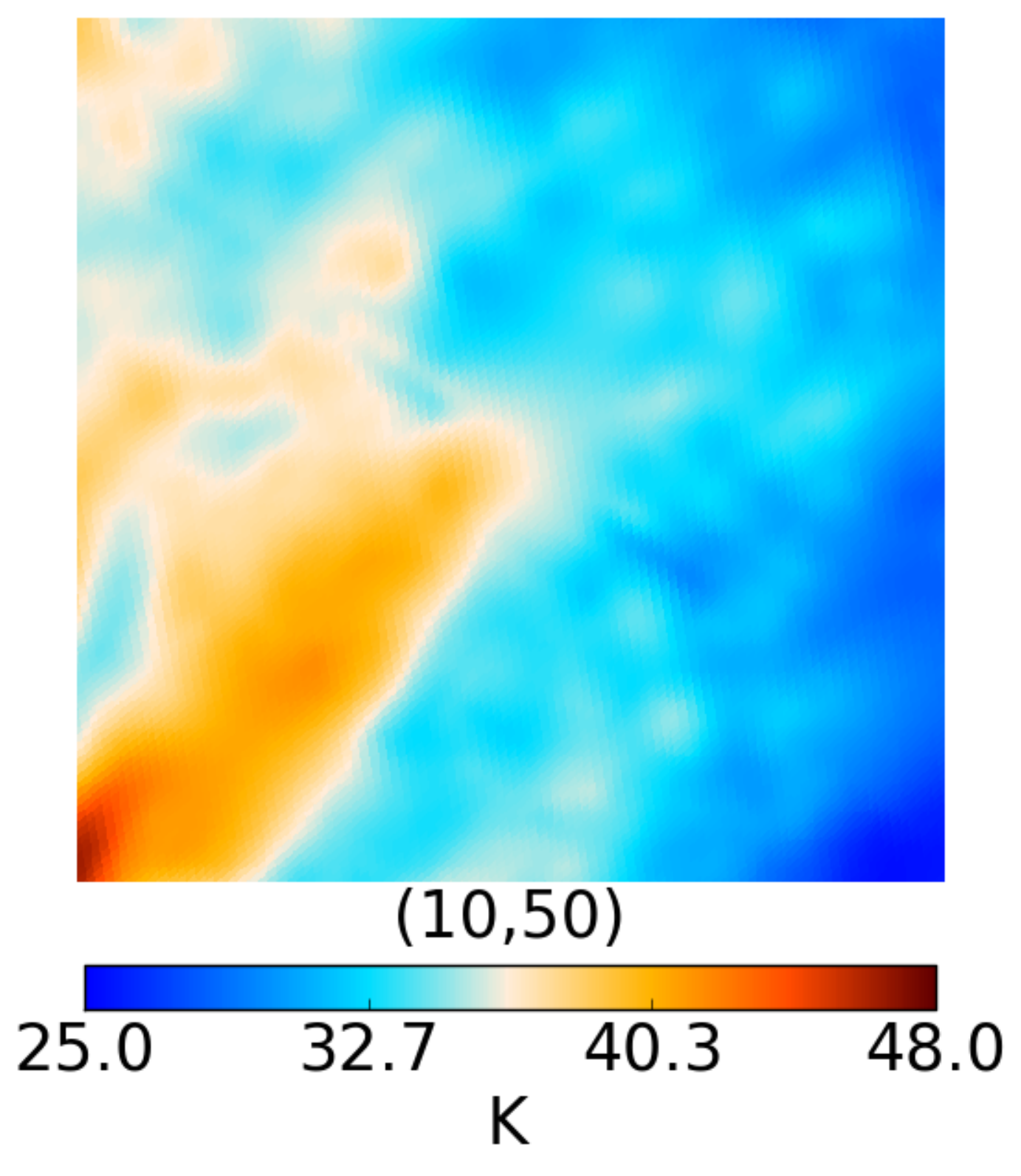}~  
    \\
    \includegraphics[width=0.33\textwidth]{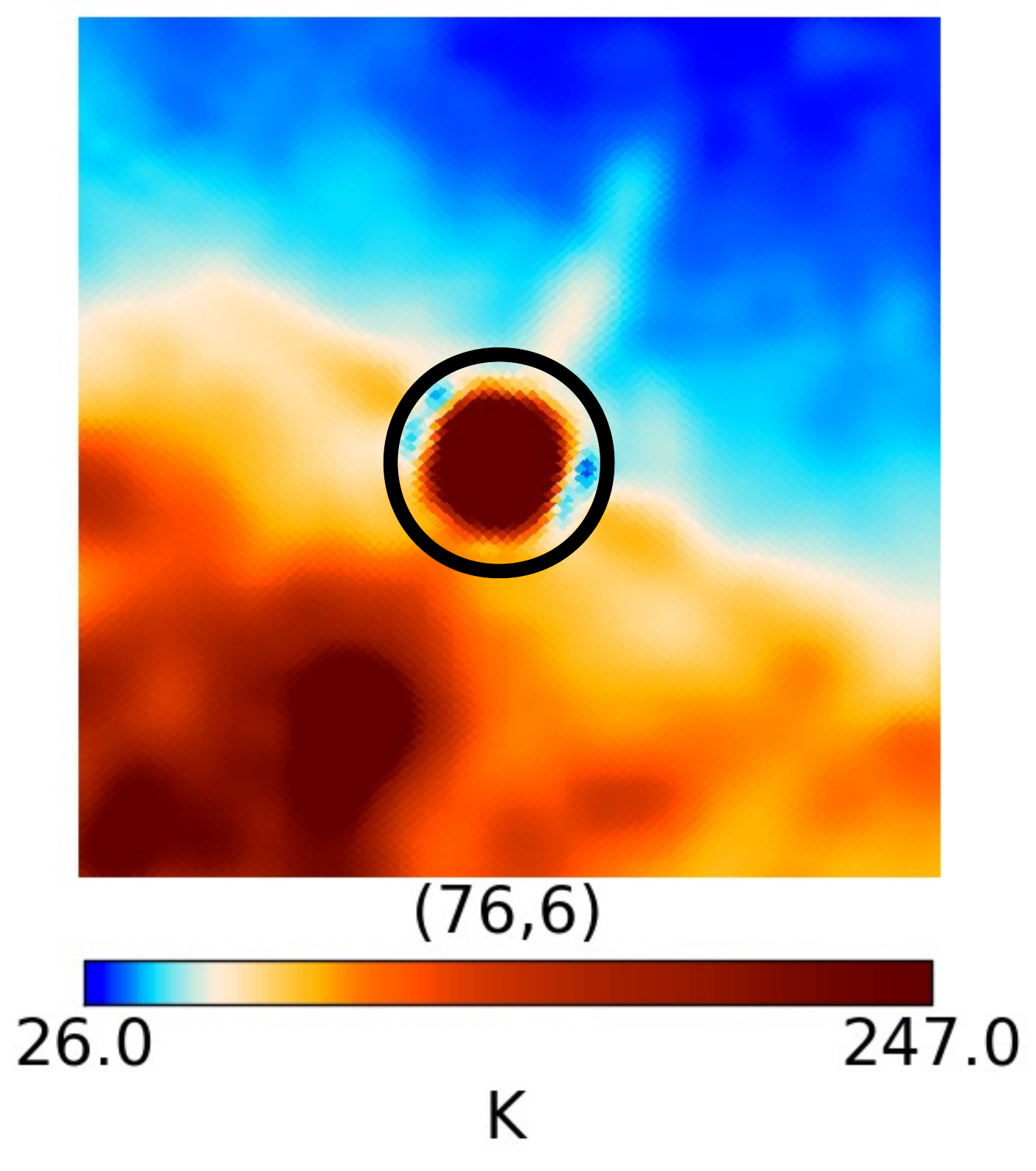}~    
    \includegraphics[width=0.33\textwidth]{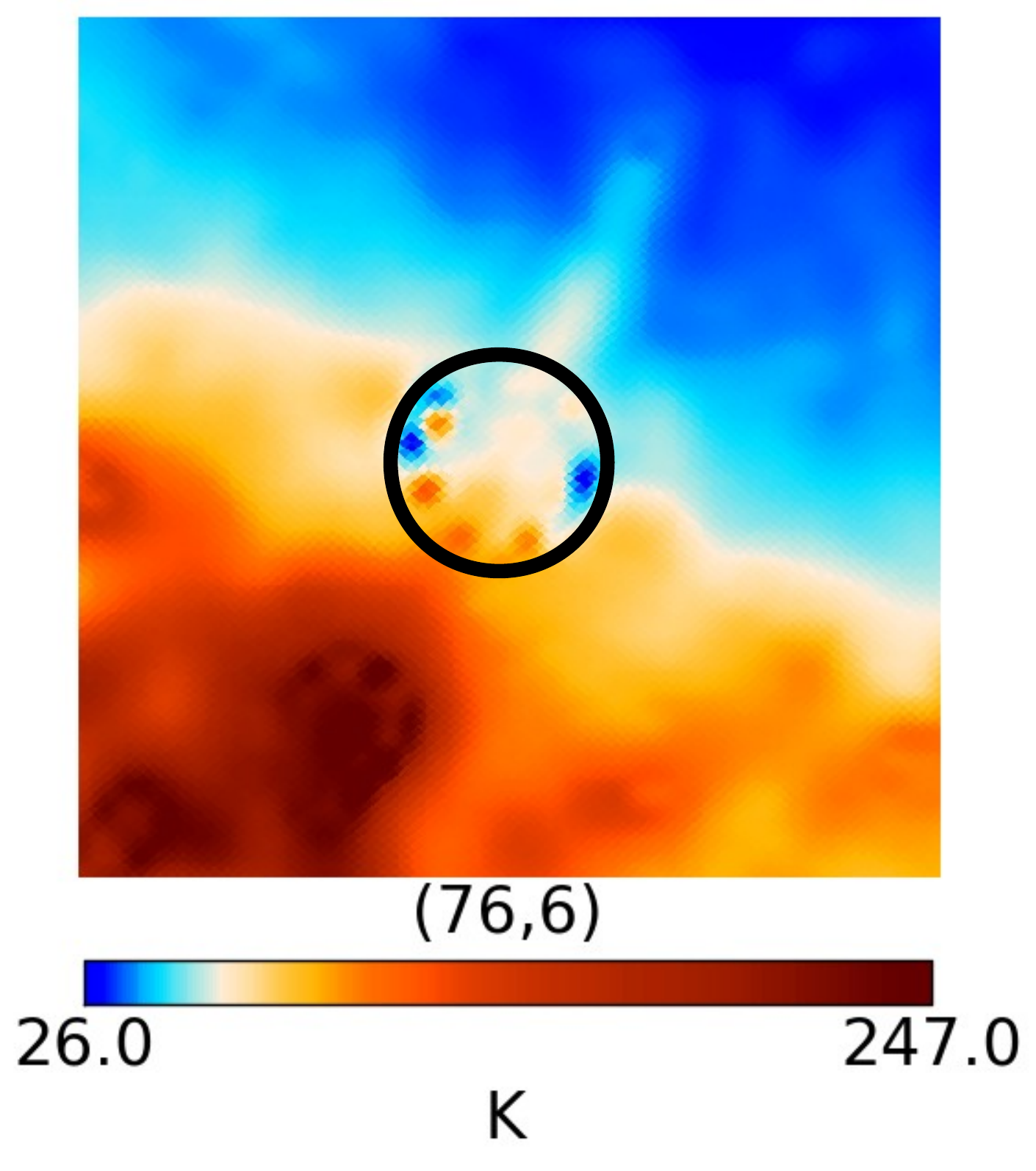}~
    \includegraphics[width=0.33\textwidth]{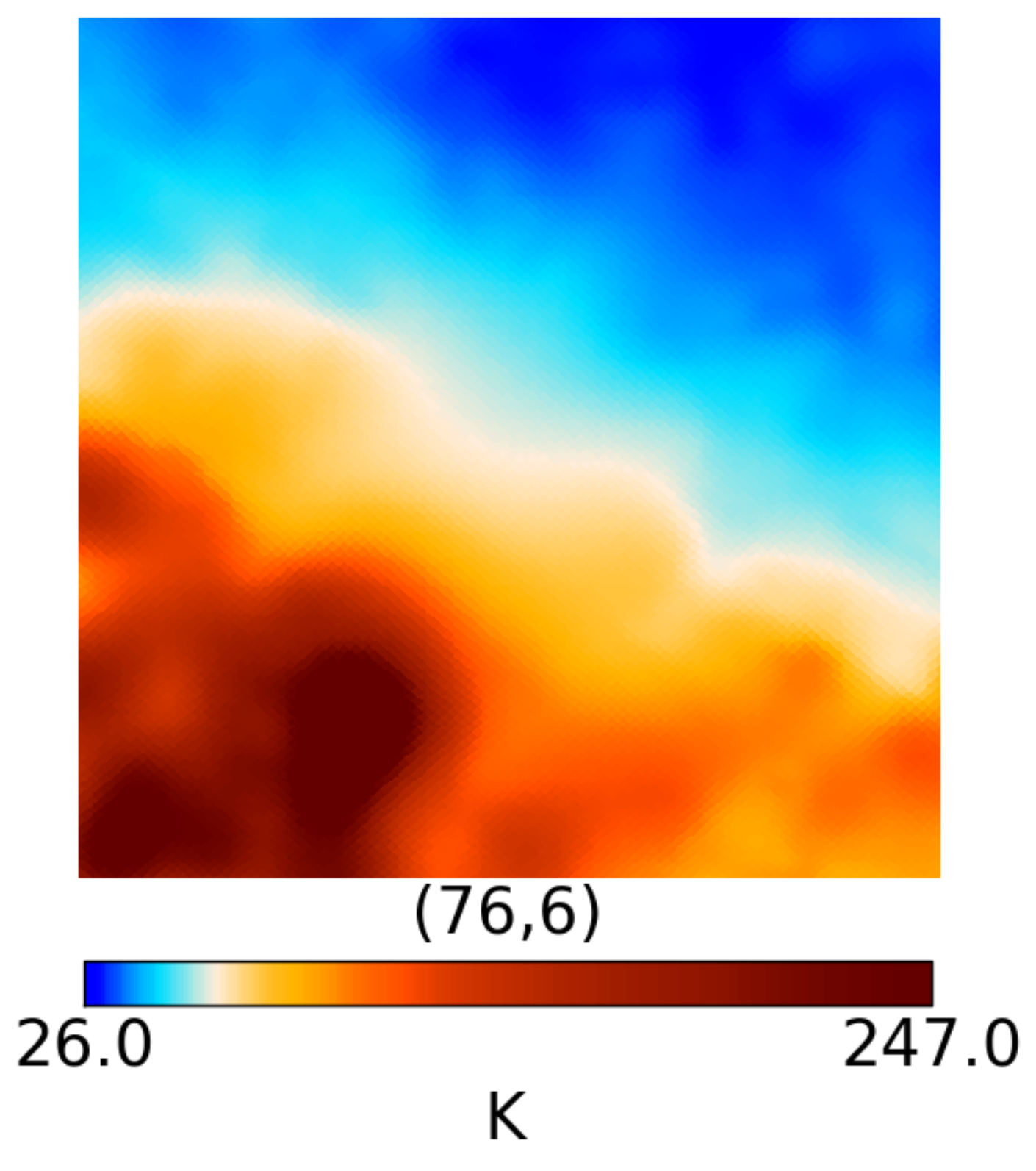}~
  \end{center}
\caption{Three more $12.5^\circ\times 12.5^\circ$ gnomonic projections
  (\emph{first row} to \emph{third row}) of the raw Haslam map
  (\emph{left column}), the HAS03 post-processed version
  (\emph{middle column}), and of our reprocessed 408\,MHz sky map (HAS14)
  (\emph{right column}).}
\label{Fig:final2}
\end{figure*}

\begin{figure*}
    \includegraphics[width=0.33\textwidth]{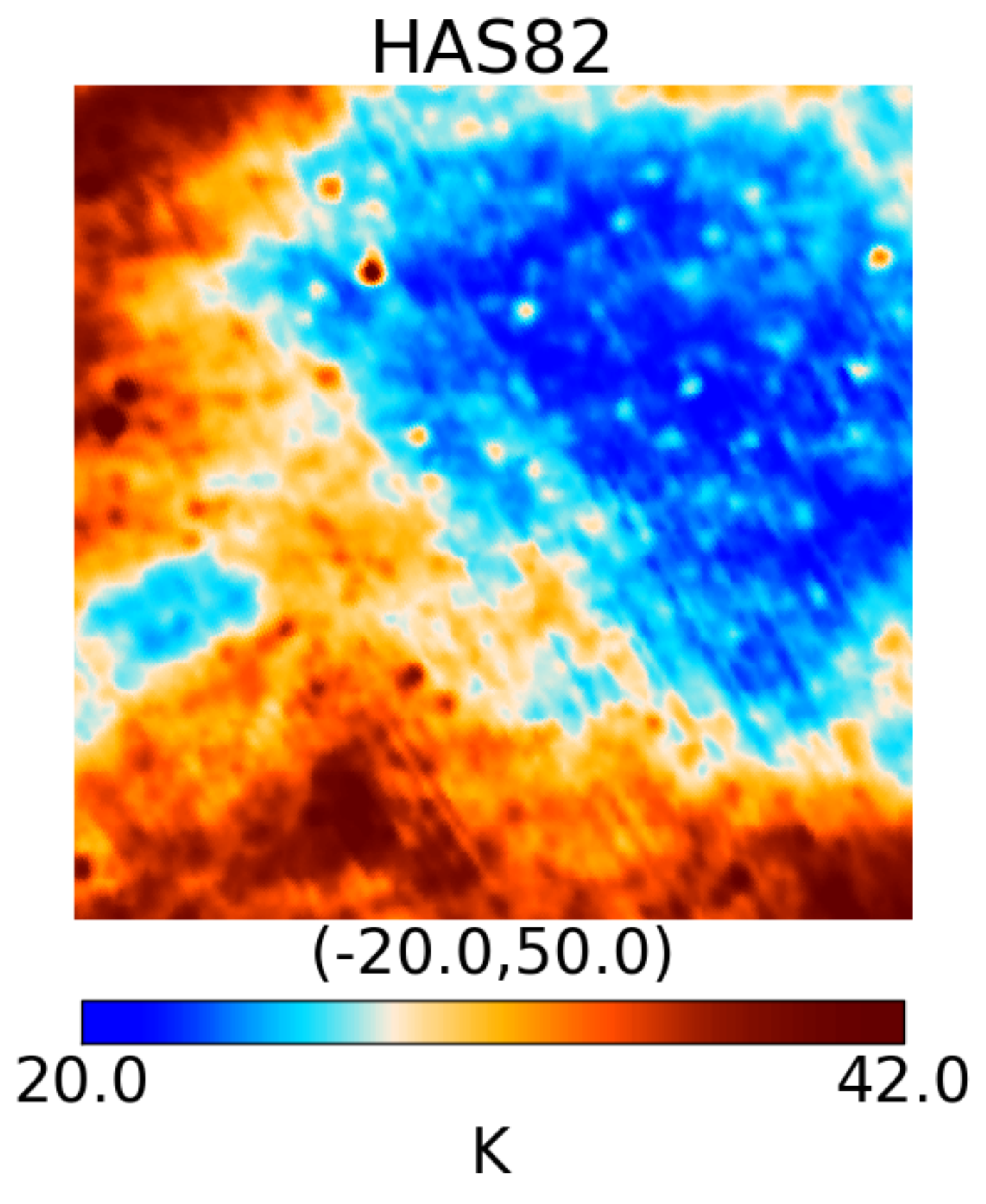}~    
    \includegraphics[width=0.33\textwidth]{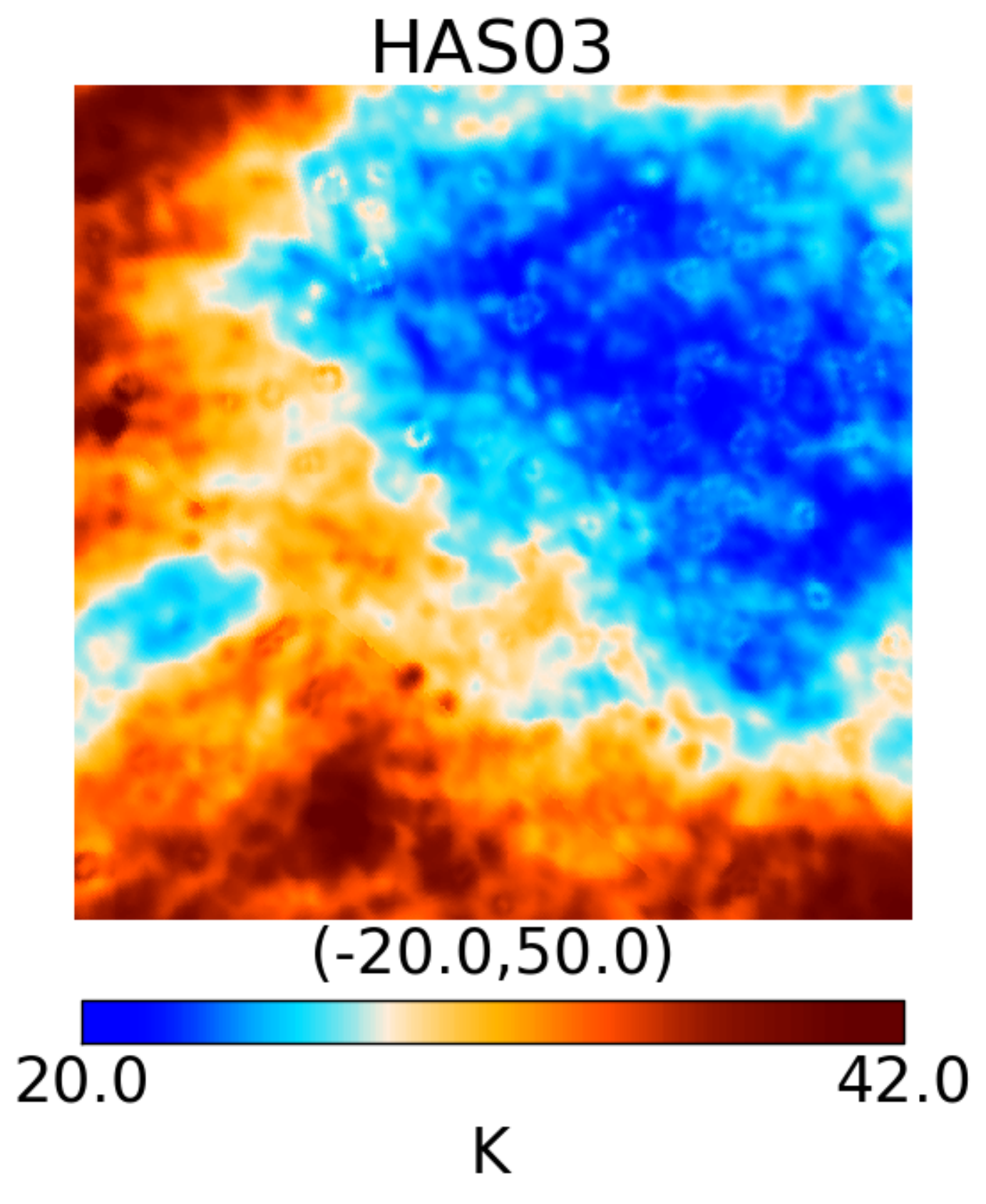}~
    \includegraphics[width=0.33\textwidth]{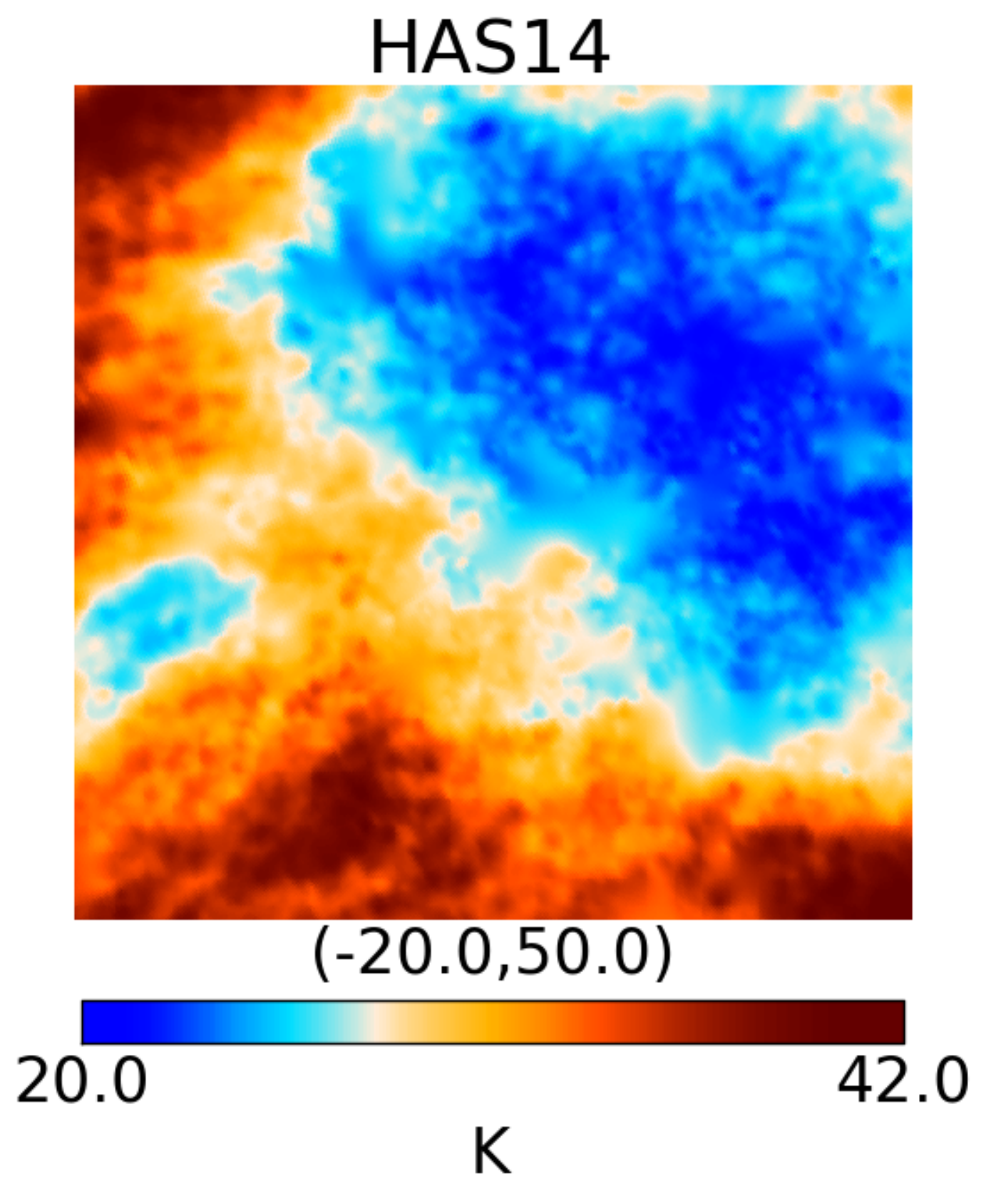}~   
\caption{A $43^\circ\times 43^\circ$ gnomonic projection of the
  408\,MHz map centred at high latitude ($l,b)=(340^{\circ},
  +50^{\circ}$) for the raw ({\it left}), HAS03 ({\it middle}) and
   HAS14 ({\it right}) versions of the map.  Both striations
  and numerous extragalactic radio
  sources are are evident across the raw map. The HAS03 version also contains
  artefacts, including source residuals, unremoved
  sources, and line features due to remapping to a {\tt quadcube}
  pixelization. In our map, the striations and source residuals have
  been minimised.}
\label{Fig:hl-large-area}
\end{figure*}

The new filtered 408\,MHz sky map (HAS14) is presented for different patches
of the sky and compared to the HAS03 version in
Figs.~\ref{Fig:final}, \ref{Fig:final2} and
\ref{Fig:hl-large-area}. The first column panels of
Figs.~\ref{Fig:final}, \ref{Fig:final2} and \ref{Fig:hl-large-area}
show the source contaminated ECP raw map, where circles are used to
highlight the sources. The HAS03 map is shown in the second
column, which clearly shows many artefacts, including edges due to the
{\tt quadcube} projection, residual source artefacts, and unsubtracted
sources. The third column shows the new (HAS14) map using the
combination of inpainting and Gaussian fitting. We can see a clear
improvement by comparison to the HAS03 version. The {\tt quadcube}
artefacts are no longer present and there are no strong residual
source artefacts.

We can make an estimate of the depth (in terms of flux density) to
which we have removed sources from the Haslam map at high Galactic
latitudes where the background confusion is minimised. We do this
using the source flux catalogue of \citet{Kuehr1981}, which lists all
the radio sources detected with a flux density larger than $\sim
1$\,Jy at $5$\,GHz. We select a low-background area of the sky in the
new Haslam map, $\mbox{R.A.}\in[07^h,18^h]$ and
$\mbox{Dec.}\in[30^\circ,90^\circ]$. In this area, the catalogue lists
$2$ radio sources having a flux density $S$ larger than $20$\,Jy at
$408$\,MHz, $3$ sources with $S > 10$\,Jy, $11$ sources with
$S > 5$\,Jy, $18$ sources with $S > 3$\,Jy, $46$ sources $S > 2$\,Jy,
and a total of $62$ sources with $S > 1$\,Jy. We have identified in
this area of the sky the sources that have been removed in the new
Haslam map and we have computed the completeness of our processing:
$100$\% of the radio sources having a flux density $S > 5$\,Jy at
$408$\,MHz have been removed, $89\%$ of the sources with $S > 3$\,Jy,
$83$\% of the sources with $S > 2$\,Jy, and $68\%$ of the sources with
$S > 1$\,Jy.

Using the completeness we also give an estimate of the confusion noise
in the new Haslam map. We use the $408$\,MHz source count (i.e., the
cumulative distribution of the number of sources, $dN/dS$, brighter
than a given flux density) of \citet{1984ApJ...287..461C} to fit a
power law $N(S) =\alpha S^{-\gamma}$ in the range $2$\,Jy--$10$\,Jy,
where $S_{\rm lim} =2$\,Jy is roughly the completeness limit ($\ge
80\,\%$ over this flux density range). We find $\alpha\sim 67$ and
$\gamma\sim -2.8$. We then apply the formula for the confusion noise in
the Gaussian beam approximation \citep{1974ApJ...188..279C}
\begin{align}\label{eq:noise}
\sigma_c = {\alpha\over 3-\gamma}{\pi\theta_b^2\over 4\ln 2}\left(S_{\rm lim}\right)^{3-\gamma},
\end{align}
where $\theta_b\sim 56$\,arcmin is the beam FWHM and $S_{\rm lim}$ is
the upper limit of flux density. Based on completeness,
the upper limit of flux density is $S_{\rm lim} = 2$\,Jy, and we find
that the confusion noise is $\sigma_c \sim 0.1$\,K in the new Haslam
map.

\begin{figure}
  \begin{center}
    \includegraphics[width=\columnwidth]{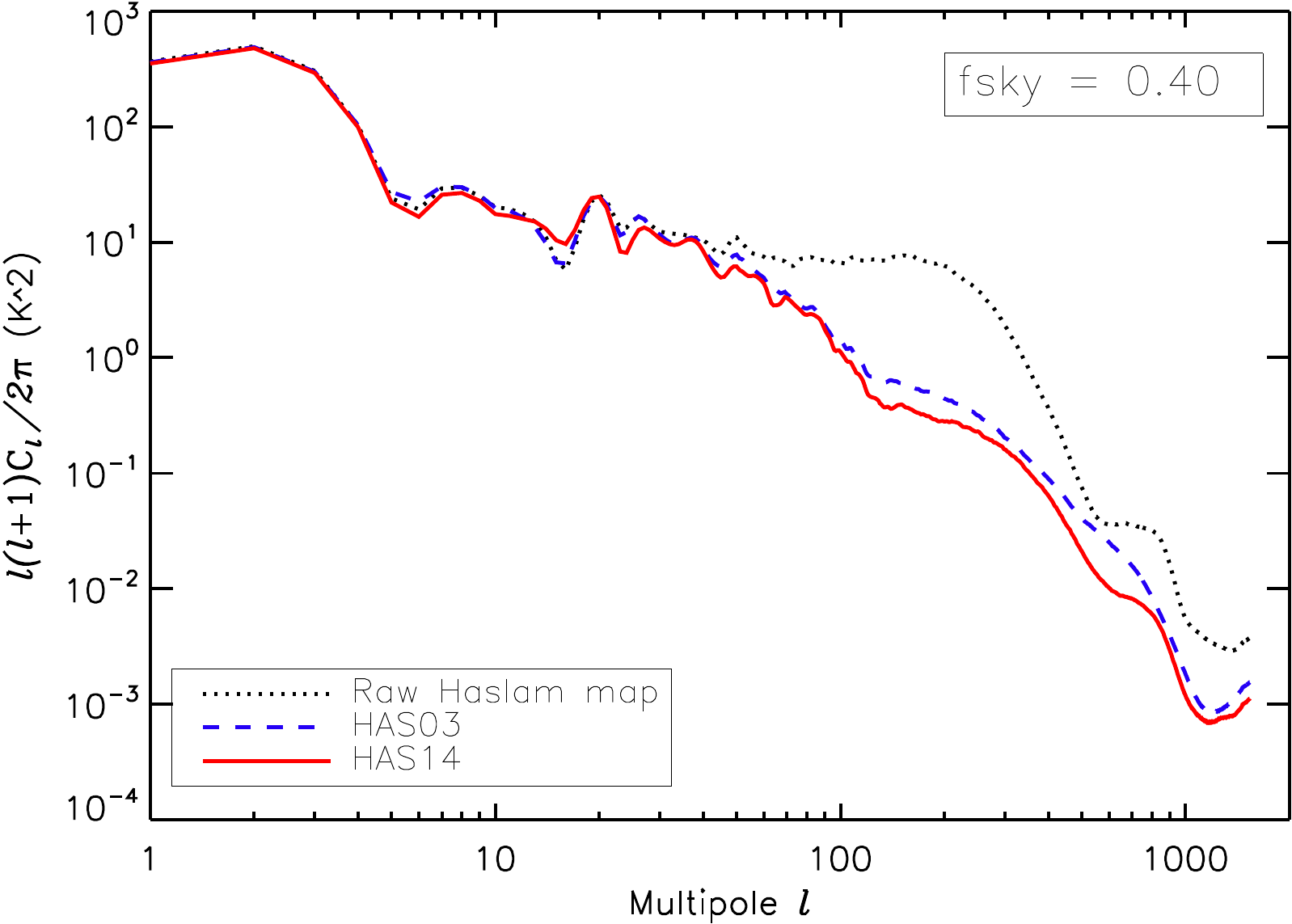}
    \includegraphics[width=\columnwidth]{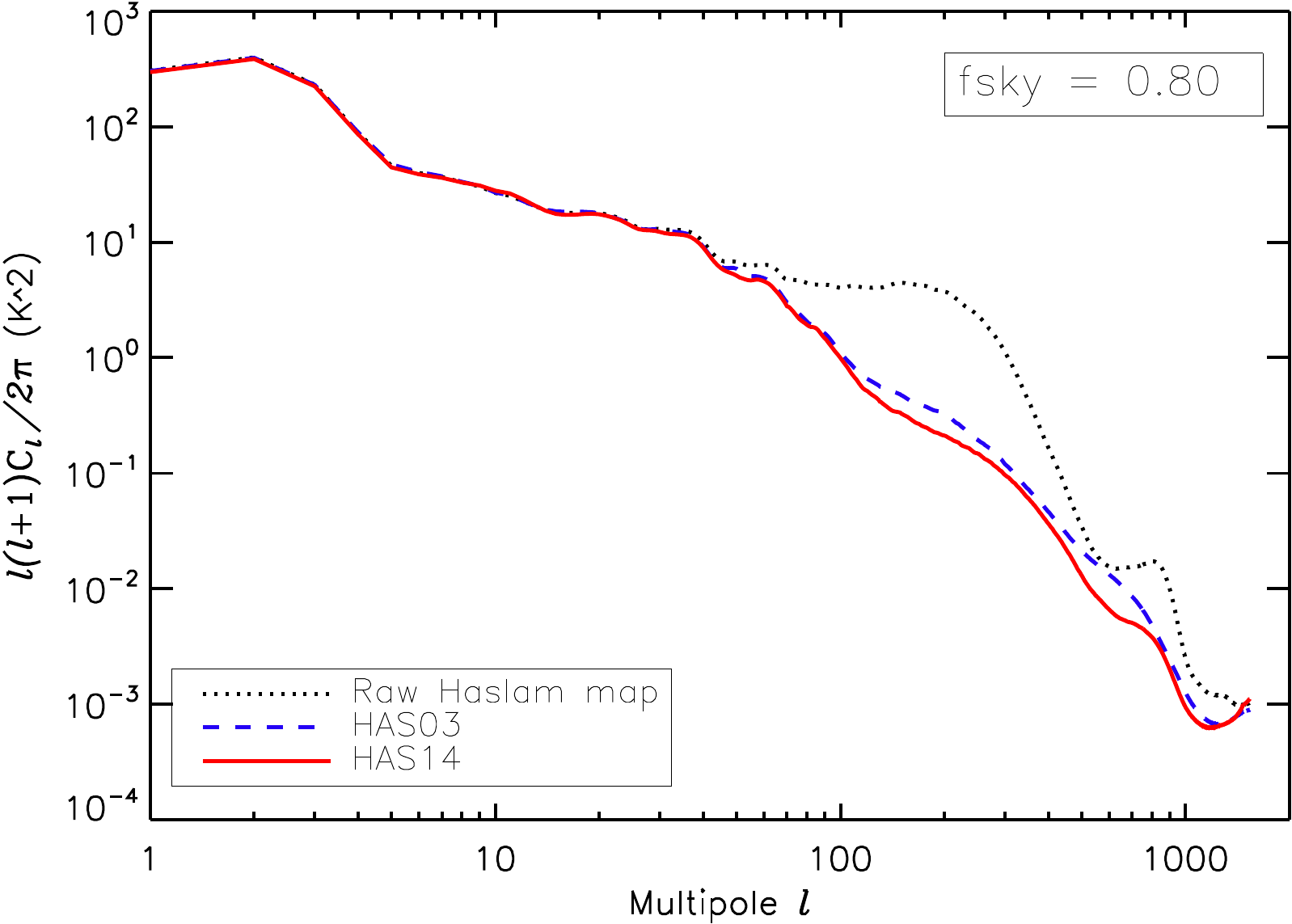}
    \includegraphics[width=\columnwidth]{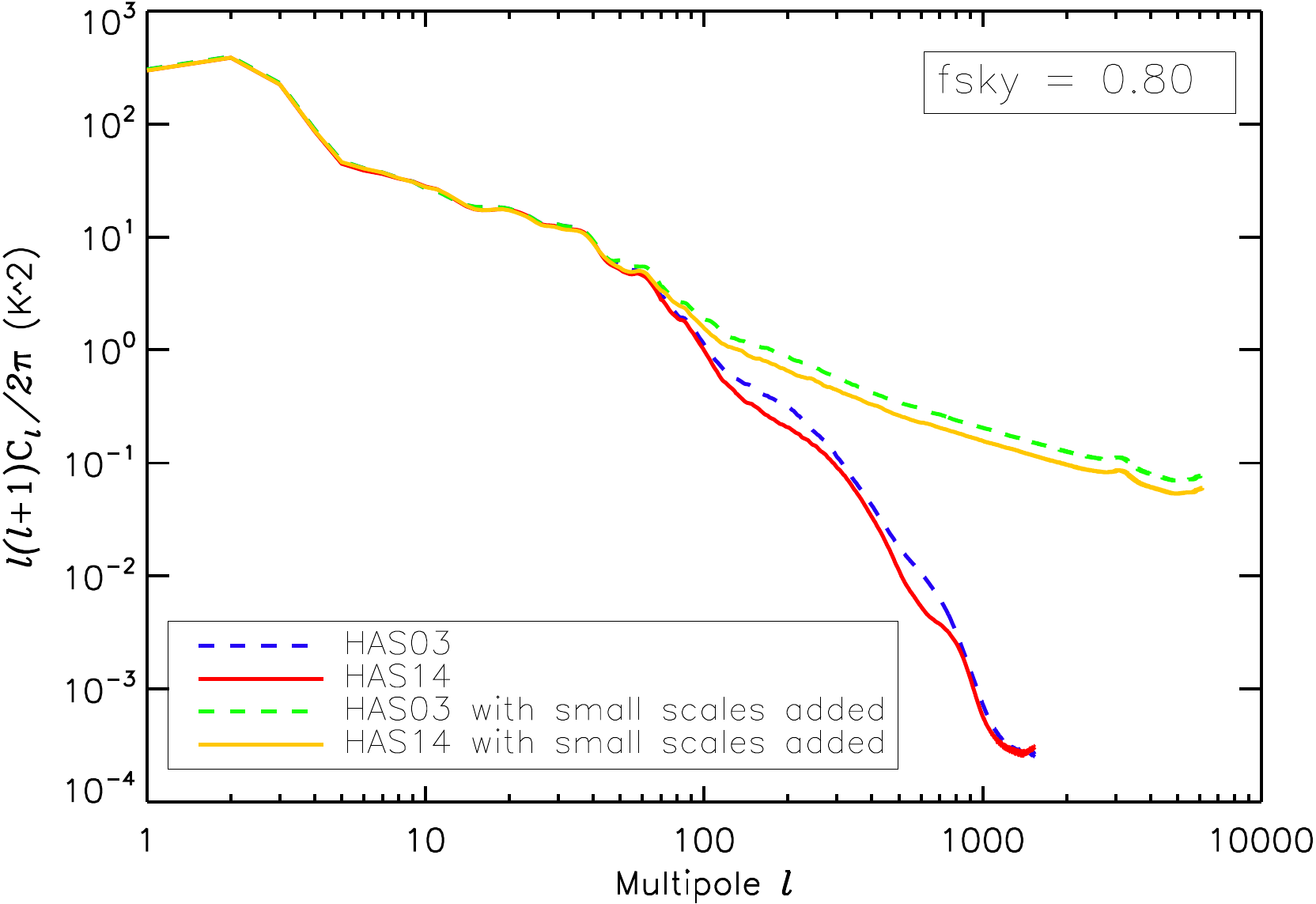}
 \end{center}
\caption{Angular power spectrum of $408$\,MHz sky map computed on both
  the HAS03 map (dashed blue line) and HAS14 map (solid red line) for different fractions of the sky:
  $f_{sky}=0.4$ ({\it top} panel) and $f_{sky}=0.8$ (middle panel). The
  power spectrum of the raw Haslam map is overplotted (dotted black
  line). {\it Bottom} panel: the power spectrum on $80$\% of the sky of the
  Haslam map with the small-scale fluctuations added.}
\label{Fig:pshaslam}
\end{figure}

\section{Discussion}\label{sec:discuss}

\subsection{Angular power spectrum}\label{ps} 

We compute the angular power spectrum of the Haslam map, both before
and after desourcing, using the {\tt PolSpice} code
on $80$\% and $40$\% of the
sky defined using intensity thresholds in the
HAS03 map. Rather than providing an accurate measure of the synchrotron power spectrum, our aim is to highlight, by doing a side-by-side comparison of the different versions of the Haslam map, how the imperfect desourcing generates a spurious excess of power at small scales in the angular power spectrum of the synchrotron template. In the top panels of Fig.~\ref{Fig:pshaslam} we
plot the power spectrum of the raw Haslam map, i.e. the unfiltered
LAMBDA version (dotted black), the power spectrum of the HAS03
destriped desourced map (dashed blue), and the power spectrum of our
newly reprocessed HAS14 map (solid red).  The excess of power at
$\ell \sim 200$ (beam scale) in the raw Haslam map (dotted black line)
is due to contamination by extragalactic radio sources.  At the smallest scales ($\ell > 500$), where the beam suppresses the power, there are additional features in the power spectrum, which could be due to beam sidelobes and/or artefacts from the various repixelization procedures that have been applied to the data. The power spectrum should be, to first approximation, corrected for by a Gaussian beam window function with \mbox{FWHM$=56$ arcmin}. However, at $\ell > 500$, the true window function is not well-modelled due to the repixelization effects. The
excess of small-scale power due to the source contamination in the raw
Haslam map (dotted black) is clearly suppressed in both versions of the filtered data.
However, the angular power spectrum of the HAS03 map still shows an excess of power on scales
smaller than the $56$ arcmin beam scale, i.e. for $\ell > 200$. This
excess power is due to numerous residual source
artefacts in the HAS03 map that we have mentioned earlier
in this paper (e.g., top left panel in
Fig.~\ref{Fig:lambda-artefacts}). Conversely, the angular power
spectrum of the newly reprocessed HAS14 map (solid red line) in the
top panels of Fig.~\ref{Fig:pshaslam} shows less power at $\ell > 200$.

\subsection{Small-scale fluctuations} 

A high-resolution template of synchrotron emission would be
particularly useful for preparing for upcoming $21$\,cm experiments,
e.g. SKA and HI intensity mapping experiments, in particular to
allow the study of foreground cleaning techniques for such projects (see e.g. \citealt{Wolz2014,Shaw2014}). However, the $408$\,MHz sky map is limited in resolution to a beam
FWHM of $56$\,arcmin, thus cannot, by itself, be directly used as a template of
Galactic synchrotron emission for high-resolution sky
simulations. Moreover, there are still no full-sky astronomical data
sets corresponding to synchrotron emission on degree scales or
smaller. It is therefore useful to add small-scale fluctuations to the
Haslam map in order to produce a high-resolution template of
synchrotron for simulation studies. Our newly processed Haslam map
with such fluctuations added is available for these purposes.  This
high-resolution template is produced at $N_{\rm side} = 2048$\footnote{Higher $N_{\rm side}$ maps with small-scales added are available on request from the authors.} beaming smoothing effects removed (i.e. a FWHM of zero arcmin).

\begin{figure}
  \begin{center}
    \includegraphics[width=0.5\columnwidth]{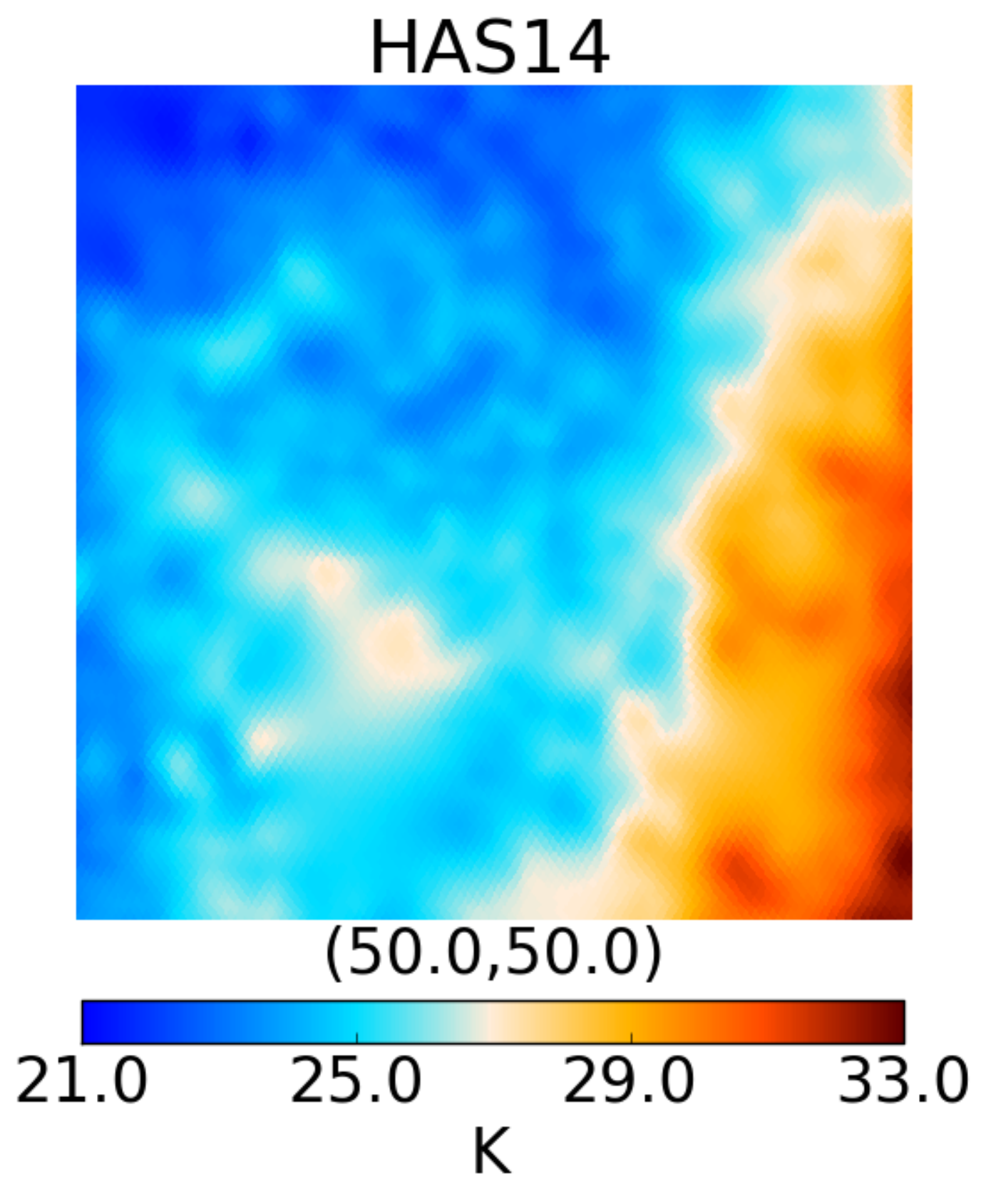}~
    \includegraphics[width=0.5\columnwidth]{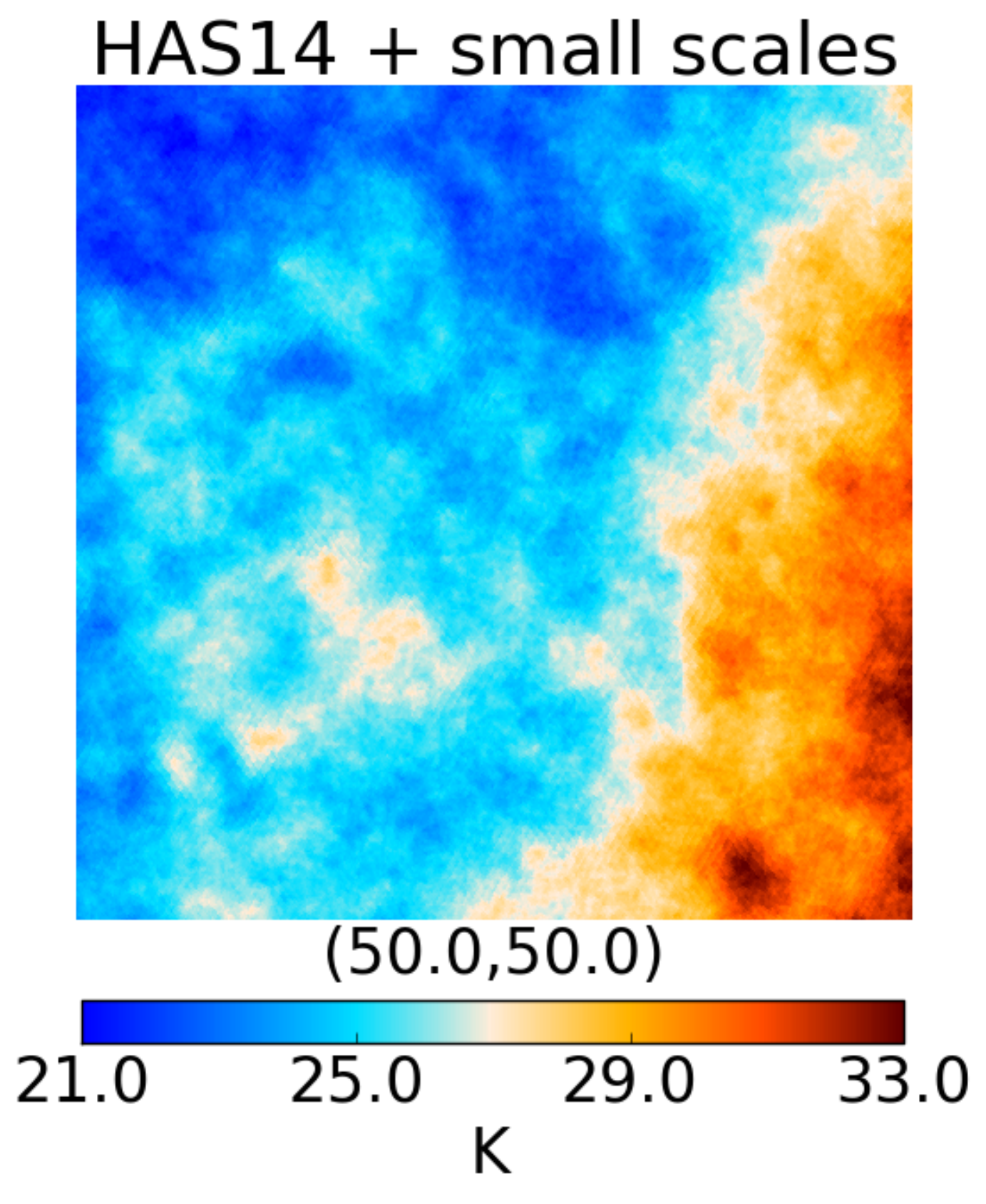}
 \end{center}
\caption{A $12.5^\circ\times 12.5^\circ$ gnomonic projection centred
  at $(l,b)=(50^{\circ}, 50^{\circ})$ showing the addition of
  small-scale fluctuations to the newly processed Haslam map. Left
  panel: the map at its native beam resolution. Right
  panel: the map with small-scale structure added.
}
\label{Fig:ss_map}
\end{figure}

To achieve this, we follow the approach used in
\citet{2013A&A...553A..96D}.  We first simulate a Gaussian random field $G_{ss}$
having the following power spectrum:
\begin{align}
\mathcal{C}_{\ell} = \ell^{\gamma}\left[1 - \exp\left(-\ell^2 \sigma_{\rm temp}^2\right)\right],
\end{align} 
where $\sigma_{\rm temp} = 56$\,arcmin is the original beam resolution of the Haslam map that we have estimated in Section~\ref{subsec:beam}. The index $\gamma$ is obtained by fitting a power-law to the power spectrum of the original Haslam map between $\ell = 30$ and $\ell = 90$. We find a best-fitting value of $\gamma = -2.703$, similar to values estimated by other analyses on similar angular scales ($\ell \lesssim 100$; \citealt{LaPorta2008}). However, we note that recent higher resolution observations suggest that the spectrum is flatter on smaller angular scales ($\ell \gtrsim 1000$; \mbox{\citealt{Bernardi2009}}). The Gaussian field $G_{ss}$ is whitened to have zero mean and unit variance, then multiplied by the template map $I_{\rm temp}$ as follows:
\begin{align}
I_{ss} = \alpha G_{ss} I_{\rm temp}^\beta.
\end{align} 
The small-scale fluctuations $I_{ss}$ are then added to the original
template $I_{\rm temp}$:
\begin{align}
I_{\rm temp}^{'} = I_{\rm temp} + I_{ss}.
\end{align}

The parameters $\alpha$ and $\beta$ in the small-scale map $I_{ss}$ have been determined such that the new template map with the small-scales added, $I_{\rm temp}^{'}$, conserves the same statistics (mean, variance, and skewness) as the original template map, $I_{\rm temp}$. The best fit is found for $\alpha = 0.0599$ and $\beta =0.782$. We constrained these two parameters in order to match the statistical moments of the final map $I_{\rm temp}^{‘}$ with the ones of the input map $ I_{\rm temp}$, after subtracting the small-scale fluctuation map by its mean and dividing by the standard deviation. This is slightly different from the approach used in \citet{2013A&A...553A..96D} where the
parameters $\alpha$ and $\beta$ were fixed for each template at a given value in the simulation.

The addition of small scales to the Haslam map is illustrated in
Fig.~\ref{Fig:ss_map}. The small-scale fluctuations can be seen and are superposed onto the large-scale structure defined by the Haslam map. The angular power spectrum of the Haslam map
with the small scales added is computed on $80$\% of the sky, and
plotted in the bottom panel of Fig.~\ref{Fig:pshaslam}. We generate
small scales both in the HAS03 map and in our newly processed HAS14 map. It is obvious from the bottom panel of the figure that the excess
power in the HAS03 map due to residual source artefacts generates
extra spurious power in the high-resolution template after the
addition of small scales (dashed green line). Any spurious power is
reduced when we generate small scales from our newly reprocessed
Haslam map, where the desourcing is much cleaner (solid yellow line).

Although we believe the HAS14 is more reliable due to the reduction in small-scale artefacts, we cannot be entirely sure that this is the correct power spectrum on these scales since our fitting procedure removes small-scale power. Furthermore, the power spectrum may depart from a simple power-law at high $\ell$- values.High-resolution observations of the diffuse synchrotron sky are needed to measure arcmin scales accurately (e.g., \citealt{Bernardi2009}).

\begin{table}
\caption{Template fit coefficients between the K-, Ka-, and Q-bands of \emph{WMAP} data and the $408$\,MHz data, used as a proxy for Galactic 
emission due to synchrotron radiation, in units of $\mathrm {\mu K\ K^{-1}_{408~MHz}}$. The last column lists the template fit coefficients between the $408$\,MHz data and the K-Ka difference map. The mean of the synchrotron coefficient is computed from all sky regions.}.
\label{tab:coeff_synch}
\scriptsize
\begin{tabular}{cccccc}
\hline
\hline
Region & Map & K &  Ka & Q & K-Ka \\
\hline
   1 &   HAS03   &     4.4$^{\pm     0.9 }$&     0.4$^{\pm     0.9 }$&    -0.1$^{\pm     0.9 }$&     2.4$^{\pm     0.1  }$\\
     &   HAS14   &     3.4$^{\pm     1.3 }$&    -0.5$^{\pm     1.3 }$&    -0.6$^{\pm     1.3 }$&     2.3$^{\pm     0.1  }$\\
   2 &   HAS03   &     6.9$^{\pm     0.7 }$&     3.4$^{\pm     0.8 }$&     2.1$^{\pm     0.8 }$&     3.8$^{\pm     0.0  }$\\
     &   HAS14   &     8.7$^{\pm     0.9 }$&     4.2$^{\pm     1.0 }$&     2.9$^{\pm     1.0 }$&     3.7$^{\pm     0.0  }$\\
   3 &   HAS03   &     5.4$^{\pm     0.4 }$&     1.6$^{\pm     0.4 }$&     0.8$^{\pm     0.4 }$&     3.0$^{\pm     0.0  }$\\
     &   HAS14   &     5.6$^{\pm     0.5 }$&     2.0$^{\pm     0.5 }$&     1.1$^{\pm     0.5 }$&     2.9$^{\pm     0.0  }$\\
   4 &   HAS03   &     6.1$^{\pm     1.3 }$&     5.3$^{\pm     1.9 }$&     3.0$^{\pm     1.8 }$&     4.0$^{\pm     0.2  }$\\
     &   HAS14   &     7.2$^{\pm     1.7 }$&     2.2$^{\pm     2.2 }$&     1.1$^{\pm     2.2 }$&     3.9$^{\pm     0.2  }$\\
   5 &   HAS03   &     4.1$^{\pm     1.9 }$&     4.5$^{\pm     2.1 }$&     4.8$^{\pm     2.6 }$&     3.1$^{\pm     0.6  }$\\
     &   HAS14   &     9.4$^{\pm     2.9 }$&     6.8$^{\pm     3.0 }$&     7.5$^{\pm     3.2 }$&     3.4$^{\pm     0.7  }$\\
   6 &   HAS03   &     3.9$^{\pm     0.9 }$&     1.3$^{\pm     1.0 }$&     1.0$^{\pm     1.0 }$&     3.3$^{\pm     0.1  }$\\
     &   HAS14   &     5.6$^{\pm     1.2 }$&     1.7$^{\pm     1.3 }$&     0.9$^{\pm     1.3 }$&     3.6$^{\pm     0.1  }$\\
   7 &   HAS03   &     4.9$^{\pm     0.9 }$&     0.7$^{\pm     1.1 }$&     0.0$^{\pm     1.0 }$&     3.2$^{\pm     0.1  }$\\
     &   HAS14   &     5.1$^{\pm     1.1 }$&     0.9$^{\pm     1.3 }$&     0.2$^{\pm     1.3 }$&     3.2$^{\pm     0.1  }$\\
   8 &   HAS03   &     5.6$^{\pm     0.5 }$&     2.4$^{\pm     0.6 }$&     1.1$^{\pm     0.6 }$&     4.5$^{\pm     0.0  }$\\
     &   HAS14   &     6.4$^{\pm     0.6 }$&     2.2$^{\pm     0.6 }$&     0.9$^{\pm     0.6 }$&     4.4$^{\pm     0.0  }$\\
   9 &   HAS03   &     7.8$^{\pm     0.9 }$&     3.1$^{\pm     0.9 }$&     2.3$^{\pm     1.0 }$&     2.2$^{\pm     0.1  }$\\
     &   HAS14   &     8.2$^{\pm     1.1 }$&     3.5$^{\pm     1.1 }$&     2.2$^{\pm     1.2 }$&     1.7$^{\pm     0.1  }$\\
  10 &   HAS03   &     3.8$^{\pm     1.0 }$&     3.7$^{\pm     1.4 }$&     3.0$^{\pm     1.4 }$&     4.1$^{\pm     0.1  }$\\
     &   HAS14   &     5.8$^{\pm     1.3 }$&     3.3$^{\pm     1.7 }$&     3.1$^{\pm     1.7 }$&     4.0$^{\pm     0.1  }$\\
  11 &   HAS03   &     2.8$^{\pm     0.7 }$&     0.2$^{\pm     0.8 }$&    -0.4$^{\pm     0.8 }$&     2.2$^{\pm     0.0  }$\\
     &   HAS14   &     4.7$^{\pm     1.0 }$&     1.2$^{\pm     1.1 }$&     0.4$^{\pm     1.1 }$&     2.4$^{\pm     0.0  }$\\
  12 &   HAS03   &     1.6$^{\pm     0.9 }$&     0.2$^{\pm     1.4 }$&     0.5$^{\pm     1.3 }$&     4.4$^{\pm     0.1  }$\\
     &   HAS14   &     4.9$^{\pm     1.5 }$&     0.3$^{\pm     1.7 }$&     0.3$^{\pm     1.7 }$&     4.1$^{\pm     0.1  }$\\
  13 &   HAS03   &     5.5$^{\pm     0.9 }$&     1.8$^{\pm     1.2 }$&     0.6$^{\pm     1.2 }$&     3.4$^{\pm     0.1  }$\\
     &   HAS14   &     6.1$^{\pm     1.2 }$&    -0.6$^{\pm     1.4 }$&    -1.3$^{\pm     1.4 }$&     3.3$^{\pm     0.1  }$\\
  14 &   HAS03   &     4.8$^{\pm     1.2 }$&    -1.4$^{\pm     1.3 }$&    -0.7$^{\pm     1.3 }$&    -0.8$^{\pm     0.1  }$\\
     &   HAS14   &     1.0$^{\pm     1.4 }$&    -1.2$^{\pm     1.5 }$&    -0.2$^{\pm     1.5 }$&    -1.0$^{\pm     0.1  }$\\
  15 &   HAS03   &     8.8$^{\pm     0.8 }$&     2.0$^{\pm     0.9 }$&     1.6$^{\pm     0.9 }$&     5.4$^{\pm     0.1  }$\\
     &   HAS14   &    21.1$^{\pm     0.9 }$&     8.2$^{\pm     0.9 }$&     5.6$^{\pm     0.9 }$&     6.7$^{\pm     0.1  }$\\
  16 &   HAS03   &     3.5$^{\pm     0.5 }$&     0.9$^{\pm     0.6 }$&    -0.1$^{\pm     0.6 }$&     3.9$^{\pm     0.0  }$\\
     &   HAS14   &     4.2$^{\pm     0.6 }$&     1.0$^{\pm     0.6 }$&    -0.0$^{\pm     0.6 }$&     3.8$^{\pm     0.1  }$\\
  17 &   HAS03   &     3.0$^{\pm     0.8 }$&     2.3$^{\pm     1.0 }$&     1.5$^{\pm     1.0 }$&     5.9$^{\pm     0.1  }$\\
     &   HAS14   &     4.0$^{\pm     1.0 }$&     2.5$^{\pm     1.2 }$&     1.5$^{\pm     1.1 }$&     5.9$^{\pm     0.1  }$\\
  18 &   HAS03   &     6.0$^{\pm     1.1 }$&     2.1$^{\pm     1.5 }$&     1.1$^{\pm     1.4 }$&     5.0$^{\pm     0.1  }$\\
     &   HAS14   &     7.2$^{\pm     1.3 }$&     3.8$^{\pm     1.7 }$&     3.0$^{\pm     1.7 }$&     4.8$^{\pm     0.1  }$\\
  19 &   HAS03   &     2.6$^{\pm     1.2 }$&    -1.2$^{\pm     1.9 }$&    -2.9$^{\pm     2.0 }$&     3.3$^{\pm     0.2  }$\\
     &   HAS14   &     1.8$^{\pm     1.8 }$&    -3.7$^{\pm     2.6 }$&    -4.4$^{\pm     2.7 }$&     3.2$^{\pm     0.2  }$\\
  20 &   HAS03   &     4.9$^{\pm     1.9 }$&     1.3$^{\pm     1.6 }$&     2.3$^{\pm     1.8 }$&    -0.3$^{\pm     0.2  }$\\
     &   HAS14   &    -0.5$^{\pm     2.5 }$&    -1.2$^{\pm     2.2 }$&     0.1$^{\pm     2.2 }$&    -1.6$^{\pm     0.2  }$\\
  21 &   HAS03   &     5.1$^{\pm     1.1 }$&     1.8$^{\pm     1.2 }$&     0.9$^{\pm     1.2 }$&     4.4$^{\pm     0.0  }$\\
     &   HAS14   &     3.2$^{\pm     1.4 }$&     0.5$^{\pm     1.5 }$&    -0.7$^{\pm     1.5 }$&     4.5$^{\pm     0.0  }$\\
  22 &   HAS03   &     4.3$^{\pm     1.1 }$&     2.5$^{\pm     1.4 }$&     2.4$^{\pm     1.5 }$&     4.0$^{\pm     0.2  }$\\
     &   HAS14   &     5.3$^{\pm     1.4 }$&     3.3$^{\pm     1.7 }$&     3.9$^{\pm     1.8 }$&     3.5$^{\pm     0.2  }$\\
  23 &   HAS03   &     3.0$^{\pm     1.2 }$&     2.2$^{\pm     1.4 }$&     2.0$^{\pm     1.4 }$&     5.0$^{\pm     0.1  }$\\
     &   HAS14   &     2.0$^{\pm     1.5 }$&     0.3$^{\pm     1.7 }$&     0.2$^{\pm     1.7 }$&     4.8$^{\pm     0.1  }$\\
  24 &   HAS03   &     3.1$^{\pm     0.9 }$&     1.1$^{\pm     1.2 }$&     0.2$^{\pm     1.2 }$&     1.3$^{\pm     0.1  }$\\
     &   HAS14   &     5.0$^{\pm     1.3 }$&     3.3$^{\pm     1.6 }$&     2.0$^{\pm     1.6 }$&     1.1$^{\pm     0.1  }$\\
  25 &   HAS03   &     3.5$^{\pm     1.2 }$&    -1.1$^{\pm     1.7 }$&     0.2$^{\pm     1.7 }$&     1.5$^{\pm     0.2  }$\\
     &   HAS14   &     3.7$^{\pm     1.8 }$&    -3.5$^{\pm     2.4 }$&    -1.8$^{\pm     2.5 }$&     1.3$^{\pm     0.3  }$\\
  26 &   HAS03   &     3.1$^{\pm     1.4 }$&    -0.1$^{\pm     2.1 }$&     0.1$^{\pm     2.3 }$&     3.8$^{\pm     0.3  }$\\
     &   HAS14   &     4.0$^{\pm     2.0 }$&    -1.3$^{\pm     2.7 }$&    -0.7$^{\pm     2.8 }$&     3.5$^{\pm     0.4  }$\\
  27 &   HAS03   &     3.8$^{\pm     1.4 }$&     1.7$^{\pm     1.9 }$&     0.5$^{\pm     1.9 }$&     3.5$^{\pm     0.3  }$\\
     &   HAS14   &     3.7$^{\pm     1.8 }$&     2.7$^{\pm     2.4 }$&     1.6$^{\pm     2.4 }$&     1.6$^{\pm     0.3  }$\\
  28 &   HAS03   &    -0.5$^{\pm     1.6 }$&    -0.6$^{\pm     1.6 }$&     1.1$^{\pm     2.0 }$&     3.3$^{\pm     0.5  }$\\
     &   HAS14   &    -1.4$^{\pm     2.5 }$&     1.1$^{\pm     2.4 }$&     0.7$^{\pm     2.8 }$&     1.2$^{\pm     0.5  }$\\
  29 &   HAS03   &     0.9$^{\pm     1.2 }$&    -1.4$^{\pm     1.8 }$&     2.4$^{\pm     2.0 }$&     2.6$^{\pm     0.5  }$\\
     &   HAS14   &    -1.5$^{\pm     2.8 }$&    -2.1$^{\pm     3.2 }$&     1.2$^{\pm     3.4 }$&     2.5$^{\pm     0.5  }$\\
  30 &   HAS03   &     6.5$^{\pm     1.7 }$&     6.1$^{\pm     2.1 }$&     4.6$^{\pm     2.3 }$&     4.3$^{\pm     0.3  }$\\
     &   HAS14   &     7.1$^{\pm     2.3 }$&     4.3$^{\pm     2.5 }$&     3.2$^{\pm     2.7 }$&     4.2$^{\pm     0.3  }$\\
  31 &   HAS03   &     1.4$^{\pm     1.1 }$&     1.4$^{\pm     1.4 }$&     1.0$^{\pm     1.4 }$&     1.4$^{\pm     0.2  }$\\
     &   HAS14   &     4.1$^{\pm     1.4 }$&     5.0$^{\pm     1.8 }$&     4.3$^{\pm     1.7 }$&     0.8$^{\pm     0.2  }$\\
  32 &   HAS03   &     5.2$^{\pm     0.6 }$&     2.4$^{\pm     0.6 }$&     1.2$^{\pm     0.6 }$&     4.3$^{\pm     0.0  }$\\
     &   HAS14   &     5.1$^{\pm     0.8 }$&     2.5$^{\pm     0.8 }$&     1.5$^{\pm     0.8 }$&     4.2$^{\pm     0.0  }$\\
  33 &   HAS03   &     4.4$^{\pm     1.0 }$&     2.8$^{\pm     1.2 }$&     2.8$^{\pm     1.2 }$&     4.4$^{\pm     0.1  }$\\
     &   HAS14   &     5.6$^{\pm     1.4 }$&     3.8$^{\pm     1.5 }$&     4.2$^{\pm     1.5 }$&     4.2$^{\pm     0.1  }$\\
  \hline
  Mean &   HAS03   &     4.3$^{\pm     0.3 }$&     1.6$^{\pm     0.3 }$&     1.2$^{\pm     0.3 }$&       3.3$^{\pm     0.3}$               \\
  &   HAS14   &     5.0$^{\pm     0.7 }$&     1.7$^{\pm     0.5 }$&     1.3$^{\pm     0.4 }$&     3.1$^{\pm     0.3}$                 \\
  \hline 
\end{tabular}
\end{table}

\subsection{Applications of the Haslam map} 

The Haslam $408$\,MHz map still remains as one of the few total-power
full-sky surveys made at radio frequencies. This map has been used
numerous times for studying a wide range of phenomena. A recent
example of this is the analysis of the power spectrum of diffuse
synchrotron emission by \cite{Mertsch2013}. They found that an
ensemble of shell supernova remnants could explain the shape of the
power spectrum at $\ell \sim 200$. Given that the residual source
artefacts add power on small-scales ($\ell \gtrsim 100$), this
analysis should be re-evaluated in light of the new data.

It has also been used extensively as a template for synchrotron emission in CMB
studies at microwave wavelengths. Although we do not expect a significant change
in any CMB results obtained using the new template (since the synchrotron
is relatively weak at frequencies $\gtrsim 30$\,GHz and the brightest
areas are typically masked), it may have some implications for the
inferred properties of the synchrotron emission at these frequencies.

We have performed an analysis following the standard template fit
approach from \citet{Davies2006}.
In particular, each of the 5 \wmap\ frequency maps is
fitted simultaneously with a constant offset and three external
templates, namely the $408$\,MHz survey, the \citet{Dickinson2003} (DDD)
H$\alpha$ survey, and the \citet{Finkbeiner1999} (FDS8) 94 GHz map,
corresponding respectively to tracers of the synchrotron,
free-free and dust emission. 
The \wmap\ data and the external templates are smoothed to a common Gaussian full-width half-maximum (FWHM) beam resolution of 1\deg and reprojected onto a {\tt HEALPix} map with a pixel resolution of $N_{\rm side}=128$. We then analyse the set of 33 sky regions defined in \citet{Ghosh2012}. We use full pixel-pixel noise covariance matrix for each region, taking into account 1\deg\ beam smoothing applied to the \wmap\ data. It contains contributions from the correlated CMB emission and the uncorrelated instrumental noise. 

We compare results derived using either the HAS03 map or the HAS14 map as the
synchrotron template, while keeping fixed the free-free and dust
templates. Any significant difference between the HAS03 and HAS14
templates at 1\deg\ resolution is evident in the cross-correlation
coefficients.

Table~\ref{tab:coeff_synch} presents the $408$\,MHz correlated (or
synchrotron) coefficients at K-, Ka- and Q-bands, expressed in antenna temperature units,
derived using the HAS03 and HAS14 templates. The synchrotron
coefficients for most of the regions are consistent at the 1-2$\sigma$
level, with the exception of region 15. This region lies close to the
Galactic plane where the Cygnus Loop, a large supernova remnant from our Galaxy located at $(l,b) = (74.0^\circ, -8.6^\circ)$, has been subtracted in the HAS03 template while it has been deliberately kept in our HAS14 template because it is an extended Galactic radio source. That is the reason why a discrepancy between HAS03 and HAS14 synchrotron coefficients is observed in region 15.

We do not find any significant
difference on the mean synchrotron coefficient, derived from all sky
regions, with respect to use of different $408$\,
MHz templates. Note that the uncertainties on the synchrotron coefficients are large
due to the presence of a dominant CMB cosmic variance term.

The synchrotron coefficients for a given sky region include a contribution from the random chance correlation between the CMB and the $408$\,MHz template. This term can either be positive, negative, or zero, but is constant in units of thermodynamic temperature. For the mean of the 33 sky regions, the CMB chance correlation term averages out, without introducing any bias on the mean synchrotron coefficient. To convert the coefficients into a synchrotron spectral index ($\beta_{\rm s}$), we  assume a simple single power-law model of the synchrotron emission that applies from $408$\,MHz to the WMAP frequencies. However, such conversion is not trivial due to the large correlated error bars introduced by the CMB emission and the presence of the CMB chance correlation term. Here we present an approach to circumvent these problems and derive $\beta_{\rm s}$ for each of the sky regions. 

We remove the CMB emission directly at the map level by taking the difference between the K- and Ka- bands in thermodynamic units. We choose to work with the K--Ka map as it has the highest signal-to-noise ratio in terms of the synchrotron emission. We apply the template-fit analysis on this K--Ka map in a similar fashion as discussed before. We add quadratically the instrumental noise coming from the K- and Ka- bands and include it in the noise covariance matrix. In this way, we remove the contribution of the correlated CMB uncertainty as well as the CMB chance-correlation term. The correlation plot between the synchrotron coefficients using K--Ka map with the use of HAS03 and HAS14 templates is shown in Fig. ~\ref{fig:corr_synch}. The error bars on the plot come from the instrumental noise contribution of the K- and Ka- bands. Most of the sky regions return consistent fit coefficients (or $\beta_{\rm s}$ values) using the two Haslam templates, except sky regions 9, 27, and 28, in which the HAS03 template still shows a large area of either unremoved or imperfectly removed extra-galactic radio sources. In general, the use of HAS14 template on average returns slightly lower synchrotron coefficients for the K--Ka map as compared to the HAS03 template. In terms of the template-fit analysis, both the Haslam templates provides similar synchrotron spectral indices at 1\deg\ resolution. 
Reassuringly, the change of the $408$\,MHz template does not significantly affect the dust or free-free coefficients.

 \begin{figure}
   \includegraphics[width=0.5\textwidth]{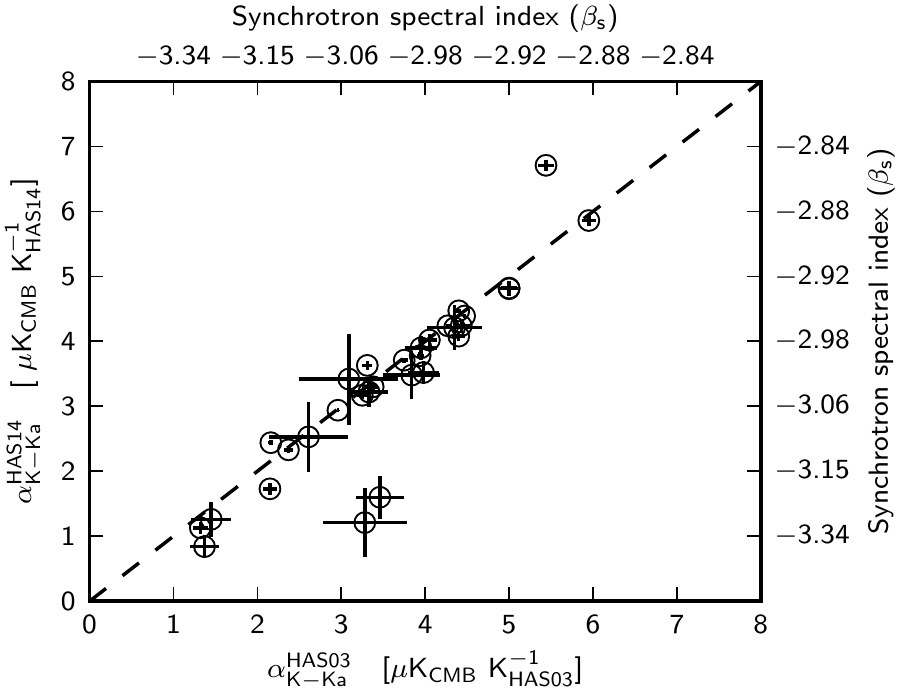} 
 \caption{The correlation plot between the synchrotron coefficients derived 
   using the K--Ka map, in thermodynamic units, with the two $408$\, MHz templates (HAS03 and HAS14). 
   Here the uncertainties on each synchrotron coefficient comes from the 
   uncorrelated instrumental noise present in K- and Ka- bands.
   A given coefficient is written as $\alpha_{\rm{K-Ka}}^{\rm{template}}$, where 
   $\rm{template}$ refers to the specific synchrotron template used.  
}
 \label{fig:corr_synch}
 \end{figure}

\subsection{Calibration and fidelity of the Haslam map}

The calibration of the Haslam map is not very well
understood. Furthermore, the map is often used without any knowledge
of some of the issues related to calibration, such as the zero-level,
overall temperature scale, and full-beam to main-beam
calibration. Here we briefly describe our current understanding of
these issues.

The $408$\,MHz \cite{Haslam1982} map has a quoted zero-level uncertainty
of $\pm 3$\,K. This is a significant uncertainty given that high
latitude areas have typical brightness temperatures of tens of K (the
minimum temperature is 11.4\,K). The definition of the zero-level in
this case includes the contribution of the CMB ($T_{\rm
  CMB}=2.73$\,K). A recent analysis by \cite{Wehus2014} have
re-estimated the zero-level in conjunction with a foreground model
fitted to {\it WMAP}/{\it Planck} and \cite{Reich2001} 1420\,MHz
maps, and determined a map monopole of $8.9 \pm 1.3$\,K, which would
include CMB and any other isotropic component (Galactic or
extragalactic). However, this result is somewhat dependent on the
foreground model and the data being fitted. We therefore leave the
zero-level untouched, with a minimum temperature on the
map of 11.4\,K.

The absolute calibration uncertainty is often quoted as $\pm 10\,\%$, based
on the scatter of correlation slopes between the $408$\,MHz data and the
absolutely calibrated survey of \cite{Pauliny-Toth1962}. However,
careful reading of the individual surveys suggests that the overall
calibration scale is better than this ($\sim 5\,\%$). The
major calibration uncertainty comes from the fact that the brightness
temperature scale is not constant as a function of angular scale. This
occurs because there is significant power outside the main beam (in
the sidelobes), which means that there is more sensitivity to large
angular scale emission (full beam calibration) while point-like
sources are only sensitive to the main beam (main beam
calibration); see \cite{Jonas1998} for a clear description and example of this effect. For some telescopes this can be as much as a $\sim
50\,\%$ correction \citep{Reich1986} with more typical values of $\sim
30\,\%$ \citep{Jonas1998}. Very carefully designed and
under-illuminated telescopes can reduce this to $\sim 10\,\%$
\citep{Holler2013}. For the \cite{Haslam1982} data these information
do not seem to be readily available. \cite{Haslam1974} discuss a
full-beam to main-beam factor of 1.05 up to a radius of $5^{\circ}$. However, this is unlikely to be
the case (it is too small). Since the calibration is based on the comparison of a large
area survey \citep{Pauliny-Toth1962} the calibration is expected to be
on the \lq\lq full-beam" scale, where this is somewhat arbitrarily
defined. This is the (approximately) correct scale for studying
diffuse synchrotron emission. For studying point sources such as
external galaxies, the brightness is likely to be
under-estimated.

\section{Conclusions}\label{sec:concl}

The $408$\,MHz all-sky map of \cite{Haslam1982} is one of the most
important radio surveys to-date. We have re-evaluated and quantified
its properties in detail. The positional accuracy is found to
be good to $\approx 7$\,arcmin. The effective beamwidth FWHM is
estimated to be $56.0\pm1.0$\,arcmin on average, significantly larger
than the claimed value of 51\,arcmin. This is presumably due to the
processing and gridding of the data into a pixelised sky map. 

We have re-processed the rawest data available to produce an improved
version of the Haslam map (HAS14). We have reduced the effect of scanning
artefacts, visible as striations in Declination, using a Fourier-based
filtering technique. The amplitude of the striations have been reduced
by a factor of $\approx 5$.

The most important improvement over the widely-used HAS03 version is
in the removal of extragalactic sources. This map contains obvious
artefacts around bright sources, which add small-scale power to the
map. We used a combination of Gaussian fitting and inpainting
techniques to remove the brightest ($>2$\,Jy) from the map. The
results are visually appealing and we estimate that we have removed
the majority of sources above 2\,Jy at high Galactic latitudes.

We have made our new and improved Haslam HAS14 map publicly
available\footnote{The new Haslam maps can be downloaded from
  \url{http://lambda.gsfc.nasa.gov/product/foreground/2014_haslam_408_info.cfm} or alternatively,
  \url{http://www.jb.man.ac.uk/research/cosmos/haslam_map/}. A bit mask of the removed extra-galactic sources is also released.}. This new map will be useful for
detailed studies of the Galactic synchrotron radiation, CMB component
separation and HI intensity mapping experiments. A limitation of the
Haslam map is the relatively low angular resolution ($\sim 1^{\circ}$)
which is not good enough for high resolution CMB and intensity mapping
experiments. For simulation work, we have also produced a map at
higher resolution, by artificially adding small-scale fluctuations. We
assume a power-law for the angular power spectrum for the synchrotron
emission and extrapolate this to the pixel scale, in this case {\tt
  HEALPix} $N_{\rm side}=2048$ ($\approx 1.7$\,arcmin).

\section*{Acknowledgements}
MR acknowledges support from the ERC Grant no. 307209. CD acknowledges funding from an STFC Advanced Fellowship, an EU Marie-Curie IRG grant under the FP7, and an ERC Starting Grant (no.~307209). TG acknowledges support from the MISTIC ERC Grant no. 267934. We thank the anonymous referee for useful comments on the paper which helped to clarify several issues. We thank Janet Weiland of the {\it WMAP} team for providing the code used to produce the LAMBDA HAS03 version of the $408$\,MHz map, which we use for destriping. We thank Paddy Leahy and Rod Davies for useful discussions about the fidelity of the Haslam map. The authors acknowledge the use of the {\tt HEALPix} package \citep{gorski05} and IDL astronomy library. This research has made use of the NASA/IPAC Extragalactic Database (NED) which is operated by the Jet Propulsion Laboratory, California Institute of Technology, under contract with the National Aeronautics and Space Administration. We also made extensive use of the SIMBAD database, operated at CDS, Strasbourg, France.

\bibliography{optimization_408MHz_mr}
\bsp


\label{lastpage}

\end{document}